\documentclass[nohyper]{JHEP3}

\usepackage{cite}
\usepackage{epsfig}

\newcommand{\mycomm}[1]{\hfill\break $\phantom{a}$\kern-3.5em{\tt===$>$ \bf#1}\hfill\break}
\newcommand{\mycommA}[1]{\hfill\break $\phantom{a}$\kern-3.5em{\tt   $>$ \bf#1}\hfill\break}
\newcommand{\be}{\begin{equation}}
\newcommand{\ee}{\end{equation}}
\newcommand{\ba}{\begin{eqnarray}}
\newcommand{\ea}{\end{eqnarray}}

\def\eq#1{Eq.~(\ref{#1})}

\def\MSbar{\hbox{\tiny ${\overline{\rm MS}}$}}

\def\tot{\hbox{\tiny total}}
\def\opt{\hbox{\tiny opt}}
\def\tHooft{\hbox{\tiny 't Hooft}}

\def\eff{\hbox{\tiny eff}}
\def\PT{\hbox{\tiny PT}}
\def\LO{\hbox{\tiny LO}}
\def\NLO{\hbox{\tiny NLO}}
\def\PV{\hbox{\tiny PV}}

\def\RS{\hbox{\tiny RS}}

\catcode`\@=11 
\def\lsim{\mathrel{\mathpalette\@versim<}}
\def\gsim{\mathrel{\mathpalette\@versim>}}
\def\@versim#1#2{\vcenter{\offinterlineskip
        \ialign{$\m@th#1\hfil##\hfil$\crcr#2\crcr\sim\crcr } }}
\catcode`\@=12 

\title{Taming the $\bar{B}\longrightarrow X_s\gamma$ spectrum by Dressed Gluon Exponentiation}

\author{Jeppe R. Andersen  and  Einan Gardi\\
Cavendish Laboratory, University of Cambridge\\
Madingley Road, Cambridge, CB3 0HE, UK}

\abstract{We show that the $\bar{B}\longrightarrow X_s\gamma$ photon
energy ($E_\gamma$) spectrum can be reliably computed by resummed
perturbation theory. Our calculation is based on Dressed Gluon
Exponentiation (DGE) incorporating Sudakov and renormalon
resummation. It is shown that the resummed spectrum does not have
the perturbative support properties: it smoothly extends to the
non-perturbative region $E_\gamma>m/2$, where $m$ is the quark pole
mass, and tends to zero near the physical endpoint. The calculation
of the Sudakov factor, which determines the shape of the spectrum in
the peak region, as well as that of the pole mass, which sets the
energy scale, are performed using Principal--Value Borel summation.
By using the same prescription in both, the cancellation of the
leading renormalon ambiguity is respected. Furthermore, in computing
the Sudakov exponent we go beyond the formal
next--to--next--to--leading logarithmic accuracy using the
large--order asymptotic behavior of the series, which is accurately
determined from the relation with the pole mass. Upon matching the
resummed result with the next--to--leading order expression we
compute the spectrum, obtain its moments as a function of a minimum
photon energy cut, analyze sources of uncertainty and show that our
predictions are in good agreement with Belle~data. }


\preprint{Cavendish-HEP-05/05}

\begin{document}

\section{Introduction}

Inclusive measurements of radiative, $\bar{B}\longrightarrow
X_s\gamma$, and semileptonic decays in the B factories have a great
potential in putting stringent constraints on short--distance
physics beyond the Standard Model and in providing an accurate
determination of the CKM parameters. However, exploiting this
potential crucially depends on our ability to extrapolate from the
experimentally accessible kinematic domain to the full phase space.
This requires precise theoretical predictions for inclusive decay
spectra.

The main obstacle in the QCD calculation of inclusive decay
spectra~\cite{Neubert:1993um,Bigi:1993ex,Falk:1993vb,KS,Bauer:1997fe,Kagan:1998ym,
Leibovich:1999xf,Ligeti:1999ea,Bauer:2000ew,Bauer:2003pi,Bosch:2004th,UG,Aglietti:2004fz,Neubert:2004dd,Benson:2004sg}
--- as opposed to total rates~--- is their sensitivity to the
momentum distribution of the heavy quark in the
meson\cite{Neubert:1993um,Bigi:1993ex}, which is defined by
\begin{equation}
\label{definition}
   f(z,M;\mu)=\int_{-\infty_{\,}}^{\infty}\frac{dy^{-}}{4\pi}\, {\rm
   e}^{-iz\,P^{+}y^{-}}\,
   \left< B(P) \right\vert\bar{\Psi}(y) \Phi_{y} (0,y) \gamma_{+} \Psi(0)
    \left\vert B(P)
    \right>_{\mu};\qquad\ P^2=M^2,
\end{equation}
where $M$ is the meson mass, $\mu$ is the renormalization scale of
the operator, $y$ is a lightlike vector in the ``$-$'' direction and
$\Phi_{y} (0,y)$ is a path--ordered exponential in this direction
connecting the points $y$ and $0$. Being a property of the bound
state, this distribution is of course non-perturbative. Although it
can be analyzed using the Operator Product Expansion (OPE) and
written as an infinite sum over forward matrix elements of local
operators, quantitative information on this distribution is very
limited.

The quark distribution function is directly relevant in the
experimentally most accessible region near the endpoint of the
spectrum, where the invariant mass of the hadronic jet is small. In
$\bar{B}\longrightarrow X_s\gamma$ decays this corresponds to the
photon energy $E_{\gamma}$ (measured in the $B$ rest frame) being
close to its maximal value $M/2$. This limit is inherently important
as the leading--order partonic process corresponds to a photon and
an s-quark recoiling back to back, with $E_\gamma=m/2$ ($m$ is the
quark pole mass). This results in a $\delta(1-x)$ distribution,
where $x\equiv 2E_{\gamma}/m$. This distribution is smeared by
perturbative and non-perturbative effects but it still peaks near
$x=1$. Because of the singular nature of this distribution, it is
convenient to consider the photon--energy moments. The perturbative
expansion of these moments is well defined to any order in
perturbation theory, but it is dominated by Sudakov logs, $\ln N$
($N$ is the moment index) and it therefore requires resummation. The
dominant non-perturbative contributions, which are associated with
the momentum distribution function of the heavy quark in the meson,
appear as powers of~$N\Lambda/m$. This parametric enhancement of
perturbative and non-perturbative contributions at large $N$ is
important even if one considers only the first few moments that are
measured experimentally: the information encoded in high moments is
absolutely essential to recover the correct dependence of the
partial decay rate on the minimal photon energy cut $E_\gamma>E_0$.
Such a cut is experimentally unavoidable.

Being motivated by the OPE, most theoretical approaches to decay
spectra have relied on introducing a {\em factorization scale},
which is either a hard cutoff or a dimensional--regularization
scale, distinguishing between soft interaction that is associated
with the momentum distribution in the meson and harder interaction
that depends on the details of the decay process at hand. While the
latter is described by perturbation theory the former can be
parametrized given sufficient experimental constraints. This
analysis is analogous to that of deep inelastic structure functions,
but the analogy is restricted to a certain kinematic domain near the
endpoint; moreover, it is incomplete since decay spectra contain an
additional source of double logarithmic corrections that is absent
in structure functions. Unfortunately, in practice the choice of
factorization scale, and even more so the procedure by which
factorization (the separation between the perturbative and
non-perturbative regimes) is implemented, strongly affects the final
answer for observable quantities. A good example is provided by the
first few moments of the photon energy in $\bar{B}\longrightarrow
X_s\gamma$ decays with experimentally relevant cuts, where the
factorization prescription has been the center of a long lasting
controversy, see e.g.~\cite{Neubert:2004dd,Benson:2004sg}.

In our approach~\cite{BDK,Gardi:2004gj} separation between
perturbative and non-perturbative corrections is implemented without
introducing any factorization scale. Instead, we strongly rely on
the resummation of the perturbative expansion. The starting point is
the fact that at the partonic level, moments of decay spectra are
infrared and collinear safe; infrared sensitivity (confinement
effects) shows up only through power corrections. Applying Dressed
Gluon Exponentiation
(DGE)~\cite{DGE_thrust,Gardi:2002bg,Gardi:2001di,CG,BDK} we resum
Sudakov logarithms as well as running--coupling effects. The
conceptual step we make is to regard the perturbative calculation of
the Sudakov exponent (in moment space) as an asymptotic series. This
series is summed using Principal--Value Borel summation\footnote{It
was shown in Ref.~\cite{Gardi:1999dq} that Principal--Value Borel
summation is equivalent in principle to a hard cutoff on some
Euclidean momentum. On the other hand, it makes much less dramatic
an impact on the distribution. In particular, it is closer to
truncating the perturbative expansion at the minimal term.}. This
amounts to parameter--free power-like separation between
perturbative and non-perturbative corrections. Non-perturbative
parameters controlling powers of $N\Lambda/m$ are then {\em defined}
in this prescription, and they loose their immediate interpretation
as local matrix elements\footnote{The locality measure in a hard
cutoff based approach is the scale. Here there is no such measure.}.
This, however, does not alter the fact that these corrections are
associated with the quark distribution function nor does it
undermine their universal nature. On the other hand, it typically
makes them numerically small compared to any cutoff--based
separation.

Sudakov resummation in inclusive decay spectra~\cite{KS} closely
parallels threshold resummation in hard scattering
processes~\cite{Sterman,CT,CSS,CTTW}: perturbative corrections from
real gluon emission are singular in the soft and the collinear
limits; in inclusive observables such as the spectral moments these
singularities cancel out by virtual corrections, leaving behind
finite but large contributions~$\sim C_{n,k} \, \alpha_s^n\, \ln^k
N$ with $1 \leq k\leq 2n$ at any order in perturbation theory. These
corrections exponentiate. Standard techniques facilitate identifying
their origin in phase space and then systematically resumming them
to a given logarithmic accuracy. In inclusive decays Sudakov
logarithms are associated with two independent subprocesses
(see~Refs.~\cite{KS,BDK}, and Fig.~1 in Ref.~\cite{Gardi:2004gj});
each of them can be defined and computed to all orders in a
process--independent manner:
\begin{enumerate}
    \item  $S_N(m;\mu)$, the {\em soft function}, which is the
    Sudakov factor of the
    {\em heavy quark distribution function}, summing up radiation
 with momenta ${\cal O}(m/N)$ --- often referred to
as the `soft scale' ---  that influences the momentum of the heavy
quark prior to its decay. This function is defined by considering
the $z\longrightarrow 1$ singular terms in
\begin{equation}
\label{definition_PT}
   f_{\PT}(z,m;\mu)=\int_{-\infty_{\,}}^{\infty}\frac{dy^{-}}{4\pi}\, {\rm
   e}^{-iz\,p^{+}y^{-}}\,
   \left< b(p) \right\vert\bar{\Psi}(y) \Phi_{y} (0,y) \gamma_{+} \Psi(0)
    \left\vert b(p)
    \right>_{\mu};\qquad p^2=m^2,
\end{equation}
which is the perturbative analogue of \eq{definition} where the
external state $\left\vert b(p) \right>$ is an on-shell heavy quark.
Upon taking moments one obtains
\begin{equation}
F_N^{\PT}(m;\mu)\equiv \int_0^1dz f_{\PT}(z,m;\mu) z^{N-1}
\,=\,H(\alpha_s(m))\,S_N(m;\mu)+{\cal O}(1/N), \label{soft_mom_def}
\end{equation}
where $S_N(m;\mu)$ sums up the log-enhanced terms to all orders and
$H(\alpha_s(m))$ incorporates the finite terms at $N\longrightarrow
\infty$. See Ref.~\cite{QD} for further details.
\item $J_N(m,\mu)$, the {\em jet function},
summing up radiation that is associated with an unresolved
final--state quark jet of invariant mass squared ${\cal O}(m^2/N)$.
This function is directly related to the large-$x$ limit of deep
inelastic structure functions. See Ref.~\cite{GR} for further
details.
\end{enumerate}
The hard interaction mediating the decay is totally irrelevant for
the Sudakov factor.

Ref.~\cite{BDK} has generalized the large-$x$ factorization approach
 beyond the perturbative (logarithmic) level. It has
been shown that, when considered to all orders, the moments
corresponding to each of the above subprocesses contain infrared
renormalons. The quark distribution Sudakov factor $S_N$ has
renormalon ambiguities scaling as integer powers of $N\Lambda/m$,
while the jet function $J_N$ contains ones that scale as integer
powers of $N\Lambda^2/m^2$. This implies that certain power
corrections are inherent to these subprocesses. As soon as these
power corrections are important, Sudakov resummation ceases to be
purely perturbative. Ref.~\cite{BDK} has shown that
running--coupling effects constitute a significant contribution to
the Sudakov exponent, and that their resummation necessarily links
the calculation of the Sudakov factor with the parametrization of
power corrections. When treated separately each of these ingredients
is inherently ambiguous, but when combined, the ambiguities cancel
out.

The leading non-perturbative corrections to inclusive decays are
those associated with the quark distribution function,
\eq{definition}. The moments of this non-perturbative function can
be expressed as~\cite{BDK}:
\begin{equation}
F_N(M;\mu)\simeq
F_N^{\PT}(m;\mu)\times\exp\left\{-\frac{(N-1)\bar{\Lambda}}{M}\right\}\,\times\,{\cal
F}((N-1)\Lambda/M); \qquad \quad \bar{\Lambda}\equiv M-m,
\label{F_N_be_approx}
\end{equation}
where $F_N^{\PT}(m;\mu)$ are the moments of the {\em perturbative}
quark distribution in an {\em on-shell} heavy
quark,~\eq{definition_PT}, the exponential factor stands for the
``binding energy'' effect, and ${\cal F}$ sums up additional,
quark--mass independent non-perturbative power corrections on the
soft scale $M/N$ to all orders. Power corrections on the hard scale
$m$ are neglected here.

The exponential factor in \eq{F_N_be_approx} has an important role
in computing decay spectra: it converts spectral moments from
partonic kinematics, where moments are defined with respect to
powers of $x=2E_{\gamma}/m$, to hadronic kinematics. It is well
known that the pole mass $m$ has an infrared  renormalon
corresponding to a linear ${\cal O}(\Lambda)$
ambiguity~\cite{Beneke:1994sw,Bigi:1994em}. In Ref.~\cite{BDK} it
was shown that this ambiguity cancels out between the exponential
factor $\exp\left\{-{(N-1)\bar{\Lambda}}/{M}\right\}$ and the
Sudakov factor $S_N$ in the perturbative quark distribution
$F_N^{\PT}(m;\mu)$. Thus, in the product in \eq{F_N_be_approx} one
recovers an unambiguous result for the quark distribution in the
meson. In this work we make use of this observation in two ways:
\begin{itemize}
\item{} When computing the Sudakov factor we use the Principal--Value
Borel sum. The same prescription is then implied for
$\bar{\Lambda}$. This is implemented here when computing the pole
mass $m$ from the short distance mass $m_{\MSbar}$.
\item{} Our approximation for the Borel function in the Sudakov factor
is improved based on the observation that its ambiguity must cancel
against that of the pole mass: the known large--order behavior of
the relation between the pole mass and $m_{\MSbar}$ is used to fix
the large--order behavior of the Sudakov factor and in this way
improve its determination beyond what is known from fixed--order
calculations.
\end{itemize}

The main purpose of the present paper is to provide a DGE--based
prediction for the photon--energy spectrum in
$\bar{B}\longrightarrow X_s\gamma$ decays, which can be confronted
with experimental data and provide a baseline for parametrization of
power corrections when the data become precise enough. As opposed to
previous attempts to describe the spectrum, here we refrain from
making any arbitrary parametrization of non-perturbative
corrections. In \eq{F_N_be_approx} we use ${\cal F}((N-1)\Lambda/M)
=1$. We only account for the ``binding--energy'' effect through
$\bar{\Lambda}=M-m$, where the pole mass is {\em computed} from
$m_{\MSbar}$. Barring the uncertainty in modeling perturbative
corrections to all orders (see below), our prediction for the
spectrum depends {\em only} on $\alpha_s$ and on the quark
short--distance mass. The total rate depends on additional
parameters such as $M_W$ and the charm--quark mass, but the
uncertainty in these parameters makes a negligible effect on the
distribution.

Recent progress in perturbative calculations~\cite{MVV,QD}
facilitates computing the Sudakov factor with
next--to--next--to--leading logarithmic (NNLL) accuracy. We
therefore begin our study in Sec.~\ref{sec:NNLL} by presenting the
NNLL results. We investigate the convergence of the perturbative
expansion with increasing logarithmic accuracy and confirm the
prediction of Ref.~\cite{BDK} that this expansion breaks down early.
Then, in Sec.~\ref{sec:Borel_rep_DGE} we recall the formulation of
the Sudakov exponent as a Borel sum and review the results obtained
in Ref.~\cite{BDK} for the Borel function in the large--$\beta_0$
limit. In Sec.~\ref{sec:residue} we address one of the central
issues of this paper, namely the construction of an approximate
Borel function for the exponent that incorporates the exact analytic
functions in the large--$\beta_0$ limit on the one hand, and has the
exact expansion coefficients in the full theory to the NNLO on the
other. We examine in detail the sensitivity of the spectrum to the
assumptions made on the Borel function away from the origin and
incorporate the constraint on the large--order behavior of the
exponent, which we determine from the relation with the pole mass.
Sec.~\ref{sec:matching} deals with the matching of the resummed
Sudakov factor to the full NLO perturbative result that incorporates
contributions from all the relevant short--distance operators. In
Sec.~\ref{sec:PC} we address non-perturbative power corrections.
Here we compute the pole mass $m_{\PV}$ from the short--distance
mass and use it to translate the perturbative moments defined with
respect to $x=2E_{\gamma}/m_{\PV}$ into the spectrum in physical
photon--energy units. We end up with a parameter--free prediction
for the spectrum, which includes all that is currently known about
the $\bar{B}\longrightarrow X_s\gamma$ spectrum in perturbation
theory. In~Sec.~\ref{sec:pheno} we present our numerical results for
the spectrum, compute its moments as a function of the
photon--energy cut and analyze sources of uncertainty. Finally, we
compare the predictions with the available data from Belle.

\section{The Sudakov exponent by Dressed Gluon Exponentiation\label{sec:Sud_DGE}}

\subsection{Sudakov resummation with fixed logarithmic accuracy\label{sec:NNLL}}

Sudakov resummation is based on the exponentiation of
logarithmically--enhanced terms in moment space.  The spectral
moments of the $b\longrightarrow X_s\gamma$ decay occurring through
the magnetic--operator interaction~$O_7$ (see \eq{O7} below) can be
expressed as
\begin{eqnarray}
\label{BXg_logs_N_2loop} \bar{M}_N^{\PT, O_7}\equiv \int_0^1
dx\,\frac{1}{\Gamma_{\tot}^{O_7, \PT}}\frac{d\Gamma^{O_7,
\PT}(x)}{dx}x^{N-1}\, =\,C_N^{O_7}(\alpha_s(m))\,\times\,{\rm
Sud}(m,N),
\end{eqnarray}
where ${d\Gamma^{O_7, \PT}(x)}/{dx}$ is the perturbative
distribution in $x\equiv 2E_{\gamma}/m$, the superscript PT stands
for `perturbative', meaning in particular that the initial state is
an on-shell heavy quark with mass $m$, and the bar indicates that
the distribution is normalized by $\Gamma_{\tot}^{O_7, \PT}$ so
$\bar{M}_{N=1}\equiv 1$. In \eq{BXg_logs_N_2loop} all the
log--enhanced terms are resummed into the Sudakov factor ${\rm
Sud}(m,N)$. After the resummation has been performed, one can
determine the spectrum from the moments by inverting the Mellin
transform:
\begin{equation}
\label{inv_Mellin} \frac{1}{\Gamma_{\tot}^{O_7,
\PT}}\frac{d\Gamma^{O_7, \PT}(x)}{dx}= \int_{{\cal C}}\frac{dN}{2\pi
i} \,\,x^{-N}\,\, \bar{M}_N^{\PT, O_7},
\end{equation}
where the contour ${\cal C}$ extends from $c-i\infty$ to $c+i\infty$
to the right of the singularities of the integrand.

The Sudakov factor ${\rm Sud}(m,N)$ is independent of the details of
the hard interaction. Therefore, other operators contributing to the
$b\longrightarrow X_s\gamma$ process staring at NLO share the same
Sudakov factor. Their contribution will be taken into account in
Sec.~\ref{sec:matching} and in the numerical analysis; for
simplicity, in this section we consider only $O_7$.

Based on previous analysis of large-$x$ factorization in
$b\longrightarrow X_s\gamma$ \cite{KS,BDK} we know that the Sudakov
resummation formula takes the form\footnote{${\widetilde{{\rm
Sud}}}(m,N)$ differs from ${\rm Sud}(m,N)$ in \eq{BXg_logs_N_2loop}
and in Eqs. (\ref{sum_FLA}) and (\ref{Sud}) below just by finite
terms and terms that vanish in the $N\longrightarrow \infty$ limit.
These terms will eventually be discarded from the Sudakov factor and
included in the matching coefficient $C_N(\alpha_s(m))$.}:
\begin{eqnarray}
\label{Sud_tilde} {\widetilde{{\rm
Sud}}}(m,N)&=&\exp\bigg\{\int_0^1dx
\frac{x^{N-1}-1}{1-x}\,\bigg[\int_{(1-x)^2m^2}^{(1-x)m^2}\frac{d\mu^2}{\mu^2}
{\cal A}\Big(\alpha_s(\mu^2)\Big)\nonumber \\&&\hspace*{40pt}+{\cal
B}\Big(\alpha_s((1-x)m^2)\Big) -{\cal
D}\Big(\alpha_s((1-x)^2m^2)\Big)\bigg]
 \bigg\}.
\end{eqnarray}
This all--order formula depends on three anomalous dimensions:
${\cal A}$ is the universal cusp anomalous
dimension~\cite{KR87,Korchemsky:1988si,KM}, which is also the
large-$x$ limit of the quark--quark splitting function; ${\cal B}$
is the Sudakov anomalous dimension associated with an unresolved
quark jet with a given invariant mass, which (in combination with
${\cal A}$) also determines the large-$x$ limit of deep inelastic
structure functions~\cite{Sterman, CT,GR}; and ${\cal D}$ is the
Sudakov anomalous dimension that is associated with the momentum
distribution in an on-shell heavy quark~\cite{QD}. Each anomalous
dimension can be expanded in $\alpha_s^{\MSbar}$ as follows:
\begin{eqnarray}
\label{A_cusp_expansion} {\cal A}\big(\alpha_s(\mu^2)\big)&=&
\sum_{n=1}^{\infty}
A_n\left(\frac{\alpha_s^{\MSbar}(\mu^2)}{\pi}\right)^n,\nonumber
\\
{\cal B}\big(\alpha_s(\mu^2)\big)&=& \sum_{n=1}^{\infty}
B_n\left(\frac{\alpha_s^{\MSbar}(\mu^2)}{\pi}\right)^n,\nonumber
\\
{\cal D}\big(\alpha_s(\mu^2)\big)&=& \sum_{n=1}^{\infty}
D_n\left(\frac{\alpha_s^{\MSbar}(\mu^2)}{\pi}\right)^n.
\end{eqnarray}
The first few orders in these expansions are known exactly. Higher
orders are known only in the large--$\beta_0$ limit~\cite{BDK}. They
will be discussed in the next section.

The known coefficients are the following. First, the cusp anomalous
dimension was recently computed to three--loop order~\cite{MVV}:
\begin{eqnarray}
\label{A_i} A_{1} &=& {C_F} \\ \nonumber A_{2} &=& \left[\left({
\frac {67}{36}} - { \frac {\pi ^{2}}{12}} \right)\,{C_A} - { \frac
{5\, {N_f}}{18}} \right]\,{C_F}
\\ \nonumber
A_{3} &=& \left({ \frac {1}{2}} \,\zeta \left(3\right) - { \frac
{55}{96}} \right)\,{N_f}\,{C_F}^{2} + \bigg[\left({ \frac {245}{96}}
+ { \frac {11}{24 }} \,\zeta \left(3\right) - { \frac {67\,\pi
^{2}}{216}} + { \frac {11\,\pi ^{4}}{720}} \right)\,{C_A}^{2}
\\ \nonumber && + \left(
 - { \frac {209}{432}}  - { \frac {7}{
12}} \,\zeta \left(3\right) + { \frac {5\,\pi ^{2}}{108}} \right)\,
{N_f}\,{C_A} - { \frac {{N_f}^{2} }{108}}\bigg]  {C_F}.
\end{eqnarray}
Second, the anomalous dimension of the jet function is known to
two--loop order from calculations in deep inelastic scattering, see
e.g. Refs.~\cite{Vogt:2000ci,GR}:
\begin{eqnarray}
\label{B_i} B_{1} &=&  - { \frac {3\,{C_F}}{4}} \\ \nonumber B_{2}
&=& \left( - { \frac {3}{32}} - { \frac {3}{2}} \,\zeta
\left(3\right) + { \frac {\pi ^{2}}{8}} \right) \,{C_F}^{2} +
\bigg[\left( - { \frac {3155}{864}} + { \frac {11\,\pi ^{2}}{72}} +
{ \frac {5}{2}} \,\zeta \left(3\right)\right)\,{C_A} \\ \nonumber
&&+ \left({ \frac { 247}{432}} - { \frac {\pi ^{2}}{36}}
\right)\,{N_f} \bigg]\,{C_F}.
\end{eqnarray}
Finally, the anomalous dimension appearing in the quark distribution
function in an on-shell heavy quark was recently computed to
two--loop order~\cite{QD}:
\begin{eqnarray}
\label{D_i} D_{1} &=& C_F
\\ \nonumber
D_{2} &=& \bigg[\left( - { \frac {55}{108}} + { \frac {9}{4}}
\,\zeta \left(3\right) - { \frac { \pi ^{2}}{12}} \right)\,{C_A} - {
\frac {{N_f} }{54}} \bigg]\,{C_F}.
\end{eqnarray}

These coefficients facilitate computing the Sudakov exponent with
NNLL accuracy. In order to obtain a fixed--logarithmic--accuracy
formula to this order we first express the running coupling in terms
of $\alpha_s^{\MSbar}(m^2)$,
\begin{eqnarray}
\frac{\alpha_s^{\MSbar}(\mu^2)}{\pi} &=&
\left(\frac{\alpha_s^{\MSbar}(m^2)}{\pi}\right)\frac{1}{1-\sigma}
-\left(\frac{\alpha_s^{\MSbar}(m^2)}{\pi}\right)^2\frac{\beta_1}{\beta_0}
\frac{\ln(1-\sigma)}{(1-\sigma)^2}+\left(\frac{\alpha_s^{\MSbar}(m^2)}{\pi}\right)^3\times
\\&&\hspace*{-40pt}\nonumber
\bigg[\left(\frac{1}{(1-\sigma)^3}-\frac{1}{(1-\sigma)^2}\right)
\frac{\beta_2^{\MSbar}}{\beta_0}
+\left(\frac{1}{(1-\sigma)^2}-\frac{1+\ln(1-\sigma)-\ln^2(1-\sigma)}{(1-\sigma)^3}\right)
\frac{\beta_1^2}{\beta_0^2}\bigg]+\cdots
\end{eqnarray}
where \[\sigma=\frac{\alpha_s^{\MSbar}(\mu^2)}{\pi}\beta_0\ln
\frac{m^2}{\mu^2},\] and
\begin{eqnarray}
\label{beta_i} \beta_0 &=&
\frac{11}{12}C_A-\frac{1}{6}N_f\\\nonumber
\beta_1 &=&
\frac{17}{24}C_A^2-\frac{1}{8}C_FN_f-\frac{5}{24}C_AN_f\nonumber
\\\nonumber
\beta_2^{\MSbar} &=& \frac{1}{64}C_F^2N_f +\left(-\frac{205
}{1152}C_AN_f+\frac{11}{576}N_f^2\right)C_F+\frac{2857}{3456}C_A^3+
\frac{79}{3456}C_AN_f^2-\frac{1415}{3456}C_A^2N_f,
\end{eqnarray}
 and then integrate the ${\cal A}$ term over
$\mu^2$. The integration over $x$ is then done (keeping only
logarithmically--enhanced terms to NNLL accuracy) using the general
formula:
\begin{eqnarray}
\int_0^1 dx \frac{x^{N-1}-1}{1-x}\,
F\left(\frac{\beta_0\alpha_s^{\MSbar}(m^2)}{\pi}\ln\frac{1}{1-x}\right)
&=& -\frac{\pi}{\beta_0\alpha_s^{\MSbar}(m^2)}\int_0^\lambda d\omega
F(\omega) \\ \nonumber &&\hspace*{-30pt}\,-\,\gamma_E
F(\lambda)-\frac12F^\prime(\lambda)
\left(\frac{\pi^2}{6}+\gamma_E^2\right)\frac{\beta_0\alpha_s^{\MSbar}(m^2)}{\pi}+\cdots,
\end{eqnarray}
where
\[
\lambda\equiv\frac{\alpha_s^{\MSbar}(m^2)}{\pi}\beta_0 \ln N.
\]
The resulting Sudakov factor is:
\begin{equation}
{\rm Sud}(m,N)=\exp\left\{ \sum_{n=0}^{\infty}
g_n(\lambda)\left(\frac{\alpha_s^{\MSbar}(m^2)}{\pi}\right)^{n-1}\right\}
 \label{sum_FLA},
\end{equation}
where the first three coefficients $g_{n}(\lambda)$, which sum up
the logarithms to NNLL accuracy, are\footnote{An expression for
$g_2(\lambda)$ was derived a few years ago~\cite{UG} although the
corresponding coefficients ($A_3$, $D_2$ and $B_2$) were not yet
known. We find however that the terms proportional to $A_1\,\beta_2$
in that expression (Eq.~(40) there) are incorrect.}:
\begin{eqnarray}
\label{g_i}
{g_{0}}(\lambda) &=&  \frac{A_{1}}{{\beta _{0}}^{2}}
{\left(\left( 1- \lambda\right)\,\ln\left(1 - \lambda  \right)
-\frac12 \left( 1 -2 \lambda \right) \,\ln\left(
1-2\lambda\right)\right)}
\\ \nonumber
{g_{1}}(\lambda) &=& { \frac{{A_{1} \gamma_E}}{\beta _{0}} \bigg( -
\,\ln\left( 1-\lambda\right) +  \,\ln\left( 1-2\lambda\right)\bigg)
+ {B_{1}}\,\ln\left( 1-\lambda\right) - { \frac {1}{2}}
\,{{D}_{1}}\,\ln\left( 1-2\lambda\right)}
\\ \nonumber &&  + {\frac {{A_{2}}}{{\beta _{0}}^{2}} \bigg( - \ln\left( 1-\lambda\right)
 + { \frac {1}{2}} \,\ln\left( 1-2\lambda\right)
\bigg)\,}  \\ \nonumber &&  + {\frac {{A_{1}}\,{\beta _{1}}}{{\beta
_{0}}^{3}} \left( - { \frac {1}{2}} \,\ln\left( 1-2\lambda\right) -
{ \frac {1}{4}} \,\ln\left( 1-2\lambda\right)^{2} + { \frac {1}{ 2}}
\,\ln\left( 1-\lambda\right)^{2} + \ln\left(
1-\lambda\right)\right)\,}
 \\ \nonumber
{g_{2}} (\lambda)&=& {A _{1}}\left({ \frac {1}{2}}  - { \frac {1}{
1-2\lambda}}  + { \frac {1}{ - 2\, \lambda + 2}}
\right)\,\left(\gamma_E ^{2} + \frac{\pi^2}{6}\right)
\\\nonumber &&+
{B_{1}\,\gamma_E }\left(1 - { \frac {1}{ 1-\lambda}} \right)\,
+ {{D}_{1}\,\gamma_E}\left({ \frac {1}{ 1-2\lambda} }  - 1\right) \, \\
\nonumber && + { \frac{1}{{\beta _{0}}} \left\{{ \left( { \frac
{1}{1-\lambda }} - { \frac {1}{1-2\,\lambda }}
\right)\,{A_{2}\gamma_E} + \left({    1-\frac {1}{1-\lambda  }}
\right)\,{B_{2}} + \left( - { \frac {1}{2}}  - { \frac
{1}{4\,\lambda  - 2}} \right)\,{{D}_{2}}}\right\} }  \\ \nonumber &&
+ \frac{1}{\beta_{0}^{2 }}\,\bigg\{\bigg[ \bigg({ \frac {
\,\ln\left( 1-\lambda\right)}{ \lambda - 1}} - { \frac {
\,\ln\left(1 - 2 \,\lambda \right)}{2\,\lambda - 1}} + \,{ \frac
{1}{\lambda - 1}}  - { \frac {1}{2\,\lambda
 - 1}} \bigg)\,{A_{1}\gamma_E} \\ \nonumber &&
 + \left( - { \frac {1}{\lambda  - 1}}  - { \frac {\ln\left(
1-\lambda\right)}{\lambda  - 1} } - 1\right)\,{B_{1}} + \left({
\frac {1}{2}}  + { \frac {1}{4\,\lambda - 2}}  + { \frac {1}{2}} \,{
\frac {\ln\left( - 2\,\lambda
 + 1\right)}{2\,\lambda  - 1}} \right)\,{{D}_{1}}\bigg]{\beta _{1}} \\ \nonumber &&
 + \left( - { \frac {1}{4}}  + { \frac {1}{8\,\lambda - 4}} -
{ \frac {1}{2\, \lambda  - 2}} \right)\,{A_{3}}\bigg\}
 \\
\nonumber && +\,\frac{1}{\beta _{0}^{3}}\,\bigg\{\left({ \frac
{\ln\left( 1-\lambda\right)}{\lambda  - 1 }} + { \frac
{3}{2\,\left(\lambda  - 1\right)}}  + { \frac {3}{4}}  - { \frac
{1}{2}} \, { \frac {\ln\left( 1-2\lambda\right)}{2\,\lambda
  - 1}}  - { \frac {3}{4\,\left(2\,\lambda  - 1\right)}} \right)\,{A
_{2}}\,{\beta _{1}} \\ \nonumber &&  + \left(\ln\left(
1-\lambda\right) - { \frac {1 }{2}} \,\ln\left( 1-2\lambda\right)
 - { \frac {1
}{4}}  - { \frac {1}{2\,\lambda  - 2}}  + { \frac {1}{8\,\lambda  -
4}} \right)\,{A_{1}}\,{\beta _{2 }}\bigg\}  \\
\nonumber &&  + \frac{{A_{1}}\,{\beta _{1}}^{2}}{{\beta
_{0}}^{4}}\,\bigg\{ - { \frac {1}{2}} \,{ \frac {\ln^{2}\left(
1-\lambda\right)}{\lambda  - 1}}  - { \frac {\ln\left(
1-\lambda\right)\,\lambda }{ \lambda  - 1}}  - { \frac
{1}{2\,\lambda - 2}}  + { \frac {1}{4}} \,{ \frac {\ln^{2}\left(
 1-2\lambda\right)}{2\,\lambda  - 1}}    \\
\nonumber && + { \frac {\ln\left( 1-2\lambda\right)\,\lambda
}{2\,\lambda  - 1 }} + { \frac {1}{8\,\lambda  - 4}}  - { \frac
{1}{4}} \bigg\},
\end{eqnarray}
where the coefficients of the anomalous dimensions and the $\beta$
function are in the ${\overline {\rm MS}}$ scheme. They are given by
Eqs. (\ref{A_i}) through (\ref{D_i}) and \eq{beta_i}, respectively.
\begin{figure}[th]
\begin{center}
\epsfig{file=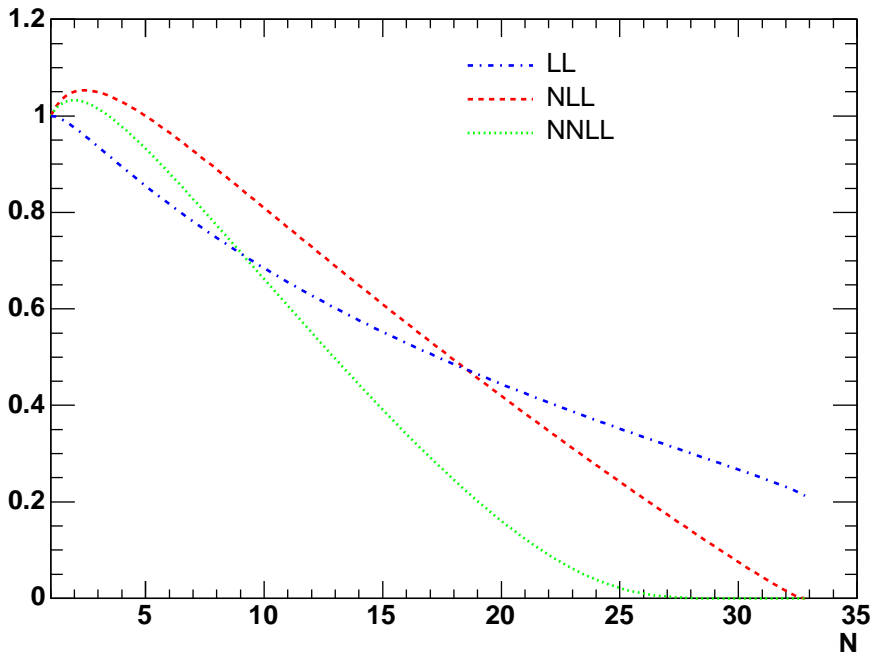,angle=0,width=13.4cm}
\caption{\label{fig:moments_FLA} The moment--space Sudakov factor
${\rm Sud}(m,N)$, computed using \eq{sum_FLA} to LL, NLL and NNLL
accuracy, plotted as dotdashes, dashes and dots, respectively. The
curves end at $N\simeq 33$, corresponding to $\lambda=\frac12$,
where the Landau singularity in $g_n(\lambda)$ appears.}
\end{center}
\end{figure}

Fig.~\ref{fig:moments_FLA} shows the Sudakov factor of \eq{sum_FLA}
with increasing logarithmic accuracy: $g_0(\lambda)$ only (LL
accuracy), then $g_0(\lambda)$ and $g_1(\lambda)$ (NLL accuracy) and
finally the first three terms (NNLL accuracy). We observe that the
convergence of the series in \eq{sum_FLA} is poor. Non-negligible
differences appear already in the first few moments. These will be
partially washed out by matching the results with the fixed--order
calculation. However, for $N\gsim 10$, where $\ln N$ starts to get
large, the NNLL contribution of $g_2(\lambda)$ is already larger
than that of the NLL term; for $N\gsim 18$ the NNLL contribution
becomes larger than the leading-log term! This clearly indicates
that the perturbative expansion breaks down. It is known from
previous studies~\cite{DGE_thrust,Gardi:2002bg,Gardi:2001di,CG,BDK}
that the asymptotic behavior of ${\rm Sud}(m,N)$ is controlled by
infrared renormalons. In Ref.~\cite{BDK} it was shown, based on the
resummation of running--coupling effects in the large--$\beta_0$
limit, that the expansion in~\eq{sum_FLA} should break down early
(see Fig.~2 there). Having at hand the full NNLL result we confirm
that this is indeed so.

The conclusions are the following: first, it becomes obvious that
the logarithmic accuracy criterion is, at best, insufficient.
Moreover, since the poor convergence of the series in \eq{sum_FLA}
is associated with running--coupling effects, these must be
resummed. Finally, since the latter brings about infrared renormalon
ambiguities that scale as powers of $(N\Lambda/m)$ the resolution of
this problem is directly linked to the question of separation
between perturbative and non-perturbative contributions: the
calculation of the Sudakov exponent cannot be considered as a purely
perturbative matter. Similar observations in different but analogous
physics problems have led to the development of
DGE~\cite{DGE_thrust,Gardi:2002bg,Gardi:2001di,CG} as an alternative
to resummation with fixed logarithmic accuracy.

Before moving on to DGE let us conclude our perturbative analysis by
summarizing the state--of--the--art fixed--order and
fixed--logarithmic--accuracy results. First, in
Appendix~\ref{sec:NNLO} we derive explicit expressions for the
log--enhanced terms at ${\cal O}(\alpha_s^2)$. These will be useful
for checking NNLO calculations. Finally, it is
straightforward\footnote{The general algorithm is explained in
Sec.~3.4 of Ref.~\cite{DGE_thrust}.} to invert the Mellin transform
of ${\rm Sud}(m,N)$ in \eq{sum_FLA} to obtain a resummation formula
with NNLL accuracy in $x$~space. The integrated spectrum with
$E_{\gamma}>E_0=(1-\Delta)m/2$,
\begin{equation}
\Gamma[b\longrightarrow
X_s\gamma]^{O_7}_{E_{\gamma}>E_0}(\Delta)\equiv \int_{1-\Delta}^1
dx\, \frac{d\Gamma^{O_7, \PT}(x)}{dx},
\end{equation}
is then:
\begin{eqnarray}
\label{NNLL_x_space} &&\frac{\Gamma[b\longrightarrow
X_s\gamma]^{O_7}_{E_{\gamma}>E_0}(\Delta)}{\Gamma_{\tot}^{O_7, \PT}}
= C_\infty^{O_7}(\alpha_s(m))
\frac{\exp\bigg\{g_0(\omega)\left(\frac{\alpha_s^{\MSbar}(m)}{\pi}\right)^{-1}+g_1(\omega)+
g_2(\omega)\frac{\alpha_s^{\MSbar}(m)}{\pi}\bigg\}}
{\Gamma\left(1-\beta_0\left(g_0^{\prime}
(\omega)+g_1^{\prime}(\omega)\frac{\alpha_s^{\MSbar}(m)}{\pi}\right)
\right)} \nonumber \\ && \hspace*{12pt}\times \left[
1+\frac12g_0^{\prime\prime}(\omega)\beta_0^2
\frac{\alpha_s^{\MSbar}(m)}{\pi}\Big(
\Psi^2\left(1-\beta_0g_0^{\prime}(\omega)\right)
-\Psi^{\prime}\left(1-\beta_0g_0^{\prime}(\omega)\right) \Big)
\right] \,+\,R_{O_7}(\Delta),
\end{eqnarray}
where
\[
\omega\equiv\frac{\alpha_s^{\MSbar}(m^2)}{\pi}\beta_0 \ln
\frac{1}{\Delta}.
\]
\eq{NNLL_x_space} was matched to agree with the exact NLO expression
upon expansion using~\eq{matching}:
\begin{equation}
\label{Const_matching} C_\infty^{O_7}(\alpha_s(m))\simeq
1-C_F\,\frac{\alpha_s^{\MSbar}(m)}{\pi}\,\left(\frac{\gamma_E^2}{2}
+\frac{\pi^2}{12}-\frac{7}{4}\gamma_E+\frac{31}{12}\right)+{\cal
O}(\alpha_s^2)
\end{equation}
incorporates finite terms in the $N\longrightarrow \infty$ limit,
while the additive contribution
\begin{equation}
R_{O_7}(\Delta)=\left[-\frac{1}{6}\Delta^3+\left(\frac{1}{4}\ln\Delta+\frac{1}{4}\right)\Delta^2
+\left(\frac{5}{2}-\ln\Delta\right)\Delta\right]
C_F\frac{\alpha_s^{\MSbar}(m)}{\pi}+{\cal O}(\alpha_s^2)
\end{equation}
completes the terms that vanish at $\Delta\longrightarrow 0$. Note
that the currently unknown ${\cal O}(\alpha_s^2)$ term in
\eq{Const_matching} influences NNLL terms in the spectrum (at ${\cal
O}(\alpha_s^3)$ and beyond) by mixing with the leading logarithms
from the expansion of the exponential function in \eq{NNLL_x_space}.
Thus, although the Sudakov factor itself is known to NNLL accuracy,
complete NNLL accuracy of the spectrum is not yet available.
\begin{figure}[t]
\begin{center}
\epsfig{file=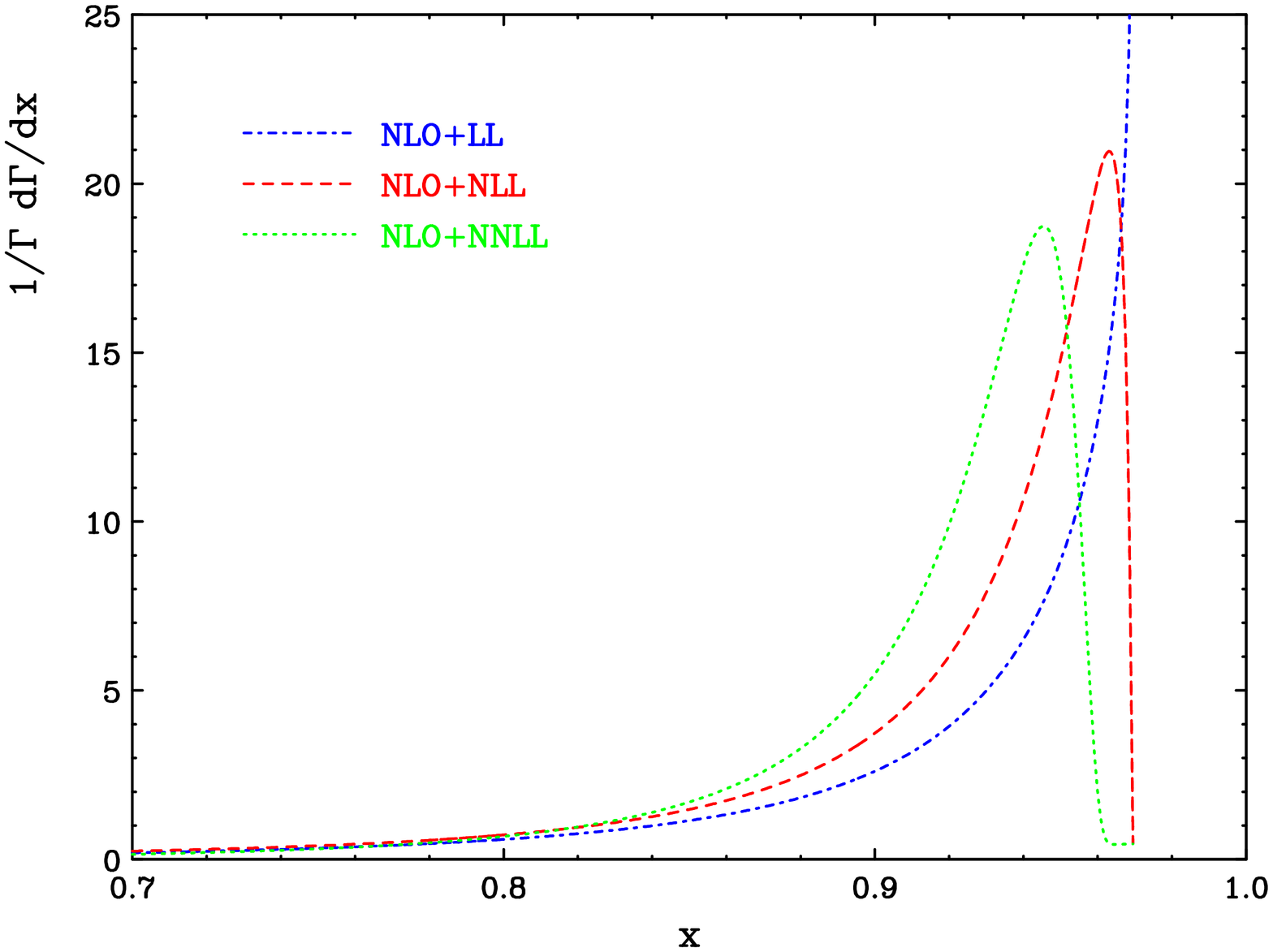,angle=90,width=11.9cm}
\epsfig{file=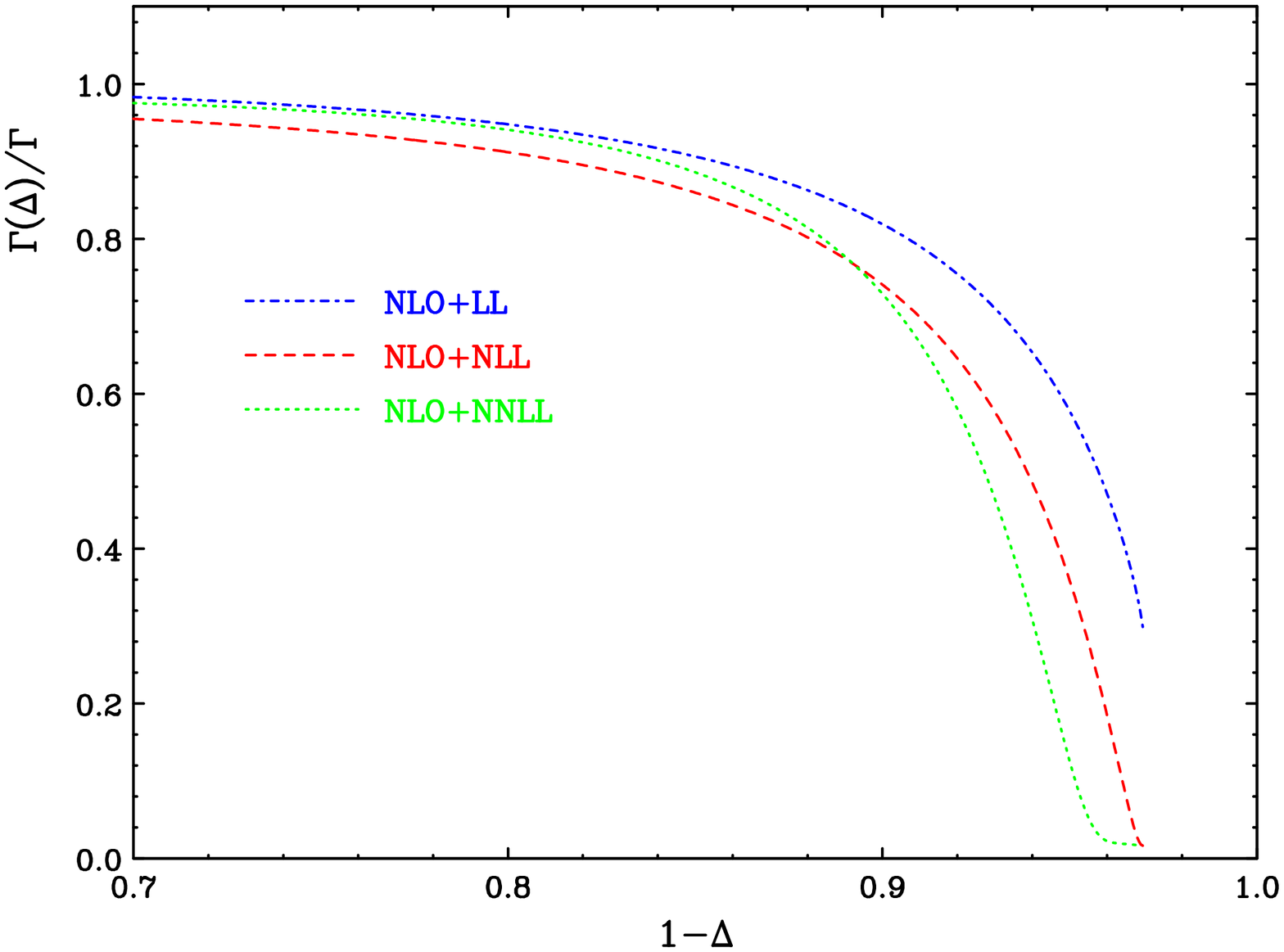,angle=90,width=11.9cm}
\caption{\label{fig:matched_NNLL_spectrum} The differential (top)
and the integrated (bottom) spectra based on the
fixed--logarithmic--accuracy formula of~\eq{NNLL_x_space}, which is
matched to~NLO. The LL, NLL and NNLL accuracy results are plotted as
dotdashes, dashes and dots, respectively. The three curves end at
$x\simeq 0.97$, where the resummed results become complex owing to
the Landau singularity at $\omega=1/2$.}
\end{center}
\end{figure}

The resulting spectrum with the Sudakov factor computed at NNLL
accuracy is shown in Fig.~\ref{fig:matched_NNLL_spectrum} together
with lower logarithmic accuracy results. We find that in spite of
matching the resummed result to the NLO expression (this compensates
for some of the difference observed in Fig.~\ref{fig:moments_FLA})
differences between the three approximations to the spectrum are
still large. Note that the three curves are not plotted beyond
$x\simeq 0.97$, where \eq{NNLL_x_space} becomes complex owing to the
presence of the Landau singularity\footnote{Ref.
\cite{Aglietti:2004fz} suggested to remove these singularities by
resumming a set of $\pi^2$ terms that arise by evaluating the
coupling at time-like momentum. These terms, similarly to other
running--coupling effects, are resummed by DGE.} at $\omega=1/2$. We
observe that with increasing logarithmic accuracy the differential
spectrum widens up and gets shifted to lower photon energies. While
these features are intuitively expected, the instability of the
perturbative result makes it hard to rely on this spectrum and use
it as a baseline for parametrization of non-perturbative effects. An
explicit separation criterion between the perturbative and
non-perturbative regimes is clearly missing here. As we shall see
below, these deficiencies are cured by DGE.

\subsection{Borel representation of the exponent and the large--$\beta_0$ limit\label{sec:Borel_rep_DGE}}

According to Ref.~\cite{BDK} the Sudakov factor in
\eq{BXg_logs_N_2loop} can be expressed as the following Borel sum:
\begin{eqnarray}
\label{Sud} && \hspace*{-20pt} {\rm Sud}(m,N)\,\equiv\,\exp \bigg\{
\frac{C_F}{\beta_0}\int_0^{\infty}\frac{du}{u} \,T(u)\,
\left(\frac{\Lambda^2}{m^2}\right)^u \bigg[ B_{\cal
S}(u)\Gamma(-2u)\left(N^{2u}-1\right)
 -\nonumber\\ &&\hspace*{240pt}
B_{\cal J}(u)\Gamma(-u)\left(N^{u}-1\right)\bigg] \bigg\}.
\end{eqnarray}
Here we use the scheme--invariant Borel
representation~\cite{Grunberg:1992hf} where $T(u)$ is the Laplace
transform of the 't~Hooft coupling:
\begin{eqnarray}
\label{tHooft_coupling}
A(\mu)&=&\frac{\beta_0\alpha_s^{\tHooft}(\mu)}{\pi}=
\int_0^{\infty}{du} \,T(u)\, \left(\frac{\Lambda^2}{\mu^2}\right)^u;
\qquad \qquad \frac{dA}{d\ln \mu^2} =-A^2(1+\delta A),
\nonumber \\
T(u)&=&\frac{(u\delta)^{u\delta}{\rm
e}^{-u\delta}}{\Gamma(1+u\delta)};\qquad \qquad \ln
(\mu^2/\Lambda^2)=\frac{1}{A}-\delta\ln\left(1+\frac{1}{\delta
A}\right)
\end{eqnarray}
with $\delta\equiv \beta_1/\beta_0^2$, where the first two
coefficients of the $\beta$ function are given in \eq{beta_i}.

The functions $B_{\cal S}(u)$ and $B_{\cal J}(u)$ are the scheme
invariant Borel representations of the anomalous dimensions of the
soft (quark distribution) and the jet functions, respectively.
Defining the Borel representation of the anomalous dimensions
introduced in the previous section, namely
\begin{eqnarray}
{\cal A}\big(\alpha_s(\mu)\big)= \frac{C_F}{\beta_0}\int_0^{\infty}
du \,T(u)
\,\left(\frac{\Lambda^2}{\mu^2}\right)^u\, B_{\cal A}(u),\nonumber \\
{\cal B}\big(\alpha_s(\mu)\big)=  \frac{C_F}{\beta_0}\int_0^{\infty}
du \,T(u) \,\left(\frac{\Lambda^2}{\mu^2}\right)^u\, B_{\cal B}(u)
,\nonumber \\
{\cal D}\big(\alpha_s(\mu)\big)=  \frac{C_F}{\beta_0}\int_0^{\infty}
du \,T(u) \,\left(\frac{\Lambda^2}{\mu^2}\right)^u\, B_{\cal D}(u),
\end{eqnarray}
we have:
\begin{eqnarray}
\label{DB_to_SJ}
B_{\cal S}(u)&= & B_{\cal A}(u) - u B_{\cal D}(u),\nonumber \\
B_{\cal J}(u)&= & B_{\cal A}(u) - u B_{\cal B}(u).
\end{eqnarray}
These are exact relations.

Using the perturbative expansions in Eqs.~(\ref{A_cusp_expansion})
through (\ref{D_i}), one obtains the following expansions for the
corresponding Borel functions at small $u$:
\begin{eqnarray}
\label{BA_expanded} B_{\cal A}(u)=1+
\left(\frac53+c_2\right)\frac{u}{1!}+\left(-\frac 13+c_3\right)
\frac{u^2}{2!}+\cdots,
\end{eqnarray}
where $c_n$ represent contributions that are subleading in
$\beta_0$; $c_2$ and $c_3$ are (see Appendix C in \cite{QD}):
\begin{eqnarray}
c_2&=&\frac{C_A}{\beta_0}\left(\frac{1}{3}-\frac{\pi^2}{12}\right)
\nonumber \\
c_3&=& \frac{1}{\beta_0}
\left[\left(\frac{649}{288}-\frac{5}{18}\pi^2
+\frac{7}{2}\zeta_3\right)C_A
+\left(\frac{23}{8}-3\zeta_3\right)C_F\right] \nonumber \\
&+&\frac{1}{\beta_0^2}
\left[\left(\frac{251}{288}+\frac{7}{144}\pi^2
-\frac{11}{4}\zeta_3+\frac{11}{720}\pi^4\right) C_A^2
+\left(-\frac{235}{96}+\frac{11}{4}\zeta_3+\frac{\pi^2}{16}\right)C_FC_A
-\frac{3}{32}C_F^2\right]\nonumber \\
&+&\frac{1}{\beta_0^3}
\left[\left(-\frac{301}{512}-\frac{7}{192}\pi^2\right)
C_A^3+\left(-\frac{11}{64}
-\frac{11}{192}\pi^2\right)C_FC_A^2+\frac{11}{128}C_F^2C_A\right].
\end{eqnarray}
and
\begin{eqnarray}
\label{B_DB} B_{\cal D}(u) &=& 1 +  \left[  \! {\displaystyle \frac
{1}{9}}  + {\displaystyle \frac{C_A}{{\beta _{0}}}
{\left({\displaystyle \frac {9}{4}} \,\zeta_3
 - {\displaystyle \frac {\pi ^{2}}{12}}  - {\displaystyle \frac {
11}{18}} \right)\,}}  \!  \right] u\,+\,{\cal O}(u^2),   \\
\nonumber B_{\cal B}(u) &=&  - {\displaystyle \frac {3}{4}}  +
\left[{\displaystyle \frac {\pi ^{2}}{6}} - {\displaystyle \frac
{247}{72}}   \,+\,
 {\displaystyle \frac{C_A}{\beta_{0}}
{\left( - {\displaystyle \frac {73}{144}}  + {\displaystyle \frac
{5}{2}} \,\zeta_3\right)\,}} + {\displaystyle \frac{C_F}{\beta_{0}}
{\left( - {\displaystyle \frac {3}{32}}
 - {\displaystyle \frac {3}{2}} \,\zeta_3 + {\displaystyle
\frac {\pi ^{2}}{8}} \right)\,}} \right] u + {\cal O}(u^2).
\end{eqnarray}

Upon expanding the entire square brackets in \eq{Sud} in powers of
$u$ using Eqs.~(\ref{DB_to_SJ}) through (\ref{B_DB}) one recovers
the same log--enhanced terms presented in the previous section to
NNLL accuracy. However, upon performing the Borel integral (barring
renormalon singularities, see below) one can resum running--coupling
effects {\em to all orders}. In the large--$\beta_0$ limit the Borel
functions are given by~\cite{BDK,CG}:
\begin{eqnarray}
\label{B_DJ_large_beta0}
B_{\cal S}(u)&=&{\rm e}^{\frac53 u}(1-u)\,+\,{\cal O}(1/\beta_0 ),\nonumber \\
B_{\cal J}(u)&=&\frac{1}{2}\,{\rm e}^{\frac53 u}
\left(\frac{1}{1-u}+\frac{1}{1-u/2}\right) \frac{\sin\pi u}{\pi
u}\,+\,{\cal O}(1/\beta_0 ).
\end{eqnarray}
Note that these are entire functions, free of singularities in the
whole complex plane. In the full theory, i.e. for finite $\beta_0$,
analytic expressions of this kind are, of course, not known. In
Sec.~\ref{sec:residue} we shall construct approximations to these
unknown functions based on their small-$u$ expansions quoted above
and some additional constraints. Because of the significant
contribution of running--coupling
effects~\cite{Gardi:2001di,Gardi:2002bg,CG,BDK} the difference
between performing the Borel integral (DGE) and extracting the
leading logarithms to a fixed logarithmic accuracy is very
significant. This will be explicitly demonstrated in the next
section.

Writing the Sudakov factor of \eq{Sud} in a factorized form, i.e.
\begin{equation}
{\rm Sud}(m,N)=S_N(m;\mu)\,J_N(m;\mu), \label{factorization_SJ}
\end{equation}
requires subtraction of logarithmic singularities:
\begin{eqnarray}
\label{Soft} &&S_N(m;\mu)=\exp \bigg\{
\frac{C_F}{\beta_0}\int_0^{\infty}\frac{du}{u} \,T(u)\,
\left(\frac{\Lambda^2}{m^2}\right)^u\,\,\times\\
\nonumber &&\hspace*{130pt} \bigg[ B_{\cal
S}(u)\Gamma(-2u)\left(N^{2u}-1\right)
+\left(\frac{m^2}{\mu^2}\right)^u B_{\cal A}(u)\ln N\bigg] \bigg\},
\end{eqnarray}
\begin{eqnarray}
\label{Jet} &&J_N(m;\mu)=\exp \bigg\{
-\frac{C_F}{\beta_0}\int_0^{\infty}\frac{du}{u} \,T(u)\,
\left(\frac{\Lambda^2}{m^2}\right)^u\,\,\times\\
\nonumber &&\hspace*{130pt} \bigg[ B_{\cal
J}(u)\Gamma(-u)\left(N^{u}-1\right)
+\left(\frac{m^2}{\mu^2}\right)^u B_{\cal A}(u)\ln N \bigg] \bigg\},
\end{eqnarray}
where, as above, $B_{\cal A}(u)$ is the Borel representation of the
cusp anomalous dimension. This anomalous dimension determines the
factorization scale ($\mu$) dependence of the separate soft and the
jet functions, while ${\rm Sud}(m,N)$ is independent of $\mu$:
\begin{eqnarray}
\frac{d\ln S_N(m;\mu)}{d\ln \mu^2} =-\frac{d\ln J_N(m;\mu)}{d\ln
\mu^2}&\equiv &-\frac{C_F}{\beta_0} {\cal A}(\alpha_s(\mu)) \ln N.
\end{eqnarray}

Finally, note that the Sudakov factors \eq{Soft} and \eq{Jet} (and
likewise \eq{Sud}) have infrared renormalon singularities. As usual
these singularities occur at integer (in both $J_N$ and $S_N$) and
half integer (in $S_N$) values of the Borel variable $u$. To
properly define these resummed Sudakov factors one needs to deform
the integration contour off the real $u$ axis or take the
Principal--Value prescription. We shall adopt the latter, which
guaranties that ${\rm Sud}(m,N)$ is a real--valued function, namely
\begin{equation}
 {\rm Sud}(m,N)=\bigg[{\rm Sud}(m,N^*)\bigg]^*,
\end{equation}
and, in particular, that it is real for real positive $N$, as it is
at any finite order in perturbation theory. By choosing this
prescription we have made an explicit separation between what will
be regarded as resummed perturbation theory and additional
non-perturbative power corrections, which we shall discuss in
Sec.~\ref{sec:PC} below.

\subsection{Renormalons beyond the large--$\beta_0$ limit\label{sec:residue}}

The all--order structure of the Sudakov exponent is summarized by
\eq{Sud}. DGE makes use of this structure in spite of the limited
knowledge of the functions $B_{\cal S}(u)$ and $B_{\cal J}(u)$. In
the large--$\beta_0$ limit these functions, given by
\eq{B_DJ_large_beta0}, are free of infrared renormalon
singularities. Being anomalous dimensions they are expected to be
free of renormalons also in the full theory. Similarly to other
cases \cite{DGE_thrust,Gardi:2002bg,Gardi:2001di,GR,CG}, infrared
renormalons appearing in the Sudakov exponent (\ref{Sud}) have a
very specific origin: the integration over the longitudinal momentum
fraction $z$ in the $z\longrightarrow 1$ limit. This integration
gives rise to the factors $\Gamma(-2u)$ and $\Gamma(-u)$ in
\eq{Soft} and \eq{Jet}, respectively.

\eq{Soft} therefore implies that the soft Sudakov exponent has
simple renormalon poles\footnote{Simple poles appear only in the
scheme invariant formulation of the Borel transform, where the Borel
variable is conjugate to the logarithm of the scale. In the standard
formulation, where the Borel variable $z$ is conjugate to $1/A$
(where $A\equiv\alpha_s\beta_0/\pi$), the singularity transforms
into a cut which is controlled by $\delta=\beta_1/\beta_0^2$ owing
to \eq{ren_int_relation}.} at all integer and half integer values
of~$u$, except where $B_{\cal S}(u)$ vanishes. Similarly, according
to \eq{Jet}, the jet function has simple renormalon poles at all
integer values of~$u$, except where $B_{\cal J}(u)$ vanishes. It
turns out that in the large--$\beta_0$ limit these functions do
vanish at some of the would-be renormalon positions: $B_{\cal S}(u)$
vanishes at $u=1$ while $B_{\cal J}(u)$ at all integers $u\geq 3$.
It is not known whether the corresponding renormalons do appear in
the full theory.

The theoretical interest aside, the renormalon structure of the
Sudakov exponent, and in particular, that of the soft function, is
important for phenomenology. Obviously, the most important
renormalon singularity is the one at $u=\frac12$. It has been shown
in Ref.~\cite{BDK} that the associated ambiguity cancels out between
the soft Sudakov exponent and the leading non-perturbative
correction $\bar{\Lambda}=M-m$. For this cancellation to be realized
--- thus avoiding a spurious ${\cal O}(N\Lambda/m)$ artifact ---
both quantities need to be computed as asymptotic expansions,
regularizing the $u=\frac12$ renormalon in the same manner. As
explained in the introduction, in this paper we shall implement this
idea. We shall use the Cauchy Principal--Value prescription in the
calculation of $\bar{\Lambda}$ from the ${\overline{\rm MS}}$ mass
(see Sec.~\ref{sec:Lambda_PV}) and in the calculation of the Sudakov
exponent.

Power accuracy is, however, not easy to achieve. The difficulty is
that in order to accurately compute the Principal Value of the Borel
integral, say in \eq{Soft}, the renormalon structure, including its
overall normalization (the residue) must be known. Standard
perturbative expansions of $B_{\cal S}(u)$ as in \eq{B_DB}, yield
power series in $u$, which may not be reliable near $u=\frac12$. In
the following we will address this problem. Higher renormalon
singularities in the soft function scaling as higher powers of
$N\Lambda/m$ are also relevant. One expects that these effects will
mainly be important in the endpoint region. However, it is hard to
know a priori how far from endpoint their influence extends. For the
jet function the situation is different: the sensitivity to the
functional form of $B_{\cal J}(u)$ away from the origin is small.
This correspond to the fact that power corrections on the scale
$m^2/N$ are smaller. In what follows we therefore concentrate on
$B_{\cal S}(u)$ and, first of all, on its value at $u=\frac12$ which
determines the corresponding renormalon residue in \eq{Soft} and in
\eq{Sud}.

We first note that direct evaluation of the available NNLO expansion
of $B_{\cal S}(u)$ at $u=\frac12$ using \eq{DB_to_SJ} with Eqs.
(\ref{BA_expanded}) and (\ref{B_DB}) is unreliable. The apparent
convergence of the series (in the Borel plane) at the first three
orders is slow: for example with $N_f=4$ the terms of increasing
powers of $u$ are: $1$, $-0.0188$ and $-0.3415$. As the expansion of
an anomalous dimension this series is expected to converge, but
since the NNLO is sizable while the large--order behavior is not
known, it appears that such a direct evaluation of the residue
cannot be accurate.

Next, we note that the exact result of the large--$\beta_0$ limit,
\eq{B_DJ_large_beta0}, constrains the anomalous dimensions to ${\cal
O}(u^0)$ only, and that the corrections to $B_{\cal S}(u)$ at ${\cal
O}(u)$ and ${\cal O}(u^2)$, which are subleading in $\beta_0$, are
large. Thus, a naive non-Abelianization approach, in which ${\cal
O}(1/\beta_0)$ terms are neglected, is not sufficiently
accurate\footnote{In general the naive non-Abelianization approach
works well for perturbative expansions that are dominated by
renormalons, not for anomalous dimensions.}. A similar conclusion
was reached before \cite{GR} considering the case of the jet
function.

Fortunately, indirect information on $B_{\cal S}(u)$ at $u=\frac12$
is available owing to the exact cancellation~\cite{BDK} of the
corresponding renormalon ambiguity with the one in $\bar{\Lambda}$,
or in the pole mass. We note that the singularity structure of the
$u=\frac12$ renormalon in the pole mass~\cite{Beneke:1994rs} --- a
simple pole in the scheme invariant Borel function --- matches
exactly the one of the soft Sudakov exponent, as it should. As
discussed in Appendix~\ref{sec:pole_mass_scheme_inv}, the
normalization of the pole--mass renormalon at $\frac12$ is well
under control. This allows us to fix the value of the Borel
function:
\begin{equation}
\left.B_{\cal S}(u)\right\vert_{u=\frac12}=\frac{q}{2}{\rm
e}^{\delta/2}. \label{Bs_ren_relation}
\end{equation}
 The normalization constant $q$ --- see \eq
{B_of_z_sing}  --- has been computed from the perturbative relation
between the pole mass and the ${\overline {\rm MS}}$ mass by several
authors \cite{Pineda:2001zq,Lee:2003hh,Cvetic:2003wk,Niko},
obtaining good numerical convergence already at the available NNLO.
In Appendix~\ref{sec:pole_mass_scheme_inv} we summarize our own
study of this renormalon. Previous work on the subject was
restricted to the standard Borel representation and to the
${\overline{\rm MS}}$ scheme. We extend this analysis using the
scheme--invariant formulation of the Borel transform. This provides
an independent check of the accuracy of this calculation and
facilitates the comparison with the Sudakov exponent.

The result for the normalization of the pole--mass renormalon at
$\frac12$ is shown in Fig.~\ref{fig:residue} in
Appendix~\ref{sec:pole_mass_scheme_inv}. The close agreement between
different calculational procedures based on the expansion of
$m/m_{\MSbar}(m_{\MSbar})$ demonstrates the reliability of this
determination. We conclude that the residue can be computed with
$\sim 2-3\%$ accuracy over a wide range of $\beta_0$ values, in
agreement with
Refs.~\cite{Pineda:2001zq,Lee:2003hh,Cvetic:2003wk,Niko}. As
anticipated, the determination that relies on the soft Sudakov
exponent is less accurate. Nevertheless, it does yield similar
values.

As explained above, in order to evaluate the Sudakov exponent of
\eq{Sud} using the Principal--Value prescription we must know
$B_{\cal S}(u)$ (and $B_{\cal J}(u)$) as a function of $u$ away from
the origin. Any uncertainty would translate into uncertainty in the
computed spectrum, and an ambiguity in the separation between
perturbative and non-perturbative corrections. To gauge the
numerical significance of this issue, let us construct a few models,
which we generically call NNLL--DGE, that all share the {\em same}
expansion in powers of $u$ up to the NNLO but differ away from the
origin.

A natural possibility~\cite{DGE_thrust,Gardi:2002bg,CG,GR} is to
start with the analytic form of the large--$\beta_0$ limit,
\eq{B_DJ_large_beta0}, and include a multiplicative correction
factor that modifies the expansion coefficients at ${\cal O}(u)$ and
at ${\cal O}(u^2)$ by terms that are subleading in $\beta_0$, so as
to match the exact coefficients given by \eq{DB_to_SJ} with
\eq{B_DB}.

At ${\cal O}(u)$ (NLL--DGE) this was already done in Ref.~\cite{BDK}
following Refs.~\cite{DGE_thrust,Gardi:2002bg,Gardi:2001di,CG} as
follows:
\begin{eqnarray}
\label{BS_NLL-DGE} B_{\cal S}^{\rm NLL-DGE}(u)&=&{\rm
e}^{\left(\frac53+c_2\right)u}(1-u)+\,{\cal O}(u^2 ),
\end{eqnarray}
and
\begin{eqnarray}
\label{BJ_NLL-DGE} B_{\cal J}^{\rm NLL-DGE}(u)&=&\frac{1}{2}\,{\rm
e}^{\left(\frac53+c_2\right)u}
\left(\frac{1}{1-u}+\frac{1}{1-u/2}\right) \frac{\sin\pi u}{\pi
u}\,+\,{\cal O}(u^2 ).
\end{eqnarray}
Proceeding to ${\cal O}(u^2)$ (NNLL--DGE) one can write:
\begin{eqnarray}
\label{B_DJ} B_{\cal S}^{(a)}(u)&=&{\rm e}^{\frac53
u}(1-u)\,\times\, \exp{ \bigg\{c_2 u+
\left[c_3-c_2^2+\frac{C_A}{\beta_0}\left(\frac{5}{18}\pi^2+\frac{7}{9}-\frac{9}{2}
\zeta_3\right)\right]\frac{u^2}{2!}+{\cal O}(u^3)\bigg\}},\nonumber \\
B_{\cal J}^{(a)}(u)&=&\frac{1}{2}\,{\rm e}^{\frac53 u}
\left(\frac{1}{1-u}+\frac{1}{1-u/2}\right) \frac{\sin\pi u}{\pi
u}\,\times  \\ \nonumber&& \hspace*{-42pt}\exp{\bigg\{c_2 u+
\left[c_3-c_2^2+\frac{C_A}{\beta_0}\left(\frac{29}{72}\pi^2
-\frac{43}{72}-5\zeta_3\right)
+\frac{C_F}{\beta_0}\left(-\frac{\pi^2}{4}+\frac{3}{16}
+3\zeta_3\right)\right]\frac{u^2}{2!}+{\cal O}(u^3)\bigg\}}.
\end{eqnarray}
It is straightforward to check that the this model coincides with
the exact analytic functions in the large--$\beta_0$ limit on the
one hand, and has the exact expansion coefficients to the NNLO on
the other.

Let us recall that a similar exercise has already been done for the
jet function in Ref.~\cite{GR}, in the context of the large--$x$
limit of deep inelastic structure functions. Since in the case of
decay spectra the soft function plays a dominant role, we shall
adopt the model of \eq{B_DJ} for $B_{\cal J}(u)$ and not consider
here other possibilities.

An alternative model for the soft function is given by
\begin{eqnarray}
\label{B_DJ_alt} \hspace*{-20pt} B_{\cal S}^{(b)}(u)={\rm
e}^{\frac53 u}(1-u)\,\times\, \bigg\{1+c_2 u+
\left[c_3+\frac{C_A}{\beta_0}\left(\frac{5}{18}\pi^2+\frac{7}{9}
-\frac{9}{2}\zeta_3\right)\right]\frac{u^2}{2!}+{\cal
O}(u^3)\bigg\}.
\end{eqnarray}
The main difference between $B_{\cal S}^{(a)}(u)$ and $B_{\cal
S}^{(b)}(u)$ is that the former inherently suppresses contributions
from the large--$u$ region, which is, in any case, not well
controlled. While in general the large--$u$ region is suppressed in
the Borel integral (\ref{Sud}) by $(\Lambda^2/m^2)^u$, at large $N$
this is replaced by $(N^2 \Lambda^2/m^2)^u$. This suppression is
relevant up to $N\sim m/\Lambda$. Beyond this region the Borel
integral still converges thanks to the suppression by the factor
$\Gamma(-2u)$, however, the perturbative result is no more a valid
approximation; the power expansion breaks down.

\begin{figure}[th]
\begin{center}
\epsfig{file=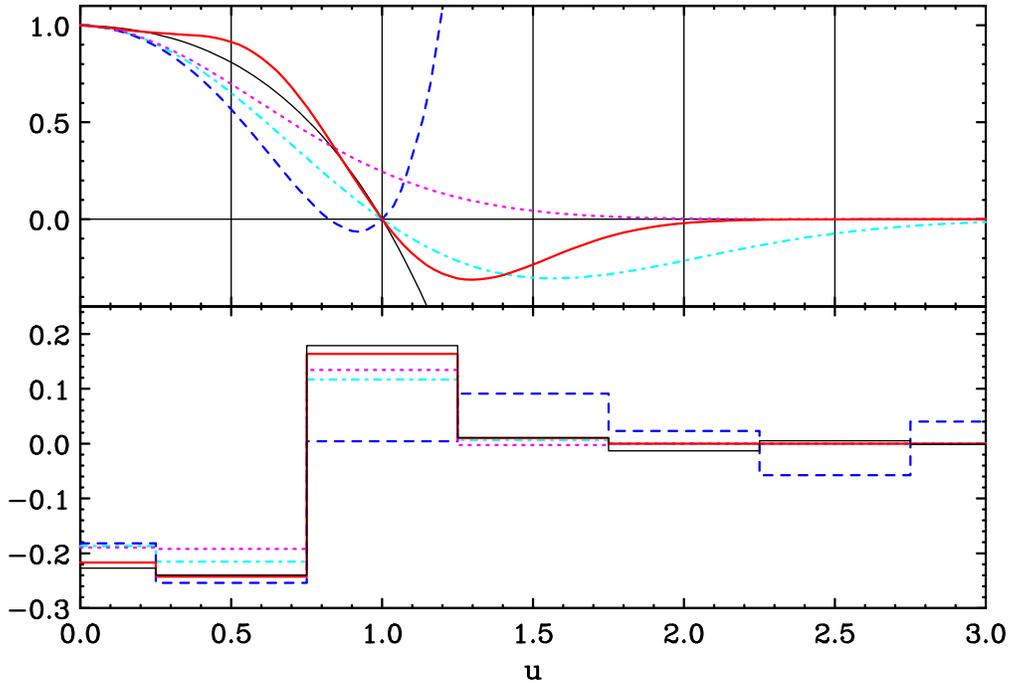,angle=90,width=13.4cm}
\caption{\label{fig:B_S_u} Top: the Borel representation of the soft
anomalous dimension, $B_{\cal S}(u)$, as a function of $u$ for
$N_f=4$. Vertical lines indicate potential renormalon locations.
Bottom: Contributions to the Principal--Value integral over $u$ in
the Sudakov exponent of \eq{Sud} for $N=20$ from the corresponding
sections, each containing one renormalon (or none). In each figure
the five curves corresponds to the different models described in the
text: NLL--DGE as thin full line and NNLL--DGE: ($a$) as dotdashes,
($b$) as dashes, ($c$) as thick full line, and ($d$) as dots.}
\end{center}
\end{figure}

For $N\lsim m/\Lambda$ the perturbative calculation by DGE is under
control. Nevertheless, non-negligible contributions may still arise
from the vicinity of the renormalon positions where the factor
$\Gamma(-2u)$ is singular. These are power suppressed terms. Indeed,
looking in Fig.~\ref{fig:B_S_u} at the contributions to the Sudakov
exponent from different sections along the Borel integration axis
for $N=20$, models~$(a)$ and $(b)$ appear rather different: while
model $(a)$ is characterized by very small contributions beyond
$u\sim 1.5$, in model $(b)$ these are not small up to $u$ of a few
tens. This directly reflects the (exponential) suppression of the
large--$u$ region in $B_{\cal S}^{(a)}(u)$ versus its enhancement in
$B_{\cal S}^{(b)}(u)$. We note, however, that the NLL--DGE model of
\eq{BS_NLL-DGE} does not generate significant contributions from $u$
values beyond $u\sim 1.5$, although it has no inherent exponential
suppression of the large--$u$ region.

\begin{figure}[th]
\begin{center}
\epsfig{file=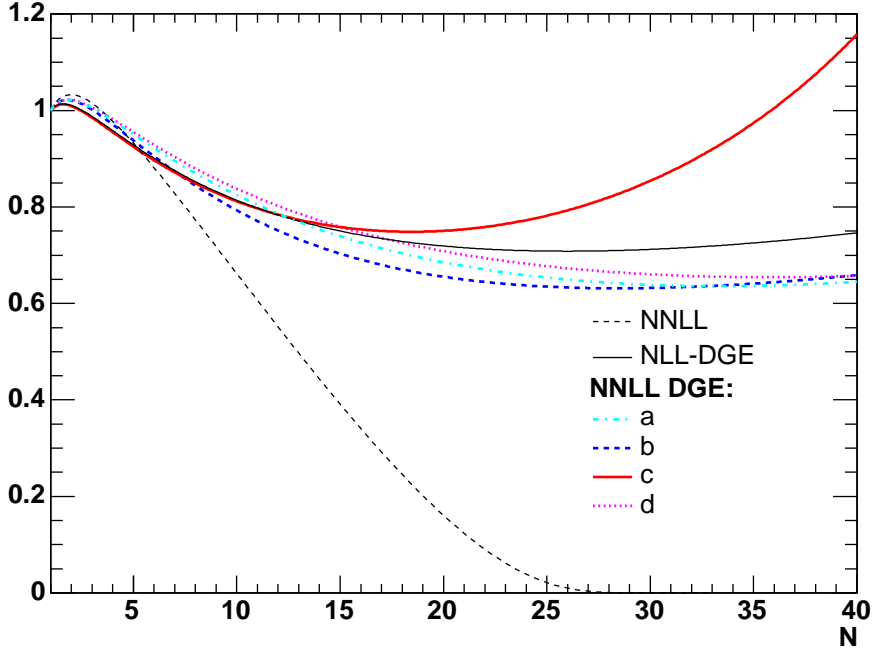,angle=0,width=13.4cm}
\caption{\label{fig:Moments} The moment--space Sudakov factor ${\rm
Sud}(m,N)$, as computed by DGE with perturbative expansions of NLL
and NNLL accuracy (in the latter case the four models $(a)$ through
$(d)$ are shown) and by conventional Sudakov resummation with NNLL
accuracy.  }
\end{center}
\end{figure}

To draw conclusions with regards to the net effect of the behavior
of the models for $B_{\cal S}(u)$ away from the origin we now
examine the value of the Sudakov factor ${\rm Sud}(m,N)$ of
\eq{Sud}. The result, shown in Fig.~\ref{fig:Moments}, is clear:
models $(a)$ and $(b)$ as well as the NLL--DGE one are close up to
very high moments. This means that the total effect of the
large--$u$ region of $B_{\cal S}(u)$, which is not under control, is
moderate; in model $(b)$ this is owing to cancellations between
contributions from the different sections in Fig.~\ref{fig:B_S_u}.
On the whole the DGE result is stable. Nevertheless, the differences
between the models are non-negligible.

Fig.~\ref{fig:Moments} also shows that some difference between DGE
and the conventional Sudakov--resummation procedure (with NNLL
accuracy) develops already at low moments $N<10$, and that it
becomes very significant for $N\sim 10-20$. One qualitative
difference is that the latter has a Landau singularity ($N\simeq
33$) whereas the former does not~\cite{CG}.

\begin{figure}[th]
\begin{center}
\epsfig{file=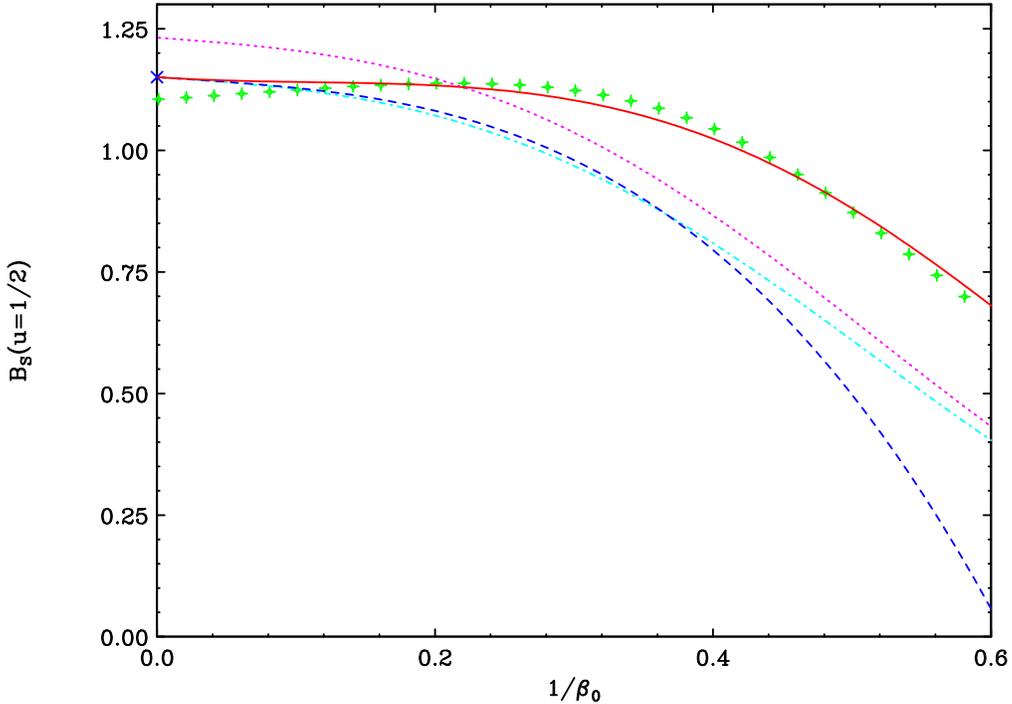,angle=90,width=13.4cm}
\caption{\label{fig:B_S} The value of the Borel representation of
the soft anomalous dimension $B_{\cal S}(u)$ at $u=\frac12$,  which
is proportional to the residue of the $u=\frac12$ renormalon in the
pole mass, plotted as a function of $1/\beta_0$. The {\em exact}
result of the large--$\beta_0$ limit is denoted by a cross. The
calculation relying on perturbative expansion of
$m/m_{\MSbar}(m_{\MSbar})$ using the scheme--invariant Borel
transform as determined using \eq{R_opt_f} is shown by plus signs.
The four curves corresponds to the different models for NNLL--DGE
described above: ($a$) dotdashes, ($b$) dashes, ($c$) full line and
($d$) dots. Model $(c)$ matches the computed value of $B_{\cal
S}(u=\frac12)$ at $N_f=4$ by construction.}
\end{center}
\end{figure}

A shortcoming of the models constructed so far is that they are
inaccurate around $u=\frac12$: \eq{Bs_ren_relation} is not
fulfilled. In order to incorporate the knowledge of $B_{\cal S}(u)$
at $u=\frac12$ we consider the following model:
\begin{eqnarray}
\label{B_DJ_c} B_{\cal S}^{(c)}(u)&=& {\rm e}^{\frac53
u}(1-u)\,\times\, \exp\left\{c_2 u+\frac12 \left[
c_3-c_2^2+\frac{C_A}{\beta_0}\left(\frac{5}{18}\pi^2+\frac{7}{9}
-\frac{9}{2}\zeta_3\right)\right]u^2\right\}\times W(u),\nonumber \\
\end{eqnarray}
which, similarly to models $(a)$ and $(b)$, has the exact
coefficients through ${\cal O}(u^2)$ and the correct
large--$\beta_0$ limit. Here, however, the additional factor
\[
W(u)\equiv {\rm e}^{w_1 u +\frac12 w_2 u^2}\, \left(1-w_1 u+\frac12
(w_1^2-w_2)u^2\right)= 1+{\cal O}(u^3),
\]
is constructed such that correct value of $B_{\cal S}(u)$ at
$u=\frac12$ would be reproduced, at least for the physically
relevant values of $N_f$. To this end we set:
\begin{eqnarray}
\begin{array}{ll}
w_{1}=1.144 C_A/\beta_0, &\hspace*{20pt} w_2=-2.8C_A/\beta_0.
\end{array}
\end{eqnarray}
The values of $B_{\cal S}(u)$ at $u=\frac12$ in the different models
are shown as a function of $1/\beta_0$ in Fig.~\ref{fig:B_S}.
Returning to Fig.~\ref{fig:Moments}, we observe that fixing the
value at $u=\frac12$ makes a large effect on high moments $N\gsim
20$. This shows that this information is relevant for the final
spectrum. Nevertheless, the fact that differences with the other
models are moderate for $N\lsim 20$ is reassuring.

Finally, let us focus on the peculiar feature of the
large--$\beta_0$ result for $B_{\cal S}(u)$, namely its vanishing at
$u=1$, leading to the absence of a corresponding renormalon
ambiguity. In the models considered so far we assumed that this
property is shared by the full theory. This, however, has not been
proven. Moreover, in Ref. \cite{Neubert:1996zy} it has been shown
that a $u=1$ renormalon appears in the kinetic--energy operator once
terms that are subleading in $1/\beta_0$ are taken into account.
This suggest that the same might occur in the Sudakov exponent,
although this should be checked explicitly. In order to estimate the
numerical significance of this issue, let us construct another
model, $(d)$, which has $B_{\cal S}^{(d)}(u=1)\neq 0$:
\begin{eqnarray}
\label{B_DJ_d}
 B_{\cal S}^{(d)}(u)&=& \exp\left\{\left(\frac23+c_2\right)u+
 \frac12 \left[c_3-c_2^2+\frac{C_A}{\beta_0}\left(\frac{5}{18}\pi^2+\frac{7}{9}
-\frac{9}{2}\zeta_3\right)-1\right] u^2\right\}.
\end{eqnarray}
While $B_{\cal S}^{(d)}(u)$ does not respect the large--$\beta_0$
limit result, it does have the correct NNLO expansion at $u=0$.
Returning again to Fig.~\ref{fig:Moments}, we observe that the
effect of this modification is moderate, although probably not
negligible.

\begin{figure}[th]
\begin{center}
\epsfig{file=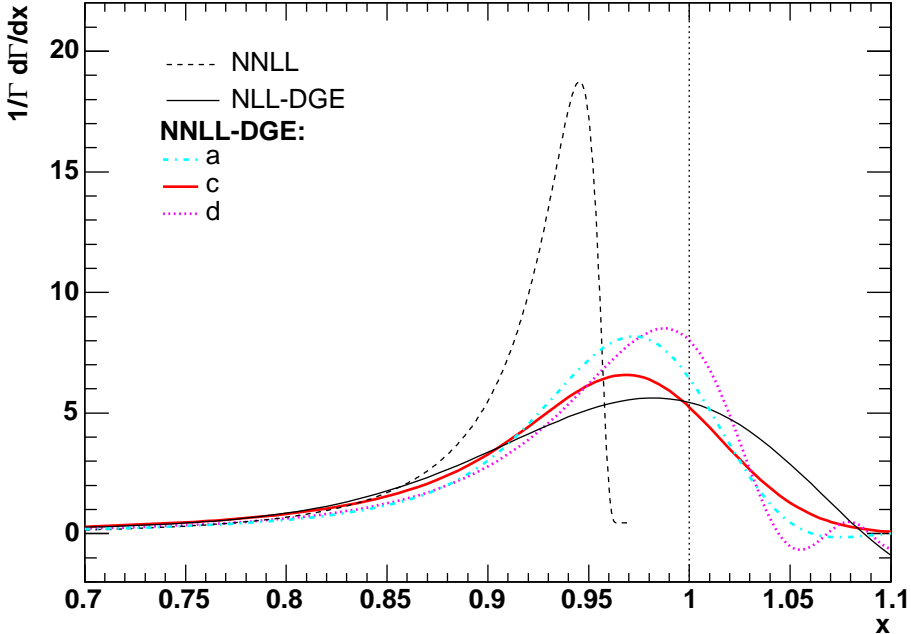,angle=0,width=13.4cm}
\caption{\label{fig:Model_1} The perturbative differential spectrum
in $b\longrightarrow X_s \gamma$ with different approximations for
the Sudakov factor as a function of $x=2E_{\gamma}/m$, normalized by
the total rate with $E_{\gamma}>m/20$. The different curves
correspond to conventional Sudakov resummation with NNLL accuracy
(thin dashed line) and DGE with perturbative expansions of NLL (thin
full line) and NNLL accuracy (thick lines) where the three models
$(a)$, $(c)$ and $(d)$ are shown by dotdashed, full and dotted
lines, respectively. }
\end{center}
\end{figure}
Finally, let us compare between the different calculations of the
resummed Sudakov exponent described above at the level of the
normalized differential spectrum. To this end we match the resummed
expression for the Sudakov factor with the full NLO result as
explained in Sec.~\ref{sec:matching} below. Both the Borel
integration in the Sudakov exponent in \eq{Sud} and the
inverse--Mellin integration in \eq{inv_Mellin} are performed
numerically, avoiding any further approximation. The results are
shown in Fig.~\ref{fig:Model_1}. Using DGE, the differences between
different models with NLL or NNLL accuracy are moderate. On the
other hand the result obtained by conventional NNLL Sudakov
resummation is entirely different in spite of having the same formal
logarithmic accuracy!

It is interesting to note that the DGE result does {\em not} have
the perturbative support properties: while the perturbative
coefficients at any given order vanish for $E_\gamma>m/2$, the
DGE--resummed spectrum does not. It peaks for $E_{\gamma}\lsim m/2$
and then smoothly crosses the perturbative endpoint $E_\gamma=m/2$
and drops to zero at $E_{\gamma}=\left(m+{\cal
O}(\Lambda)\right)/2$. The resummed expression having different
analytic properties than the coefficients is not surprising given
that, strictly speaking, the sum does not exist. Nevertheless, it is
remarkable that by taking the Principal--Value prescription in
moment space we obtain a spectrum, which qualitatively corresponds
to the decay of a higher--mass state.

\section{Matching the resummed spectrum to the full NLO result\label{sec:matching}}

In the previous section we discussed the QCD description of the
endpoint region neglecting ${\cal O}(1/N)$ corrections. In order to
recover the correct spectrum away from the endpoint region, the
resummed result must be systematically matched onto the fixed--order
perturbative expansion, which is available in full at NLO --- see
Refs.~\cite{Chetyrkin:1996vx,Buras:2002tp,Buras:2002er} and
references therein.

\subsection{The NLO result}

The NLO calculation in the Standard Model is based on the effective
Hamiltonian
\begin{equation}
{\cal H}_{\eff}=\frac{4G_F}{\sqrt{2}} V^{*}_{ts} V_{tb}
\sum_{k=1}^{8} C_{k} O_{k}
\end{equation}
where $C_k$ are the Wilson coefficients and $O_k$ are local
operators of dimension 5 or 6. The most important operator
contributing at leading order corresponds to the magnetic
interaction,
\begin{equation}
O_7= \frac{e\, m_{\MSbar}}{16\pi^2}\,
\left(\bar{s}_L\sigma^{\mu\nu}b_R\right) \,F_{\mu\nu}, \label{O7}
\end{equation}
where $m_{\MSbar}$ is the running quark mass in the $\overline{\rm
MS}$ scheme and $F_{\mu\nu}$ is the photon field strength. The
complete basis of operators can be found for example in
Refs.~\cite{Chetyrkin:1996vx,Buras:2002tp}.

Sudakov logarithms originate in the universal soft and collinear
limits in which the specific structure of the hard interaction is
irrelevant. Therefore, the calculation in Ref.~\cite{BDK} and the
formulae of the previous section, although derived starting with
$O_7$, apply to all operators. Terms that are finite (or vanish) at
large $N$, e.g. those of \eq{matching}, are different between the
different operators. These terms will now be incorporated in the
process of matching the Sudakov exponent to the full NLO result.

Our treatment of the endpoint region is based on moment space. On
the other hand contributions that are non-singular at $x=1$ may be
included either in moment space or directly in $x$ space. Since
$b\longrightarrow s\gamma g$ contributions associated with $O_8$ are
singular at $x\longrightarrow 0$ (when the photon gets soft) full
NLO analysis in moment space is excluded. We therefore choose a
mixed matching procedure where the dominant contribution at large
$x$ is taken into account in moment space, but some terms, which
vanish at $x\longrightarrow 1$, are included directly in $x$ space.

Let us express the partonic decay rate with photon energy above some
fixed energy cut $E_0=(1-\Delta){m}/{2}$ as:
\begin{eqnarray}\label{Gamma_cut}
\Gamma[b\longrightarrow X_s\gamma]_{E_{\gamma}>E_0}
=\Gamma_{\LO}\times \bigg[|{\cal
M}_V|^2\,\theta(\Delta>0)+R(\Delta)\bigg],
\end{eqnarray}
where $\theta(\Delta>0)$ is the heaviside function and
\begin{eqnarray}
\Gamma_{\LO}&=&\frac{G_F^2 m^3m_{\MSbar}^2(m)}{32\pi^4}\left|
V_{ts}^* V_{tb}\right|^2
\left(C_7^{(0)\,eff}(\mu_b)\right)^2,\label{Gamma_LO}
\end{eqnarray}
where $m\equiv m_b$, the b-quark pole mass, $m_{\MSbar}(m)$ is the
running mass evaluated at $m$ and, finally, $\mu_b\simeq m$. In
\eq{Gamma_cut} ${\cal M}_V$ corresponds to the amplitude of the
process $b\longrightarrow s \gamma$ so it begins at ${\cal
O}(\alpha_s^0)$ and includes all the purely virtual corrections at
higher orders, while $R(\Delta)$ corresponds to partonic processes
with at least one gluon in the final state, and it therefore begins
at ${\cal O}(\alpha_s)$ and depends on $\Delta$. At NLO one
has~\cite{Chetyrkin:1996vx}:
\begin{eqnarray}
\label{NLO} {\cal M}_V&=&1\,+\, \frac{\alpha_s}{4\pi}
\left(\frac{C_7^{(1)\,\eff}(\mu_b)}{C_7^{(0)\,\eff}(\mu_b)}+
\!\!\sum_{i=1..8}
\frac{C_i^{(0)\,\eff}(\mu_b)}{C_7^{(0)\,\eff}(\mu_b)}
\left[r_i+\gamma_{i7}^{(0)\,\eff}\ln\frac{m}{\mu_b}\right]\right)+\cdots
\nonumber
\\
R(\Delta) &=&\frac{C_F \alpha_s}{\pi}\Bigg(-\frac12 \ln^2\Delta-\frac74\ln\Delta\,+\!\!\sum_{\scriptsize \begin{array}{c}{i,j=1..8}\\
{ i\leq j}\end{array} }c_{ij} \,\phi_{ij}(\Delta)\Bigg)+\cdots,
\end{eqnarray}
where
\[
c_{ij}\equiv \frac{C_i^{(0)\,\eff}(\mu_b)C_j^{(0)\,
\eff}(\mu_b)}{\left( C_7^{(0)\,\eff}(\mu_b)\right)^2}.
\]
The expressions for the LO coefficients $C_i^{(0)\,\eff}$, the NLO
coefficients $C_i^{(1)\,\eff}$ and the (complex) constants $r_i$ as
well as the anomalous dimensions $\gamma_{i7}^{(0)\,\eff}$ can be
found e.g. in \cite{Chetyrkin:1996vx,Gambino:2001ew}. In $R(\Delta)$
we exhibited explicitly the terms\footnote{These specific terms
originate in $O_7$. However, at higher orders there are similarly
singular terms from all operators.} that are singular at the
phase--space boundary $\Delta\longrightarrow 0$ (or
$x\longrightarrow 1$). These terms must be resummed to all orders
incorporating the cancellation of divergences between real emission
and virtual corrections. All the functions $\phi_{ij}(\Delta)$
vanish at $\Delta=0$ --- see Appendix~\ref{sec:phi} for explicit
expressions.

\subsection{The matching procedure}

Our goal is to improve the determination of $\Gamma[b\longrightarrow
X_s\gamma]_{E_{\gamma}>E_0=(1-\Delta){m}/{2}}$ in \eq{Gamma_cut} for
small values of $\Delta$ by performing Sudakov resummation.  At the
same time we require that upon expansion the matched expression
would coincide with the fixed--order result for any~$\Delta$. To
perform the matching we first split the real--emission terms into
two parts, writing
\begin{equation}
\label{P_split} \frac{1}{\Gamma_{\LO}}\Gamma[b\longrightarrow
X_s\gamma]_{E_{\gamma}>E_0=xm/2}=|{\cal M}_V|^2\,\bigg[\theta(x<1)+
G_1(x)\bigg] \,+\, G_2(x),
\end{equation}
where both $G_{1,2}(x)$ begin at ${\cal O}(\alpha_s)$ and we require
that $G_1(x)$ contains all the singular terms for $x\longrightarrow
1$ whereas $G_2(x)={\cal O}((1-x)^2)$ so that it does not contribute
to the {\em differential rate} near $x=1$:
\begin{eqnarray}\label{dGamma_dx}
\frac{1}{\Gamma_{\LO}}\,\frac{d\Gamma^{\PT}(x)}{dx}
=|{\cal M}_V|^2\left[\delta(1-x)-\frac{dG_1(x)}{dx}\right]-
\frac{dG_2(x)}{dx}.
\end{eqnarray}
Finally, we replace the square brackets by a matched Sudakov
resummation (or DGE) formula in moment space while the $G_2$ term is
left in $x$ space,
\begin{eqnarray}\label{dGamma_dx_Gamma_N}
\frac{1}{\Gamma_{\LO}}\,\frac{d\Gamma^{\PT}(x)}{dx} =|{\cal M}_V|^2
\int_{c-i\infty}^{c+i\infty}\frac{dN}{2\pi i} x^{-N} M_N^{\PT} -
\frac{dG_2(x)}{dx},
\end{eqnarray}
where $M_N^{\PT}$ is proportional to the Sudakov factor ${\rm
Sud}(N)$ in \eq{Sud}. Here we used the inverse Mellin transform
formula of \eq{inv_Mellin}.

\eq{dGamma_dx_Gamma_N} is, by construction, consistent with both the
fixed--order expansion and the resummed expression in the
$x\longrightarrow 1$ limit. From
Eqs.~(\ref{P_split}),~(\ref{dGamma_dx})
and~(\ref{dGamma_dx_Gamma_N}) it follows that
\begin{equation}
\label{M_N_G1} M_N^{\PT}=1+G_1(x=0)-\int_0^1dx x^{N-1}
\left\{\frac{dG_1(x)}{dx}\right\}_+,
\end{equation}
where $\{\,\,\}_+$ stand for the ``+'' prescription (see
\eq{plus_presc}). Note that the contribution to the $N=1$ moment
from the integral vanishes, so $M_{N=1}=1+G_1(x=0)$. We emphasize
that, in contrast with our definition of $\bar{M}_N$ in
Sec.~\ref{sec:Sud_DGE}, $M_{N=1}\neq 1$: the distribution is not
normalized.

\subsection{Matching formulae at NLO\label{sec:Matching_form}}

This matching procedure can be applied at any order\footnote{Note
that at NLO one simply has $R(\Delta=1-x)=G_1(x)+G_2(x)$, however,
starting at NNLO the relation is more complicated.}. Specifically,
at NLO we have:
\begin{eqnarray}
\label{R12_NLO}
G_1(x) &=&\frac{C_F \alpha_s}{\pi}\Bigg(-\frac12 \ln^2(1-x)-\frac74\ln(1-x)+\!\!\!\sum_{\scriptsize \begin{array}{c}{i,j=1..8}\nonumber\\
{ i\leq j}\end{array} }\!\!\!c_{ij} \,\eta_{ij}(\Delta=1-x)\Bigg)+{\cal O}(\alpha_s^2),\\
G_2(x) &=&\frac{C_F \alpha_s}{\pi}\sum_{\scriptsize \begin{array}{c}{i,j=1..8}\\
{ i\leq j}\end{array} }c_{ij} \,\xi_{ij}(\Delta=1-x)+{\cal
O}(\alpha_s^2),
\end{eqnarray}
where $\phi_{ij}(\Delta)=\eta_{ij}(\Delta)+\xi_{ij}(\Delta)$ with
$\xi_{ij}(\Delta)={\cal O}(\Delta^2)$; the explicit expressions for
$\eta_{ij}$ and $\xi_{ij}$ appear in Appendix~\ref{sec:phi}. Using
\eq{M_N_G1} with \eq{R12_NLO} we then get:
\begin{eqnarray*}
M_N^{\PT}&=&1+\frac{C_F\alpha_s}{\pi}\Bigg\{ \int_0^1dx
x^{N-1}\Bigg[-\frac{\ln(1-x)}{1-x}-\frac74 \frac{1}{1-x}+
\!\!\sum_{\scriptsize \begin{array}{c}{i,j=1..8}\nonumber\\
{ i\leq j}\end{array} }\!\!\!c_{ij}
\,\left.\frac{d\eta_{ij}(\Delta)}{d\Delta}\right\vert_{\Delta=1-x}\Bigg]_{+}
\\
&&\hspace*{60pt}+
\sum_{\scriptsize \begin{array}{c}{i,j=1..8}\nonumber\\
{ i\leq j}\end{array} }\!\!\!c_{ij} \,\eta_{ij}(\Delta=1)
\Bigg\}\,+\,{\cal O}(\alpha_s^2)
\end{eqnarray*}
Performing the $x$-integration and extracting the single and double
$\ln N$ terms which are included in the Sudakov exponent we finally
obtain the matched Sudakov--resummed moments:
\begin{equation}
\label{M_N_full} M_N^{\PT}=
\Bigg[1+\frac{C_F\alpha_s}{\pi}\Bigg(f(N)+
\!\!\sum_{\scriptsize \begin{array}{c}{i,j=1..8}\\
{ i\leq j}\end{array} }c_{ij}
\,\mu_{ij}(N)\Bigg)\Bigg]\,\times\,{\rm Sud}(m,N),
\end{equation}
where
\[
f(N)=-\frac12 (\Psi(N)+\gamma_E)^2+\frac12 \Psi_1(N)
-\frac{\pi^2}{12} +\frac74 (\Psi(N)+\gamma_E)+\frac12 \ln^2N
-\left(\frac74-\gamma_E\right)\ln N
\]
and
\begin{equation}
\label{mu_ij_def} \mu_{ij}(N)\equiv \int_0^1 dx x^{N-1}
\left.\frac{d\eta_{ij}(\Delta)}{d\Delta}\right\vert_{\Delta=1-x}.
\end{equation}
It is straightforward to verify that, upon expanding \eq{M_N_full}
to ${\cal O}(\alpha_s)$, \eq{dGamma_dx_Gamma_N} reproduces the full
NLO result. The $x\longrightarrow 1$ singular terms in
$-\frac{dG_1(x)}{dx}$ are taken into account through the Sudakov
factor:  they coincide with the ${\cal O}(\alpha_s)$ terms
in~\eq{x_space}; the finite terms are incorporated into $M_N^{\PT}$,
and the remaining terms, which vanish at $x=1$, appear explicitly in
$-\frac{dG_2(x)}{dx}$.

The explicit expressions for $\mu_{ij}(N)$ appear in
Appendix~\ref{sec:phi}. Having extracted the factor $|{\cal M}_V|^2$
which contains the constants at ${\cal O}(\alpha_s)$ in
\eq{dGamma_dx_Gamma_N}, $\mu_{ij}(N)$ vanish at large $N$ and the
information needed to compute the ${\cal O}(\alpha_s^2\,\ln N)$ term
in $M_N^{\PT}$ is fully contained in $f(N)$ and the NNLL
contributions to the exponent in ${\rm Sud}(m,N)$.

\begin{figure}[t]
\begin{center}
\epsfig{file=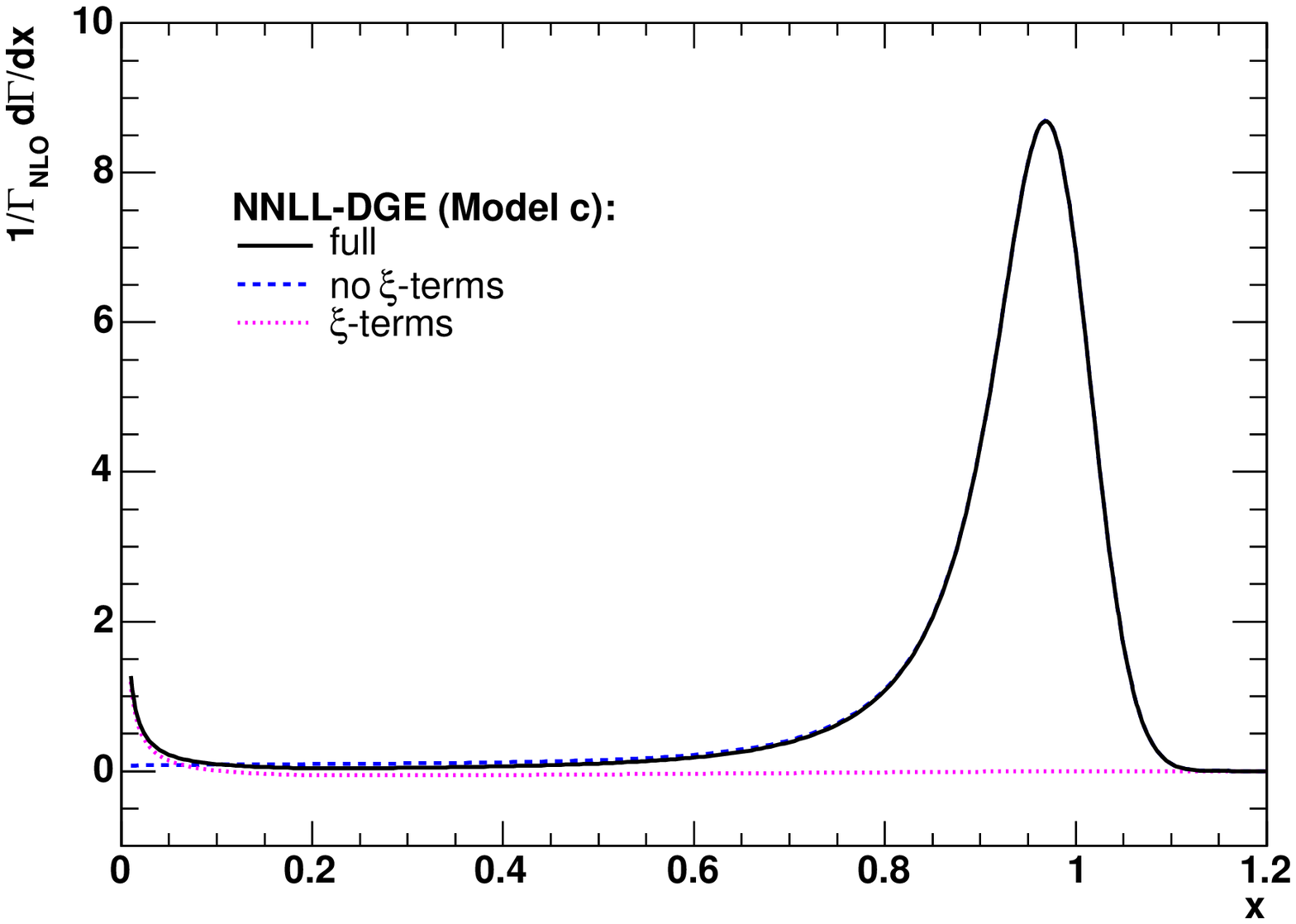,angle=0,width=13.4cm}
\caption{\label{fig:G_2_cont}The differential spectrum according to
\eq{dGamma_dx_Gamma_N_NLO} with $M_N^{\PT}$ computed by NNLL--DGE
using \eq{M_N_full} (where the Sudakov factor is evaluated by
\eq{Sud} using model ($c$) of \eq{B_DJ_c}). We show the spectrum
matched to the full NLO result (full line) together with the
separate contributions in \eq{dGamma_dx_Gamma_N_NLO} that are taken
into account in moment space (dashed line) and in $x$ space (dotted
line), respectively. The latter is significant only for
$x\longrightarrow 0$.}
\end{center}
\end{figure}

Having matched the result at NLO only, {\em non-singular} terms at
${\cal O}(\alpha_s^2)$ are not under control. These terms depend on
the details of the matching procedure. The particular choice we made
gives preference to the moment--space treatment: as reflected in
Fig.~\ref{fig:G_2_cont} the contribution of $G_2$ is very small.

To conclude this section we summarize the result. The
Sudakov--resummed differential rate, matched to the exact NLO
result, takes the form:
\begin{eqnarray}\label{dGamma_dx_Gamma_N_NLO}
\frac{1}{\Gamma_{\NLO}}\,\frac{d\Gamma^{\PT}(x)}{dx}
= \int_{c-i\infty}^{c+i\infty}\frac{dN}{2\pi i} x^{-N} M_N^{\PT} +\frac{C_F\alpha_s(m_b)}{\pi}\!\!\sum_{\scriptsize \begin{array}{c}{i,j=1..8}\\
{ i\leq j}\end{array} }\!\!\!c_{ij}
\,\left.\frac{d\xi_{ij}(\Delta)}{d\Delta}\right\vert_{\Delta=1-x}
\end{eqnarray}
where the moments $M_N^{\PT}$ are given by \eq{M_N_full} and the
normalization is fixed by $\Gamma_{\NLO}$, which is defined such
that:
\begin{eqnarray}
\label{Gamma_NLO} && \Gamma[b\longrightarrow
X_s\gamma]_{E_{\gamma}>E_0=(1-\Delta){m}/{2}}=\Gamma_{\NLO}\times
\left[\theta(\Delta>0)+R(\Delta)\right],\\
\nonumber&&\hspace*{-62pt} {\rm with}\\
\nonumber \Gamma_{\NLO}&=& \frac{G_F^2
m^3m_{\MSbar}^2(m)}{32\pi^4}\left| V_{ts}^* V_{tb}\right|^2
\left(C_{7\gamma}^{\eff}(\mu_b)\right)^2\,\times \\
\nonumber
&&\hspace*{100pt} \left\{1+\frac{\alpha_s}{2\pi} \sum_{i=1..8}
\frac{C_i^{(0)\,\eff}(\mu_b)}{C_7^{(0)\,\eff}(\mu_b)}\left[{\rm Re}
\{r_i\}+\gamma_{i7}^{(0)\,\eff}\ln\frac{m}{\mu_b}\right] \right\},
\end{eqnarray}
where
$C_{7\gamma}^{\eff}(\mu_b)=C_7^{(0)\,\eff}(\mu_b)+\frac{\alpha_s}{4\pi}
C_7^{(1)\,\eff}(\mu_b)$. We emphasize that the contribution of the
$\xi_{ij}$ terms, which is treated in $x$ space, vanishes linearly
with $(1-x)$ at large $x$ and it is numerically very small in the
experimentally relevant region of $x$.

Finally, the corresponding formula for the total rate with an energy
cut $E_{\gamma}>E_0=(1-\Delta){m}/{2}$ is:
\begin{eqnarray}
\label{Gamma_N_NLO} \hspace*{-20pt}
\frac{1}{\Gamma_{\NLO}}\Gamma[b\longrightarrow
X_s\gamma]_{E_{\gamma}>E_0}
\!=\!\!\int_{c-i\infty}^{c+i\infty}\frac{dN}{2\pi
i}\,\frac{(1-\Delta)^{1-N}}{N-1} M_N^{\PT}
\!+\!\frac{C_F\alpha_s(m_b)}{\pi}\!\!\!\!\!\!\!\sum_{\scriptsize \begin{array}{c}{i,j=1..8}\\
 { i\leq j}\end{array} }\!\!\!\!\! c_{ij} \xi_{ij}(\Delta),
\end{eqnarray}
where $c>1$. Note that at $\Delta\longrightarrow 1$ (completely
inclusive measurement) the inverse Mellin integral in
\eq{Gamma_N_NLO} becomes trivial: it equals $M_{N=1}$.

\section{Power corrections\label{sec:PC}}

The perturbative analysis of decay spectra in the endpoint
region~\cite{KS,BDK} exposes the presence of three different scales:
(1)~hard,  corresponding to virtual corrections with momenta  ${\cal
O}(m)$; (2)~final--state dynamics of the jet with invariant mass
squared ${\cal O}(m^2/N)$; and (3)~soft radiation, with momenta
${\cal O}(m/N)$, that determines the heavy--quark momentum
distribution function. Naturally, there is also non-perturbative
dynamics on each of these scales, which brings about power
corrections. The largest power corrections are clearly those on the
soft scale. Smaller power corrections, which nevertheless become
increasingly important near the endpoint, are the ones corresponding
to the jet--mass scale. Their combined effect is related to the
appearance of resonances when the jet mass is sufficiently small. In
the following we address power corrections on the soft scale only.

\subsection{Power corrections on the soft scale\label{PC_soft}}

The classical approach to the quark distribution function
\cite{Neubert:1993um,Bigi:1993ex} is based on the OPE. Expanding the
non-local lightcone operator in \eq{definition} in powers of the
lightcone separation~$y^-$ one obtains local operators with a
corresponding number of $iD_{+}$ derivatives. The matrix elements of
these operators in the meson define non-perturbative
parameters~$f_n$, which, similarly to the non-local operator in
\eq{definition}, require renormalization. In the endpoint region all
of these powers are relevant, and need to be summed.  One arrives at
the following result:
\begin{equation}
\frac 12\left< B(P) \right\vert\bar{\Psi}(y) \Phi_{y} (0,y)
\gamma_{+} \Psi(0)
    \left\vert B(P)
    \right>\simeq
\exp\left\{\frac{iP^+y^-m}{M}\right\}\,\times
\left[1+\sum_{n=2}^{\infty}\frac{f_n}{n!}(iP^+y^-/M)^n\right]
\label{F_N_OPE}
\end{equation}
where the square brackets define the so called ``shape function''.
Up to ${\cal O}(1/N)$ corrections, the Mellin integral is the
inverse of the Fourier integral in \eq{definition}, so one gets the
moment--space result by analytic continuation: $iP^+y^-
\longrightarrow N$. This leads to~\eq{F_N_be_approx}.

A complementary point of view was presented in Ref.~\cite{BDK}. One
begins by computing the quark distribution in an on-shell heavy
quark, \eq{definition_PT}, which differs from the object of
interest, \eq{definition}, by the external states. Moments of this
distribution are infrared safe: all {\em logarithmic} singularities
cancel out between real and virtual corrections. However, the
all--order result for these moments has infrared
renormalons\footnote{See Ref.~\cite{QD}  for the large--$\beta_0$
limit result.}, which indicate {\em power--like} sensitivity to
large distance physics. As discussed in Sec.~\ref{sec:Borel_rep_DGE}
above and in Refs.~\cite{BDK,QD}, the Sudakov factor $S_N$ of
\eq{Soft} has infrared renormalons with ambiguities that scale as
integer powers of $N\Lambda/m$:
\begin{equation}
S_N=S_N^{\PV} \times \exp\left\{\sum_{j=1}^{\infty} K_j\,\,
\left(\frac{N\Lambda}{m}\right)^j\right\}, \label{S_N_amb}
\end{equation}
where $K_j$ represent ${\cal O}(1)$ ambiguities. It was shown in
detail in Ref.~\cite{BDK} that $K_1$ has a special status: it is
directly related to the leading renormalon ambiguity in the
definition of the pole mass. This relation is realized in
\eq{F_N_be_approx} by the fact that the normalization of the ${\cal
O}(N/M)$ term in the exponent is the difference between the meson
mass and the quark pole mass. Higher--order terms in the exponential
factor in \eq{S_N_amb} are reflected in \eq{F_N_be_approx} by ${\cal
F}$.

Therefore, in principle the non-perturbative function ${\cal F}$ can
be thought of in terms of the OPE or from the renormalon
perspective. What is important in practice however, is the way the
separation between the perturbative and non-perturbative regimes is
implemented. Here we choose the Principal--Value prescription. We
believe that this facilitates making maximal use of perturbation
theory thus minimizing the role of power corrections.

Based on \eq{F_N_be_approx}, the physical photon--energy moments can
be computed from their perturbative counterparts by
\begin{equation}
M_N=M_N^{\PT,\PV}\times\exp\left\{-\frac{(N-1)\bar{\Lambda}_{\PV}}{M}\right\}\,\times\,{\cal
F}((N-1)\Lambda/M); \qquad \qquad \bar{\Lambda}_{\PV}\equiv
M-m_{\PV}, \label{M_N_physical}
\end{equation}
with the matched expression for $M_N^{\PT,\PV}$ given by
\eq{M_N_full}, where the Sudakov factor of \eq{Sud} is {\em defined}
in the Principal--Value prescription. This implies of course the
same prescription for $\bar{\Lambda}$ and for the higher--power
ambiguities in~${\cal F}$. In the numerical analysis that follows we
shall simply drop the unknown non-perturbative function~${\cal F}$.
Its parametrization would be worthwhile doing once stringent
theoretical constrains are available or when experimental data
become sufficiently precise.

In this approximation we get the following DGE result for the
differential spectrum in physical units:
\begin{eqnarray}\label{dGamma_dx_Gamma_N_NLO_NP}
\frac{1}{\Gamma_{\NLO}}\,\frac{d\Gamma(E_{\gamma})}{dE_\gamma} &=&
\,\frac{M}{2} \int_{c-i\infty}^{c+i\infty}\frac{dN}{2\pi i}
\left(\frac{2E_{\gamma}}{M}\right)^{-N} M_N
\\ \nonumber &\simeq &
\frac{m_{\PV}}{2}\, \int_{c-i\infty}^{c+i\infty}\frac{dN}{2\pi i}
\left(\frac{2E_{\gamma}}{m_{\PV}}\right)^{-N} M_N^{\PT,\PV},
\end{eqnarray}
where\footnote{For simplicity, we dropped the $\xi$ terms appearing
in \eq{dGamma_dx_Gamma_N_NLO}; while small, these terms are included
in all the numerical results that follow.} $M_N$ in the first line
is given by \eq{M_N_physical} with ${\cal F}= 1$ whereas
$M_N^{\PT,\PV}$ in the second is given by \eq{M_N_full}, where the
Sudakov factor of \eq{Sud} is defined in the Principal--Value
prescription. While the two lines in \eq{dGamma_dx_Gamma_N_NLO_NP}
are trivially equivalent --- the equality is violated by terms
${\cal O}(\Lambda/M)$ --- they reflect two different physical
interpretations: in the first line the moments are the physical
spectral moments which are free of any prescription dependence
whereas in the second these are resummed perturbative moments in a
given prescription.

The equality of the two formulations in
\eq{dGamma_dx_Gamma_N_NLO_NP} indicates that the accuracy at which
the {\em meson} mass is known is irrelevant for the calculation of
the spectrum. On the other hand, no matter which of the two is
chosen, an accurate determination of the pole mass in the
Principal--Value prescription is crucial.

\subsection{Calculation of $\bar{\Lambda}_{\PV}$\label{sec:Lambda_PV}}

In principle, $\bar{\Lambda}_{\PV}$, or the Principal--Value pole
mass $m_{\PV}$, can be determined from the relation with any other
well--defined mass. A natural choice is the relation with the
${\overline{\rm MS}}$ mass because it is reasonably well measured
\cite{Kuhn:2001dm}
\begin{eqnarray}
m_{\MSbar}(m_{\MSbar})=4.19\pm 0.05,  \,{\rm GeV}
\label{m_MSbar_val}
\end{eqnarray}
and the corresponding perturbative expansion is known to the
NNLO~\cite{Chetyrkin:1996vx}, while the normalization of the leading
infrared renormalon is well under control --- see Appendix
\ref{sec:pole_mass_scheme_inv} and
Refs.~\cite{Pineda:2001zq,Lee:2003hh,Cvetic:2003wk,Niko}.

Let us begin by defining $m_{\PV}$ through the Principal Value of
the standard Borel sum (see
Appendix~\ref{sec:pole_mass_scheme_inv}):
\begin{equation}
\frac{m_{\PV}}{m_{\MSbar}}=1+\frac{C_F}{\beta_0}\, {\rm
PV}\int_0^{\infty}dz B(z)\,{\rm e}^{-z/\bar{A}}, \label{m_over_m}
\end{equation}
where $\bar{A}=\beta_0\alpha_s(m_{\MSbar})/\pi$. Owing to the
dominance of the first infrared renormalon, which is apparent
already at low orders, and to the fact that the singularity
structure of this renormalon is exactly known~\cite{Beneke:1994rs},
it is possible to accurately estimate the normalization of the
renormalon
singularity~\cite{Pineda:2001zq,Lee:2003hh,Cvetic:2003wk,Niko} and
thus construct a bilocal expansion of the form:
\begin{eqnarray}
\label{B_of_z} B(z)=\sum_{n=0}^{n_{\max}} b_n z^n
+\frac{q}{(1-2z)^{1+\frac12\delta}}
\left[1+\sum_{k=1}^{k_{\max}}c_k(1-2z)^k \right],
\end{eqnarray}
where, as before, $\delta=\beta_1/\beta_0^2$. Here the first sum is
known to the NNLO and we therefore use $n_{\max} =2$. The
coefficients in the second term depend only on the coefficients of
the ${\overline{\rm MS}}$ $\beta$ function (see
e.g.~\cite{Pineda:2001zq}) which sets $k_{\max}=2$.

An accurate determination of the normalization constant $q$ is
essential for \eq{B_of_z} to be useful for the calculation of
$m_{\PV}$ from \eq{m_over_m}. Clearly, also the convergence of the
sums in \eq{B_of_z} is relevant; but assuming that $q$ is known,
both series in \eq{B_of_z} are free of any $z=\frac12$ renormalon
singularity\footnote{Higher renormalons are present in the first
sum. In \cite{Lee:2003hh} the convergence of this sum was further
improved using conformal mapping, which we do not use here.} and
thus their convergence is much better than that of the original
$m/m_{\MSbar}$ expansion. Therefore, the accuracy at which $m_{\PV}$
can be determined is comparable to the one at which $m_{\MSbar}$ is
known.

According to
Refs.~\cite{Pineda:2001zq,Lee:2003hh,Cvetic:2003wk,Niko} and
Appendix~\ref{sec:pole_mass_scheme_inv}, $q$ can be accurately
computed. Using \eq{R_opt_f} we obtain:
\begin{equation}
C_F q(N_f=3)/\pi=0.560;\qquad C_F q(N_f=4)/\pi=0.536. \label{q_det}
\end{equation}
This determination can be trusted within $\sim 2-3\%$. Proceeding to
evaluate $m_{\PV}$ from \eq{B_of_z} we obtain
\begin{eqnarray}
\left.\frac{m_{\PV}}{m_{\MSbar}}\right\vert_{N_f=3}=1.161\pm0.005;\qquad
\left.\frac{m_{\PV}}{m_{\MSbar}}\right\vert_{N_f=4}=1.164\pm0.005,
\label{mass_ratio_det}
\end{eqnarray}
where we used the values in \eq{q_det} for $q$,
$\alpha_s(m_b^{\MSbar})=0.226$, and estimated the error based on the
NNLO contribution. Taking into account the effect of the finite
charm mass we shall use ${m_{\PV}}/{m_{\MSbar}}=1.163\pm0.005$.
Thus, based on \eq{m_MSbar_val} and the measured value of the mason
mass, $M=5.279 \,{\rm GeV}$, we conclude that
\begin{equation}
m_{\PV}=4.874 \pm 0.050\,{\rm GeV};\qquad
\bar{\Lambda}_{\PV}=M-m_{\PV}=0.405\pm 0.050\,{\rm GeV},
\label{default_value_m_Lambda}
\end{equation}
where the error is dominated by that of the short--distance mass
determination, \eq{m_MSbar_val}.

\section{Numerical results and comparison with data\label{sec:pheno}}

The state--of--the--art analysis of the {\em total}
$\bar{B}\longrightarrow X_s\gamma$ branching ratio is based on the
NLO calculation~\cite{Gambino:2001ew,Buras:2002er,Buras:2002tp}.
According to Ref.~\cite{Gambino:2001ew} the integrated spectrum
above $m_b/20$ is:
\begin{eqnarray}
\label{Gambino_Misiak_BR_numeric} {\mathrm
{BR}}[\bar{B}\longrightarrow X_s\gamma]_{E_{\gamma}>E_0=m_b/20}
\simeq 3.73\,\cdot 10^{-4}.
\end{eqnarray}
This prediction has $\sim 10\%$ accuracy. It is well known that a
much higher cut on the photon energy is experimentally unavoidable.
Realistic measurements in the B factories can be done with
$E_{\gamma}>E_0=1.815\, {\rm GeV}$ \cite{Koppenburg:2004fz}, but
higher cuts are advantageous and lead to smaller experimental
errors. Moreover, it turns out to be useful to measure spectral
moments over this limited energy domain. In this section we show
that direct comparison of such measurements to theory is now
possible.

To describe the total rate we follow Ref.~\cite{Gambino:2001ew}. In
this approach, instead of evaluating $\Gamma_{\NLO}$ in
\eq{Gamma_N_NLO} directly from its definition in \eq{Gamma_NLO}, one
uses the experimental value for the $\bar{B}\longrightarrow X_c
e\bar{\nu}$ branching ratio avoiding explicit dependence on the
b-quark~pole~mass. In addition, by an appropriate choice of mass
scheme, Ref.~\cite{Gambino:2001ew} accounted for some charm--mass
effects associated with charm quark loops in the $O_2$ operator.
These effects are formally NNLO in the strong coupling, but they are
numerically important. Comparing the result of
Ref.~\cite{Gambino:2001ew} with $E_0=m_b/20$,
\eq{Gambino_Misiak_BR_numeric}, to our calculation at the
corresponding cut value $\Delta=0.9$, we fix the normalization of
the branching ratio ${\rm BR}_0$:
\begin{eqnarray} \label{BR_final}
&& \hspace*{-8pt} {\mathrm {BR}}[\bar{B}\longrightarrow
X_s\gamma]_{E_{\gamma}>E_0} \!=\!{\rm BR}_0
\Bigg[\int_{c-i\infty}^{c+i\infty}\frac{dN}{2\pi
i}\,\frac{(1-\Delta)^{1-N}}{N-1}\, \, M_N^{\PT,\PV}\!\!+\!\!
\,\frac{C_F\alpha_s(m_b)}{\pi}\!\!\!\!\!\sum_{\scriptsize
\begin{array}{c}{i,j=1..8}\\
 { i\leq j}\end{array} }\!\!\!\!\! c_{ij}\xi_{ij}(\Delta)\Bigg],\nonumber \\
 && \hspace*{-7pt} {\rm with} \,\,\,\, {\rm BR}_0=2.86\pm 0.29.
\end{eqnarray}
This formula will now be used to predict the dependence on
$\Delta=1-2E_0/m_{\PV}$. Throughout this analysis we shall use the
central value\footnote{This is phenomenologically sensible because
the sources of uncertainty in the distribution are different from
those of the total branching fraction. } of ${\rm BR}_0$ in
\eq{BR_final}.
\begin{figure}[t]
\begin{center}
\epsfig{file=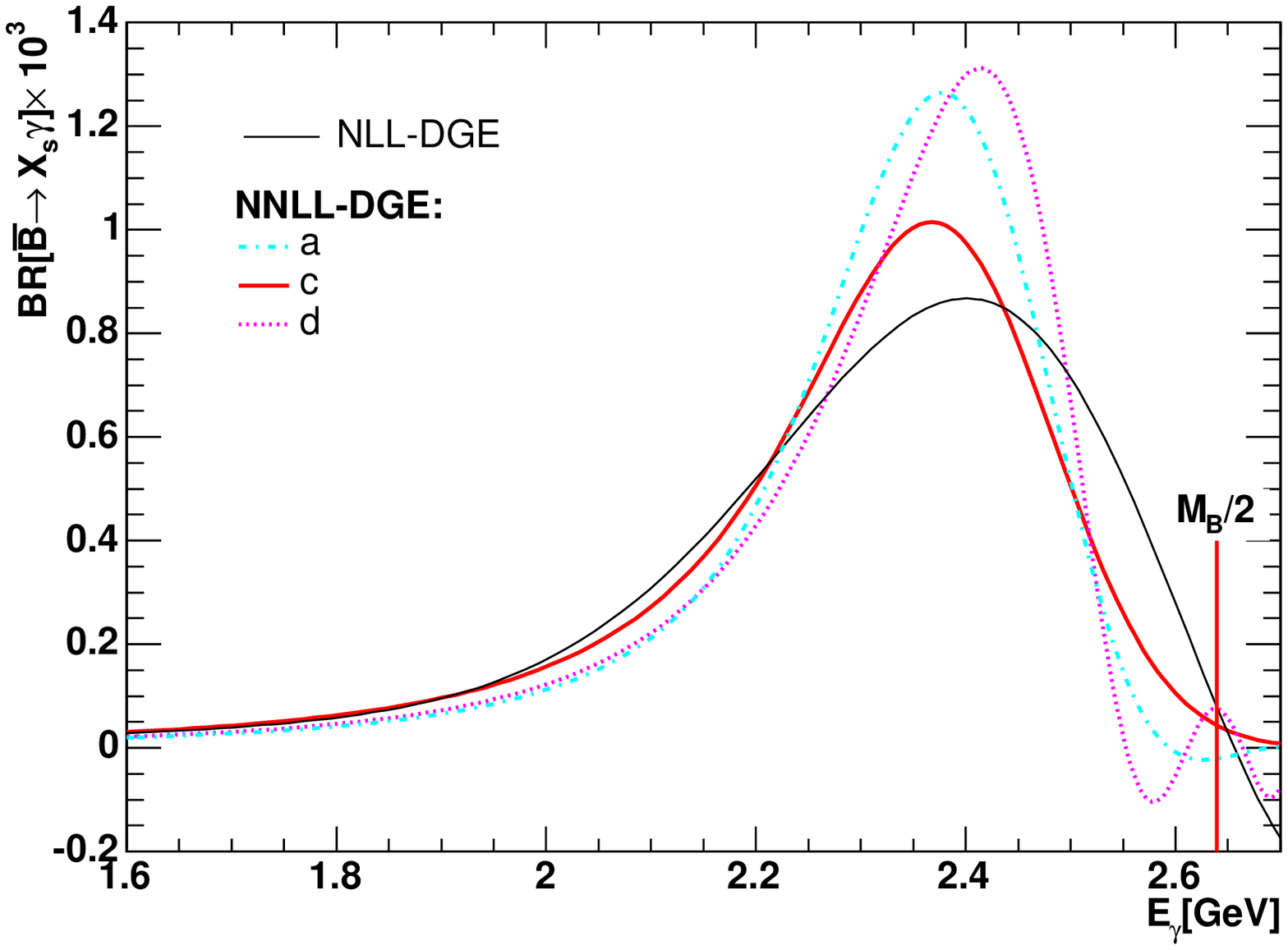,angle=0,width=12.5cm}
\epsfig{file=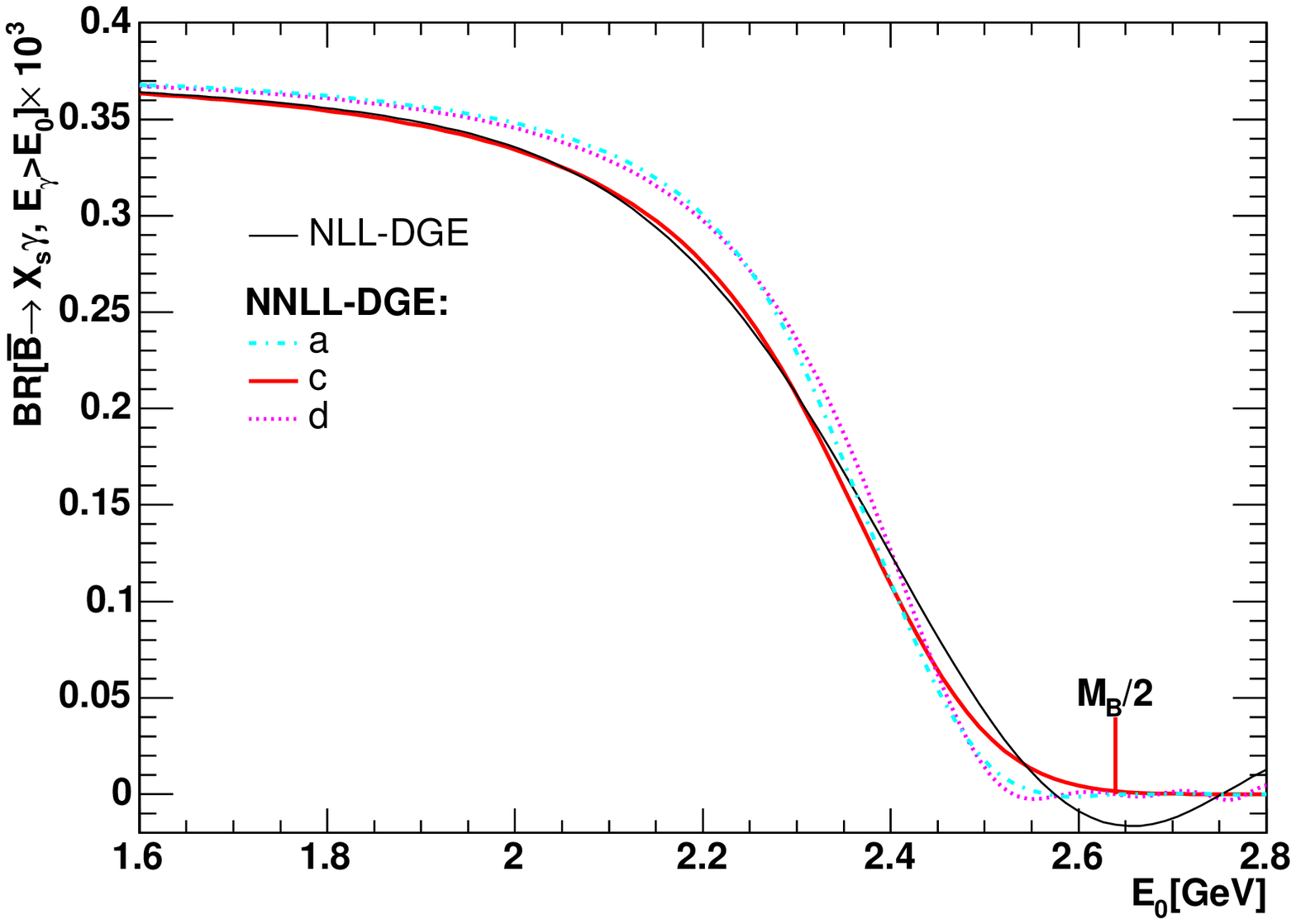,angle=0,width=12.5cm} \caption{
\label{fig:model_differences} The dependence of the differential
(top) and integrated (bottom) $\bar{B}\longrightarrow X_s\gamma$
spectrum as predicted by DGE on the model assumed for the Borel
function of the soft Sudakov exponent (see Sec.~\ref{sec:residue}).}
\end{center}
\end{figure}

\begin{figure}[t]
\begin{center}
\epsfig{file=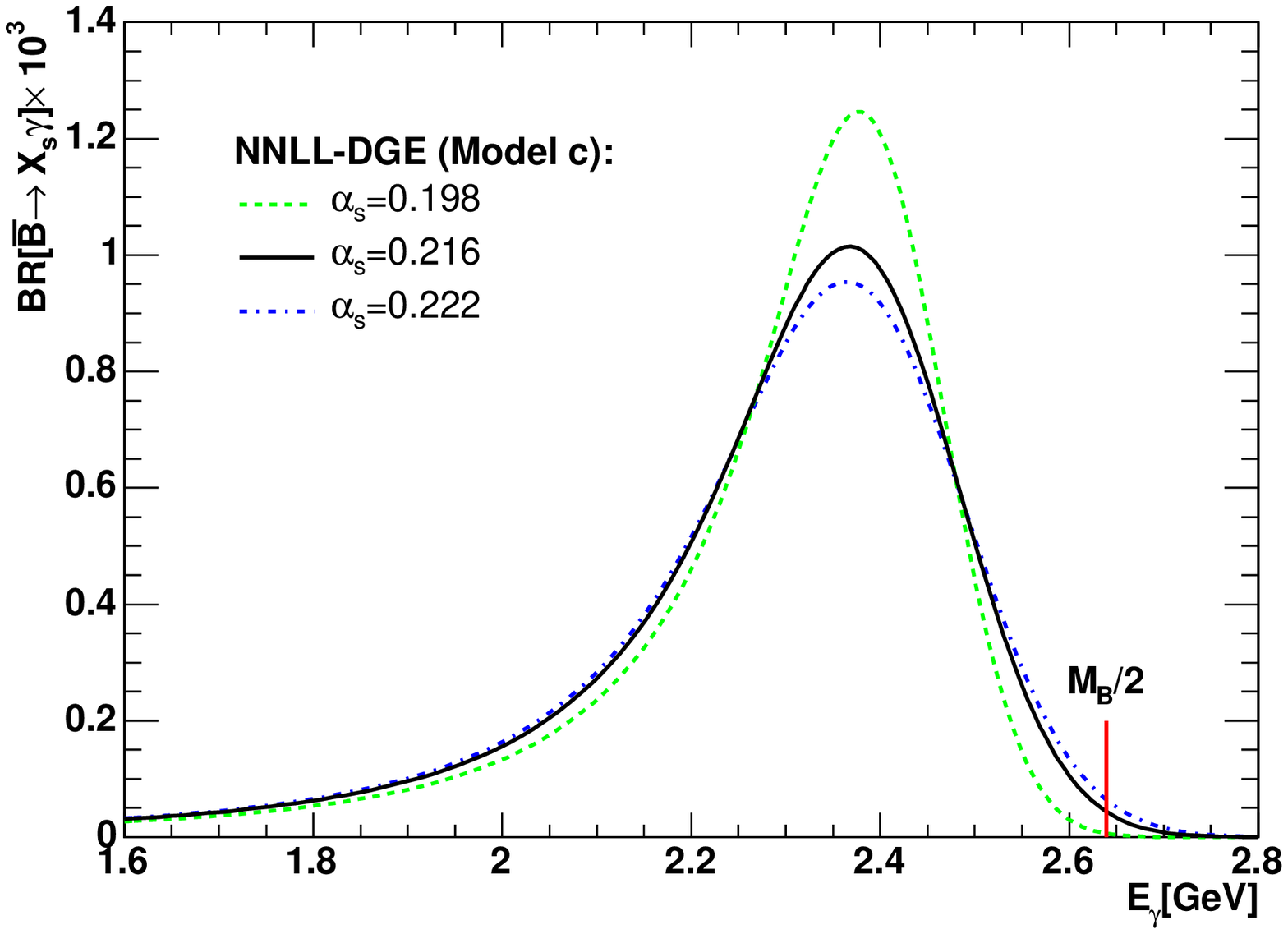,angle=0,width=12.5cm}
\epsfig{file=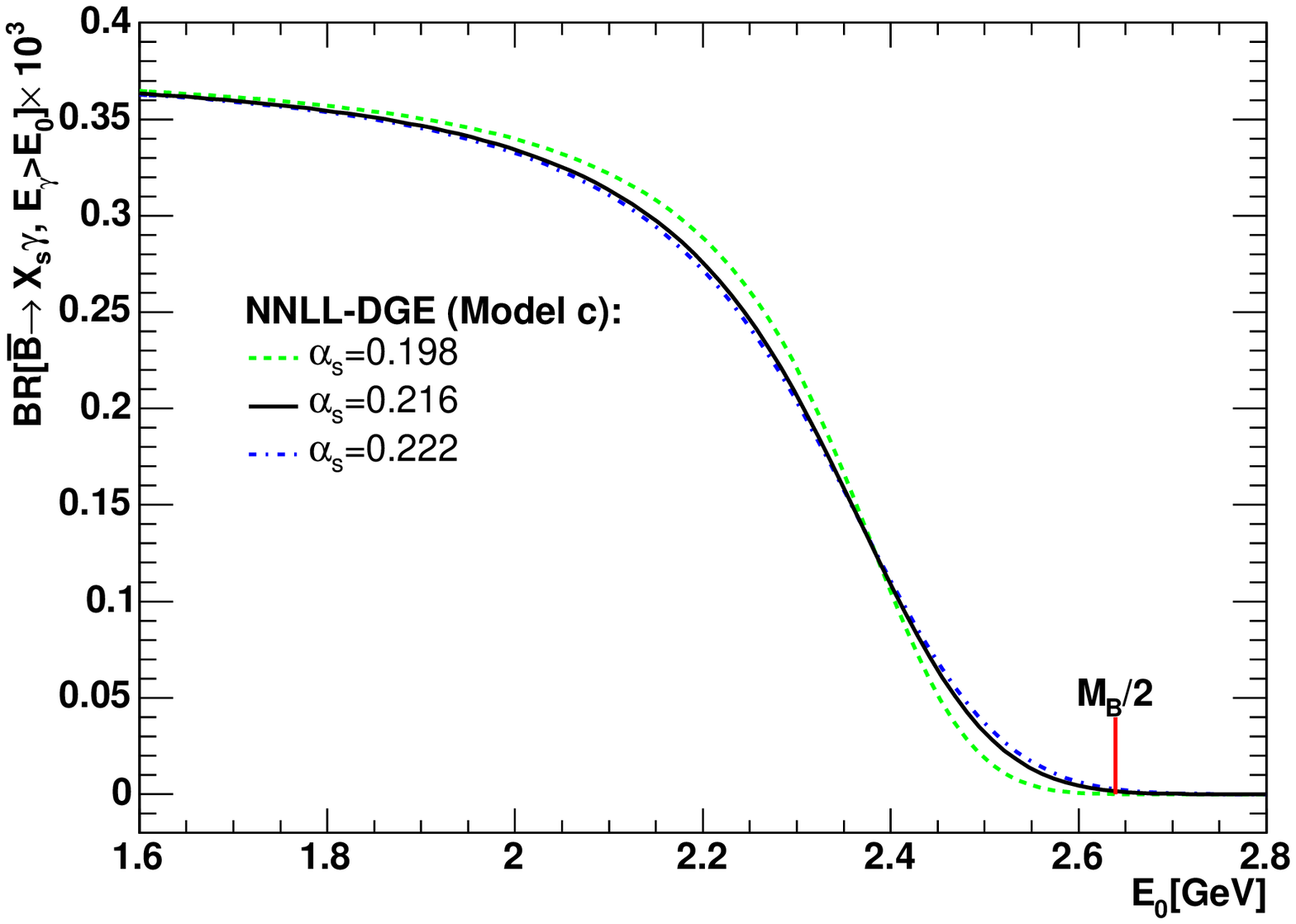,angle=0,width=12.5cm}
\caption{The dependence of the differential (top) and integrated
(bottom) $\bar{B}\longrightarrow X_s\gamma$ spectrum as predicted by
DGE on the value of the strong coupling. \label{alpha_s_dependence}
}
\end{center}
\end{figure}

\begin{figure}[th]
\begin{center}
\epsfig{file=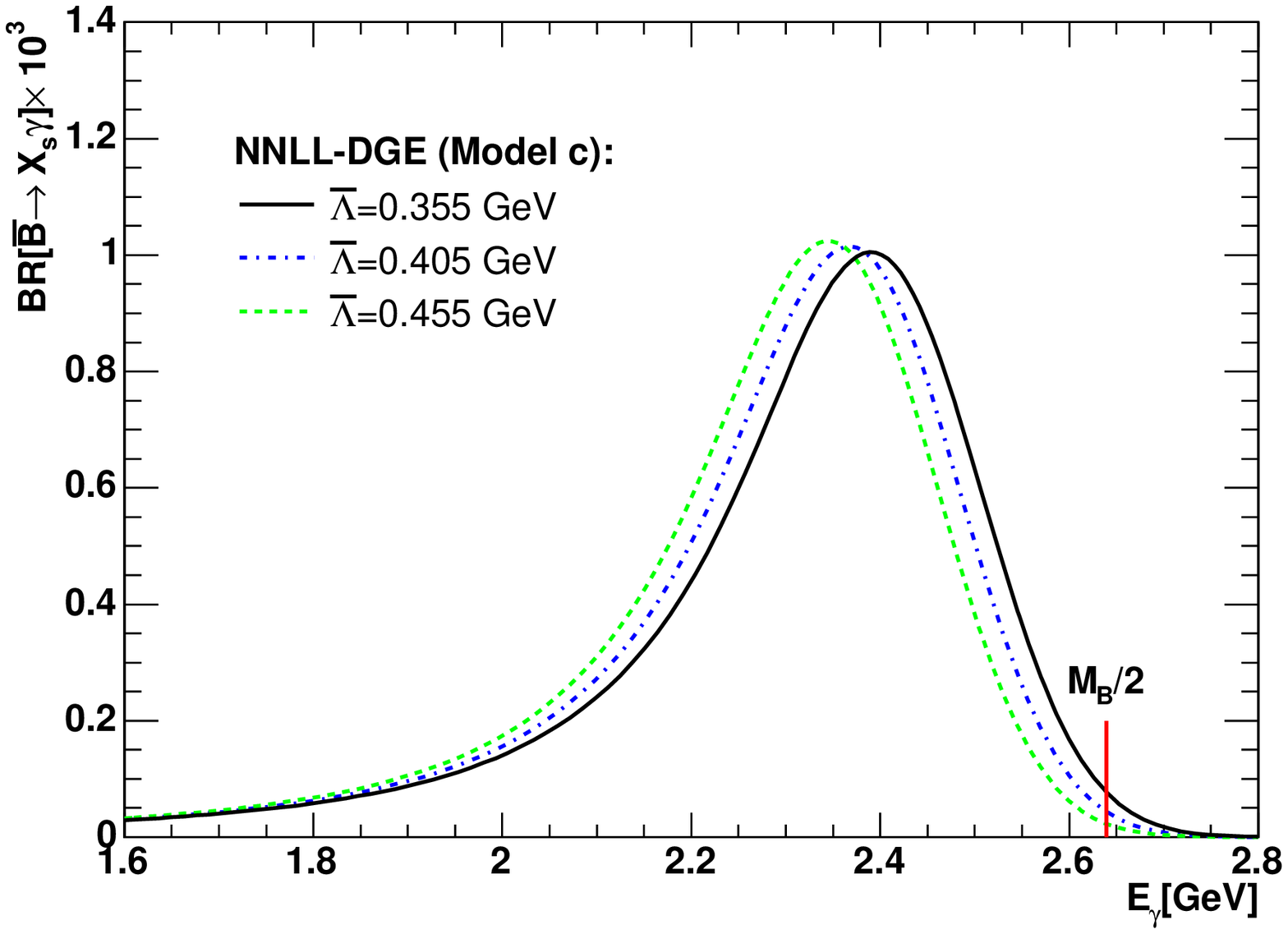,angle=0,width=12.5cm}
\epsfig{file=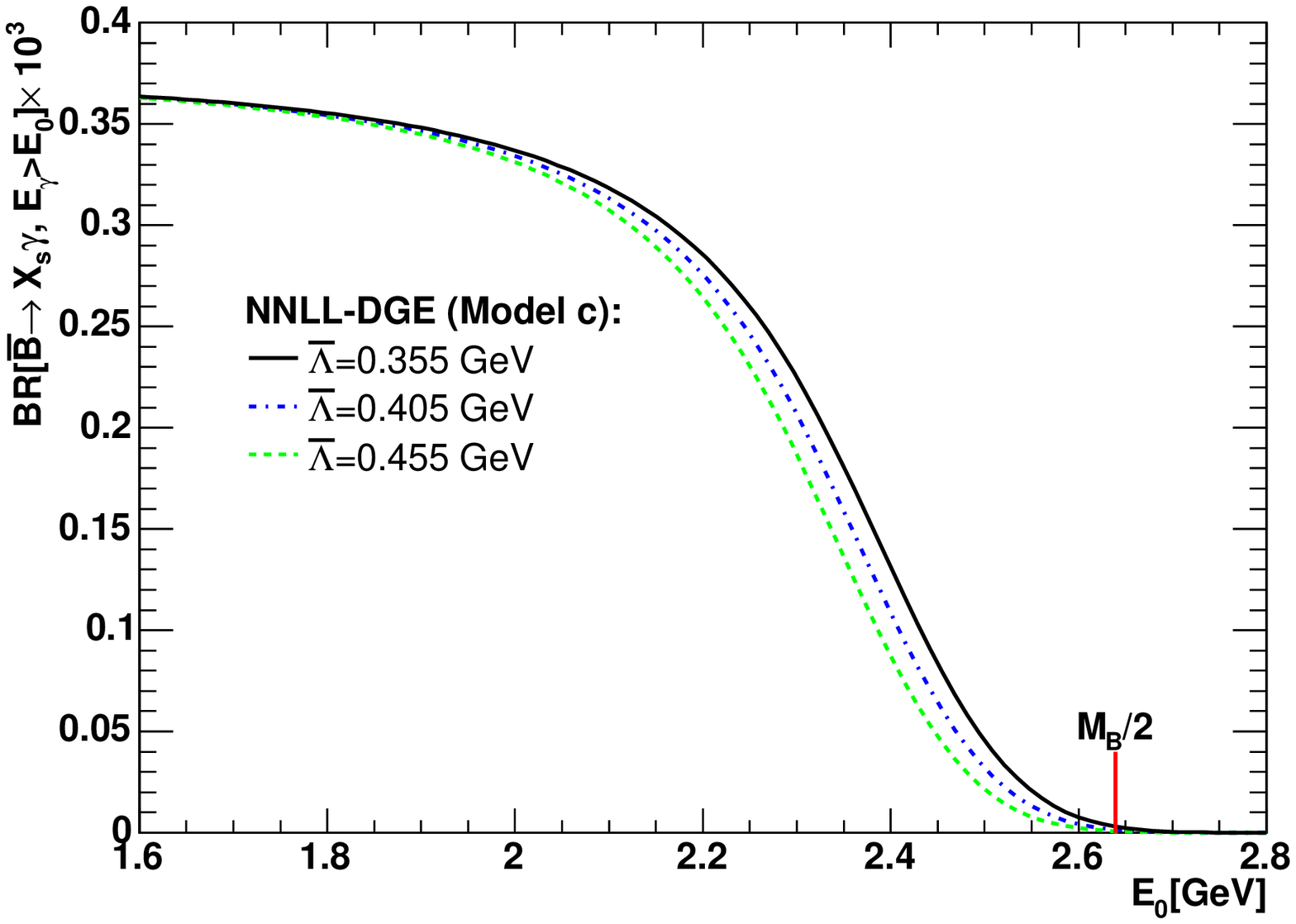,angle=0,width=12.5cm}
\caption{The dependence of the differential (top) and integrated
(bottom) $\bar{B}\longrightarrow X_s\gamma$ spectrum as predicted by
DGE on the assumed value for $\bar{\Lambda}$.
\label{fig:Lambda_bar_dependence}}
\end{center}
\end{figure}

Recall that $M_N^{\PT,\PV}$ in \eq{BR_final} is given by
\eq{M_N_full}, where the DGE Sudakov factor~(\ref{Sud}) is defined
in the Principal--Value prescription. As discussed in
Sec.~\ref{PC_soft} the use of the corresponding Principal--Value
mass $m_{\PV}$ in the definition of $\Delta$ guaranties consistency
at the level of the ``binding energy''; it effectively accounts for
the exponential factor in~\eq{M_N_physical}. On the other hand,
additional power corrections ${\cal O}\Big(((N-1)\Lambda/M)^2\Big)$
corresponding to~${\cal F}$ in~\eq{M_N_physical} are ignored. We are
assuming that they are numerically small when using DGE with the
Principal--Value prescription. Some evidence that this assumption is
sensible is:
\begin{itemize}
\item{} {\em The DGE prediction is perturbatively stable}:
as shown in Fig.~\ref{fig:model_differences} the change from
NLL--DGE to NNLL--DGE is moderate. This feature is best seen in
moment space, see Fig.~\ref{fig:Moments}. Perturbative stability is
of course only a necessary condition, not a sufficient one, as not
all sources of power corrections can be probed by perturbative
means.
\item{} {\em The predicted spectrum smoothly extends beyond the perturbative endpoint
and tends to zero for $E_{\gamma}=\left(m+{\cal
O}(\Lambda)\right)/2$, close to the physical endpoint
$E_{\gamma}=M/2$}. In this respect the differences between the
various models shown in Fig.~\ref{fig:model_differences} are less
than 100 MeV. This is quite remarkable given that in the close
vicinity of the endpoint {\em all} power corrections on the soft
scale are a priori important. On the other hand, this should not be
taken as an indication that all power corrections are small: they
may be large, yet conspire to cancel. Moreover, in the close
vicinity of the endpoint the approximation based on keeping power
corrections on the soft scale only breaks down: the non-perturbative
structure of the jet becomes important; in our formulation
resonances are simply ignored.
\end{itemize}
These are, of course, qualitative arguments. In the long run data
will hopefully allow quantifying these power corrections.

The immediate advantage of the DGE calculation of the spectrum is
the possibility to provide a reliable estimate of the partial
branching fraction and higher moments with experimentally relevant
cuts. To assess the remaining theoretical uncertainty in this
prediction let us first compare the chosen model $(c)$ for $B_{\cal
S}(u)$ (\eq{B_DJ_c}) to other models for the soft\footnote{Note that
closer to the endpoint also the behavior of $B_{\cal J}(u)$ away
from the origin becomes relevant.} Sudakov factor introduced in
Sec.~\ref{sec:residue}. Recall that the differences between these
models are associated with the structure of the Borel function away
from the origin, and are therefore equivalent in principle to the
power corrections on the soft scale $(m/N)$ that would be
parametrized by~${\cal F}$. On the other hand, using the differences
between the models as an error estimate is rather
conservative\footnote{A more precise uncertainty estimate can be
obtained by comparing different models that are consistent with the
NNLO expansion of $B_{\cal S}(u)$ as well as with $B_{\cal S}(1/2)$.
Since the differences between the models introduced in
Sec.~\ref{sec:residue} are of the same order of magnitude as other
sources of uncertainty (e.g. the values of short--distance
parameters) we postpone such detailed analysis to future work.},
since only model $(c)$ is, by construction, consistent with the
large--order behavior of the Sudakov exponent.
Fig.~\ref{fig:model_differences} shows the differential and
integrated spectra according to the various models for $B_{\cal
S}(u)$. Differences between models for the differential spectrum
near the peak exceed $20\%$. For the integrated spectrum, however,
this translates into less than $10\%$ difference for practically any
cut (excluding, of course, the immediate vicinity of the endpoint,
$E_{0}>2.4$ GeV).

\begin{table}[th]
\begin{center}
\begin{tabular}{|l|c|c|c|}
\hline $\alpha_s^{\MSbar}(M_Z)$ & $\alpha_s^{\MSbar}(m_{\PV})$ &
$\Lambda_{N_f=4}$ GeV &$\bar{\Lambda}_{\PV}=M-m_{\PV}$ GeV
\\
\hline 0.113 & 0.198 & 0.265  & 0.392
\\
\hline 0.1182& 0.216 & 0.353  & 0.405
\\
\hline 0.120 & 0.222 &  0.385 & 0.455
\\
\hline
\end{tabular}
\caption{Variation of the parameters entering the calculation of the
spectrum with $\alpha_s (M_Z)$.\label{alpha_s_value}}
\end{center}
\end{table}

\begin{figure}[th]
\begin{center}
\epsfig{angle=90,file=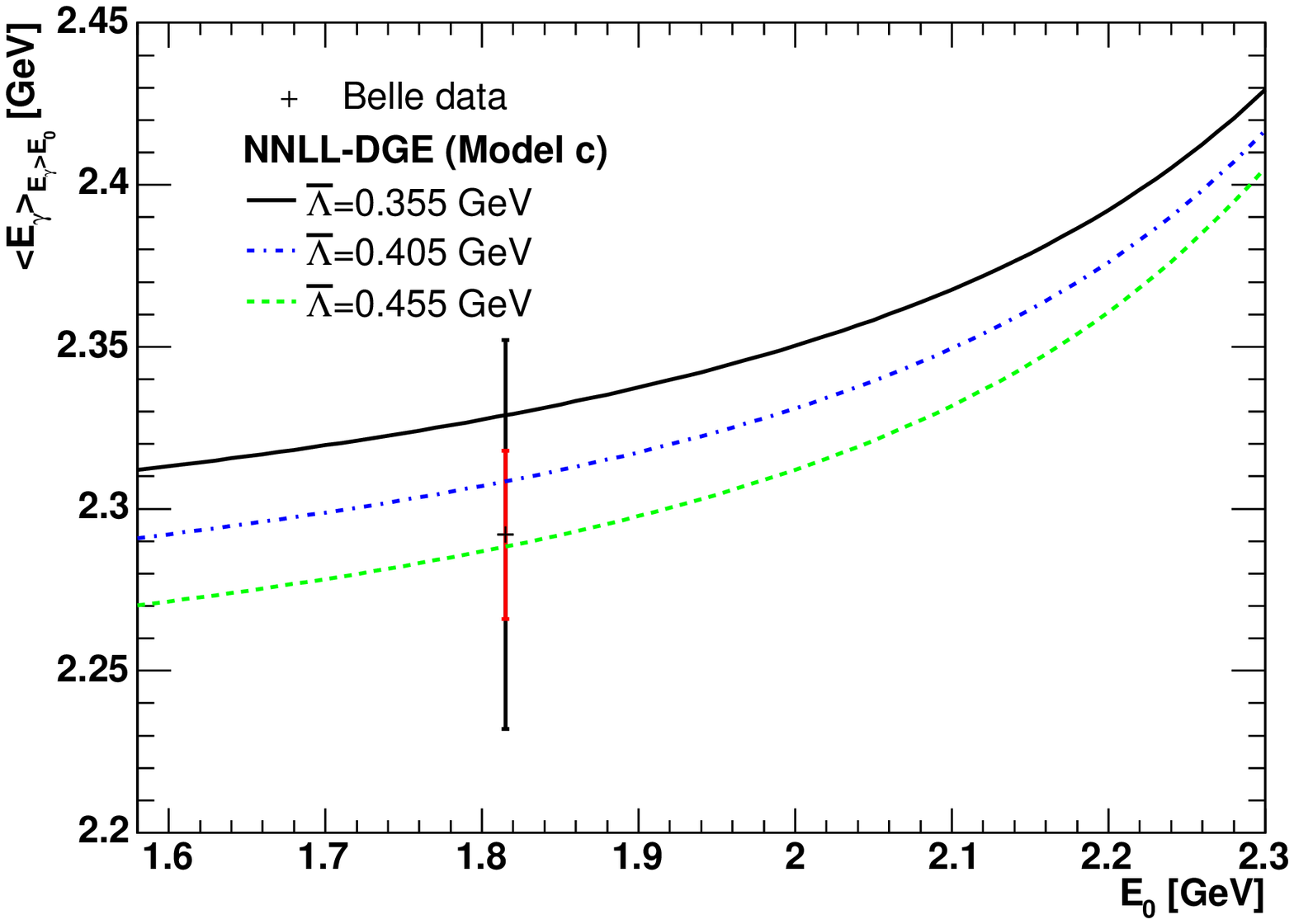,angle=-90,width=11.8cm}
\epsfig{angle=90,file=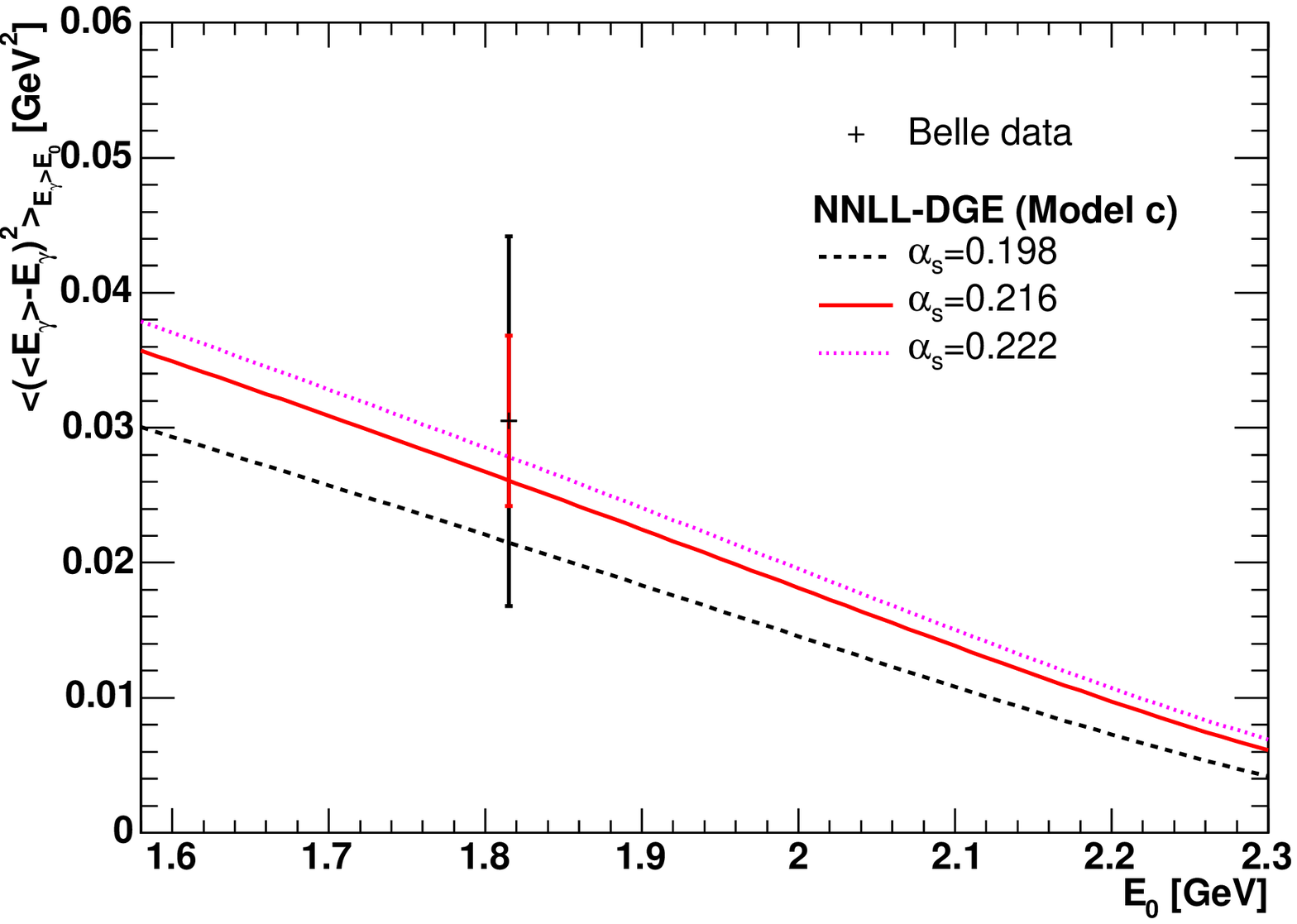,angle=-90,width=11.8cm}
\caption{The first two truncated moments of the
$\bar{B}\longrightarrow X_s\gamma$ spectrum:
$\left<E_{\gamma}\right>$ (top) and
$\left<(\left<E_{\gamma}\right>-E_{\gamma})^2\right>$ (bottom), as a
function of the minimum photon energy cut $E_0$, as calculated by
DGE using model $(c)$ of \eq{B_DJ_c} while varying $\alpha_s$ and
$m_{\MSbar}$ within their error ranges. In each plot we show a few
curves representing the dominant source of uncertainty considering
the various models for $B_{\cal S}(u)$ and the values of $\alpha_s$
and $m_{\MSbar}$. The DGE result is compared to the available data
from Belle
 \cite{Koppenburg:2004fz} with a cut $E_0=1.815$ GeV.
 \label{varying_energy_cuts} }
\end{center}
\end{figure}

\begin{figure}[th]
\begin{center}
 \epsfig{file=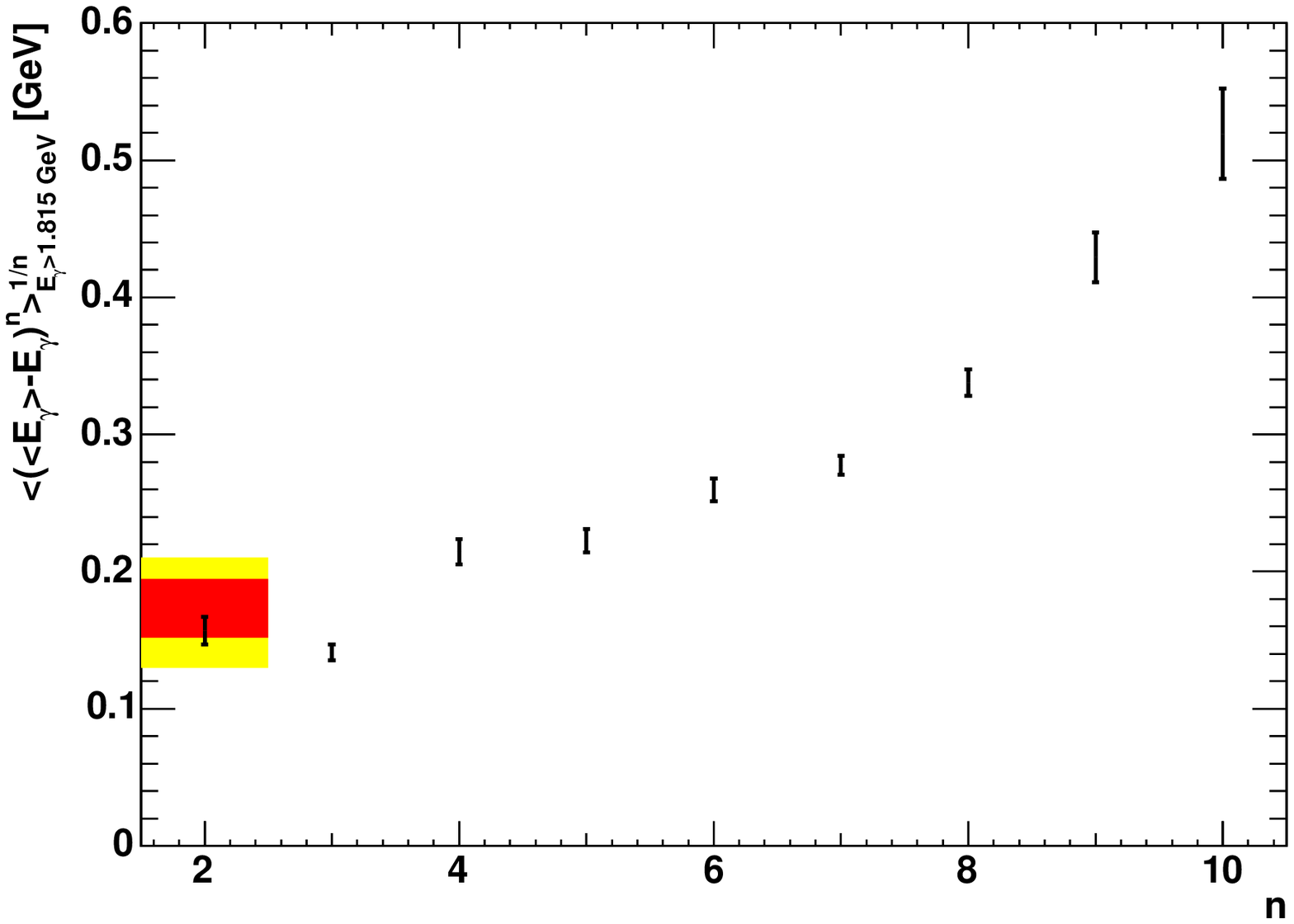,angle=0,width=13.4cm} \caption{\label{fig:moments_1_to_10} DGE
results for the first ten truncated moments of the
$\bar{B}\longrightarrow X_s\gamma$ spectrum,
$\left<(\left<E_{\gamma}\right>-E_{\gamma})^n\right>$, with a cut
$E_{\gamma}>E_0=1.815 \,{\rm GeV}$. All moments are presented in
energy units (GeV) by taking the corresponding $n$-th root. The
$n=2$ result is compared with Belle data, where the internal and
external shaded regions represent statistical and statistical plus
systematic errors, respectively.  }
\end{center}
\end{figure}

Another theoretical--uncertainty estimate can be obtained by varying
the input parameters $\alpha_s$ and $m_{\MSbar}$ within their error
ranges. The sensitivity of the DGE spectrum to these parameters is
higher than in typical short--distance calculations. The precise
value of the coupling is important since $\alpha_s$ is probed at low
scales ${\cal O}(m/N)$. The value of the short--distance mass
$m_{\MSbar}$ directly influences the pole mass $m_{\PV}$
(Sec.~\ref{sec:Lambda_PV}), which sets the scale for the spectrum in
physical units, see \eq{dGamma_dx_Gamma_N_NLO_NP}. Table
\ref{alpha_s_value} details the values of the parameters entering
the calculation for the central and the two extreme choices we made
for $\alpha_s^{\MSbar}(M_Z)$. Fig.~\ref{alpha_s_dependence} shows
the corresponding differential and integrated spectra.
Fig.~\ref{fig:Lambda_bar_dependence} shows the spectrum for the
central value of of the coupling, $\alpha_s(m)=0.216$, while varying
$\bar{\Lambda}$ by $\pm 50$ MeV. This corresponds, to a good
approximation, to varying $m_{\MSbar}$ within its error range. Note
that while the variation in the coupling influences all the moments
(the shape of the distribution gets modified) the variation of
$m_{\MSbar}$ mainly affects the average energy: it generates a
global shift of the distribution.

A word of caution is due regarding the interpretation of the
theoretical uncertainty reflected in the figures. As usual,
uncertainty estimates are based on corrections that are known, e.g.
degrading the NNLL--DGE to NLL--DGE. In this paper we have not
parametrized non-perturbative power corrections, whose influence
would increase towards the endpoint. Our analysis only reflects
these effects through the different models for $B_{\cal S}(u)$. Such
differences might not capture all possible power corrections. In
this case, the uncertainty may be underestimated near the endpoint.

Next, let us consider the average of the photon energy with a cut,
namely
\begin{equation}
\left<E_{\gamma}\right>_{E_{\gamma}>E_0}\equiv\frac{\displaystyle
\int_{E_0} dE_{\gamma}\, \frac{d\Gamma(E_{\gamma})}{dE_{\gamma}}\,
E_{\gamma}}{\displaystyle \int_{E_0} dE_{\gamma}\,
\frac{d\Gamma(E_{\gamma})}{dE_{\gamma}}} \label{ave_E}
\end{equation}
and, similarly, higher truncated moments:
\begin{equation}
\left<\left(\left<E_{\gamma}\right>_{E_{\gamma}>E_0}-
E_{\gamma}\right)^n\right>_{E_{\gamma}>E_0}\equiv\frac{\displaystyle\int_{E_0}
dE_{\gamma}\, \frac{d\Gamma(E_{\gamma})}{dE_{\gamma}}\,
\left(\left<E_{\gamma}\right>_{E_{\gamma}>E_0}-
E_{\gamma}\right)^n}{\displaystyle \int_{E_0} dE_{\gamma}\,
\frac{d\Gamma(E_{\gamma})}{dE_{\gamma}}}. \label{higher_cent_mom}
\end{equation}
Note that these moments differ from the standard Mellin moments
discussed so far in two respects: first, they are defined over a
limited photon--energy range, and second, these are ``central
moments'' in the sense that they depend of the {\em difference}
between the photon energy and the average energy in this range.

For sufficiently high $E_0$ cuts and low enough $n$ these moments
are accessible experimentally. Results for the average energy,
\eq{ave_E}, and the variance ($n=2$ in \eq{higher_cent_mom}) were
recently published by Belle~\cite{Koppenburg:2004fz,Limosani:2004jk}
with $E_0=1.815$ GeV. In Fig.~\ref{varying_energy_cuts} we compare
this experimental result to our calculation while providing
additional theoretical predictions for the dependence on the cut
value. We find very good agreement. The experimental error is
somewhat larger than the theoretical one. Note that the dominant
source of uncertainty in the theoretical prediction for the average
energy is the value of $m_{\MSbar}$, affecting the calculation of
the pole mass $m_{\PV}$ or $\bar{\Lambda}_{\PV}$. In the future it
may even be possible to get an accurate measurement of $m_{\MSbar}$
from experimental data for
$\left<E_{\gamma}\right>_{E_{\gamma}>E_0}$. On the other hand, the
dominant source of uncertainty in the variance is the value of the
strong coupling. More interesting comparison can hopefully be done
in the future with experimental data having higher cuts, where the
systematic experimental errors are expected to be smaller. The
possibility to make comparison between theory and data as a function
of the cut has an important added value: clearly, at sufficiently
high $E_0$ values the theoretical prediction, which is lacking any
non-perturbative corrections, will fail.

Finally, in Fig.~\ref{fig:moments_1_to_10} we show the theoretical
prediction for the first ten moments. Going to higher moments the
finer details of the shape become important. The expectation is, of
course, that power corrections will become increasingly important at
high $n$. Having excluded power corrections, the dominant source of
uncertainty in Fig.~\ref{fig:moments_1_to_10} is the value of
$\alpha_s$. Since for higher moments the coupling is probed at lower
scales $\sim {\cal O}(m/n)$, this uncertainty gradually increases
with $n$. Following Belle \cite{Koppenburg:2004fz} we have chosen
$E_{\gamma}>E_0=1.815 \,{\rm GeV}$. Fig.~\ref{fig:moments_1_to_10}
shows the comparison with Belle data for the variance; there is no
data to compare with for $n\geq 3$.

\section{Conclusions\label{sec:conclusions}}

Inclusive $B$--decay spectra present a special challenge to theory:
because of their inclusive nature the details of the hadronic wave
function are largely irrelevant, but on the other hand, the soft
scales involved prohibit a straightforward perturbative approach.
\begin{itemize}
\item{}
The clearest manifestation of the non-perturbative nature of the
$\bar{B}\longrightarrow X_s\gamma$ spectrum is that any fixed--order
perturbative result has support only for $E_{\gamma}<m/2$ while the
physical spectrum extends to $E_{\gamma}=M/2$. The gap between the
quark pole mass and the meson mass is related to non-perturbative
dynamic of the bound state.
\item{}
Moreover, the perturbative endpoint region is characterized by
parametrically large corrections.  The conventional approach to
Sudakov resummation, which relies on the dominance of logarithmic
corrections in the large--$N$ limit where $\alpha_s(m) \ln N$ is
fixed, fails because of large running--coupling effects. The
breakdown of the perturbative expansion with increasing logarithmic
accuracy is demonstrated in Fig.~\ref{fig:matched_NNLL_spectrum}
(see also Fig.~2 in \cite{BDK}).
\end{itemize}

In this paper we showed that DGE provides a solution to both these
problems. Although it is based on resummation of perturbation theory
the DGE--resummed spectrum {\em does not have perturbative support}.
By taking the Principal--Value prescription in moment space this
inherent limitation of any fixed--order approximation is removed.
The spectrum smoothly extends to the non-perturbative regime and
tends to zero for $E_{\gamma}=\left(m+{\cal O}(\Lambda)\right)/2$,
qualitatively as expected in the meson decay (see
Fig.~\ref{fig:Model_1}).

Of course, the DGE spectrum does not have correct non-perturbative
support properties: it does not strictly vanish for
$E_{\gamma}>M/2$. Moreover, is not necessarily positive definite for
any $E_{\gamma}<M/2$ (see e.g.~Fig.~\ref{fig:model_differences}).
Indeed, the close vicinity of the endpoint is beyond theoretical
control as it involves an infinite number of non-perturbative power
corrections on the soft scale $m/N$, and eventually also the
non-perturbative structure of the jet.

In this work we presented perturbative predictions and refrained
from making any parametrization of non-perturbative power
corrections. It is conceivable that there exist a region where {\em
a few} non-perturbative corrections on the soft scale are
phenomenologically relevant, while corrections on the jet--mass
scale can be ignored. Once experimental data are precise enough,
power corrections on the soft scale will be worthwhile parametrizing
because of their universal nature: they are related exclusively to
the quark distribution in the meson. Therefore, fixing their
magnitude can be instrumental to precision measurements of $V_{ub}$
from  charmless semileptonic B decays. The DGE spectrum presented
here provides a baseline for systematic parametrization of such
non-perturbative power corrections. Perturbative stability is
imperative in this respect. The remarkable stability of the DGE
result (compare Fig.~\ref{fig:Moments}
to~Fig.~\ref{fig:moments_FLA}) is largely thanks to the resummation
of running--coupling effects.

It should be emphasized that the difference between DGE and any
fixed--logarithmic--accuracy approach is a qualitative one. Recall,
that an accurate value of $m_{\PV}$ (or $\bar{\Lambda}_{\PV}$) is
essential for the calculation of the photon--energy spectrum in
physical units (see Sec.~\ref{sec:PC}). As opposed to the general
concept of the pole mass, $m_{\PV}$ has no linear ${\cal
O}(\Lambda)$ ambiguity. It can be determined with roughly the same
accuracy as short--distance masses ($\sim 50$ MeV). It is only
because the Sudakov exponent was defined using the Principal--Value
prescription that $m_{\PV}$ becomes the relevant mass. This is a
manifestation of the cancellation of renormalon ambiguities between
the quark--distribution subprocess in the Sudakov factor and the
pole mass~\cite{BDK}. In a fixed--logarithmic--accuracy approach the
renormalon in the Sudakov factor is hidden --- it is realized
through the divergence of the expansion in \eq{sum_FLA} --- so the
choice of mass scheme when computing the spectrum in physical units
becomes ad hoc.

Experimentally--favorable observables are moments of the
photon--energy spectrum defined over a limited range,
$E_{\gamma}>E_0$, Eqs. (\ref{ave_E}) and (\ref{higher_cent_mom}).
Having obtained a stable prediction in Mellin space over a wide
domain in the complex $N$ plane, we can reliably compute the
truncated moments over a range of experimentally--relevant cut
values. We showed that our calculation nicely agrees with the recent
data from Belle~\cite{Koppenburg:2004fz}. Furthermore, a great
variety of possibilities for comparison between data and theory is
now open, e.g. studying the dependence on the cut value and higher
truncated moments. In this way more detailed information on the
distribution can be systematically extracted. This is, in
particular, imperative to quantifying power corrections.

The qualitative success of the DGE calculation presented here, which
does not involve any non-perturbative parameters, and the good
agreement with the available data from Belle, indicate that
additional power corrections in this framework are indeed small.
This approach is therefore promising for extracting $V_{ub}$ from
charmless semileptonic B decays.

\section*{Note added}

A few weeks after the submission of this paper, the BaBar
collaboration has presented~\cite{Walsh} new preliminary data for
the partial branching ratio, as well as the average energy and the
variance as a function of the cut. Comparison of the predictions in
Figs.~\ref{fig:Lambda_bar_dependence} and~\ref{varying_energy_cuts}
above with these measurements can be found in
Ref.~\cite{Gardi:2005mf}.

\section*{Acknowledgements}

EG wishes to thank Vladimir Braun, Gregory Korchemsky and Arkady
Vainshtein for very interesting discussions and the KITP at UC Santa
Barbara for hospitality while some of this work was done. JRA
acknowledges the support of PPARC (postdoctoral fellowship
PPA/P/S/2003/00281). The work of EG is supported by a Marie Curie
individual fellowship, contract number HPMF-CT-2002-02112.

\appendix{}

\section{Singular terms at NNLO\label{sec:NNLO}}

 Let us expand the Sudakov exponent in \eq{sum_FLA}
to ${\cal O}\left(\alpha_s^2\right)$. We get:
\begin{eqnarray}
{\rm Sud}(m,N)&=&\exp\bigg\{
\left(\frac{\alpha_s^{\MSbar}(m)}{\pi}\right)\bigg[ - {\displaystyle
\frac {1}{2}} \,\ln^{2}N\, {A_{1}} + ( - \gamma_E \,{A_{1}} -
{B_{1}} + {\mathit{D}_{1}})\, \ln N \bigg]\,
\\&&\hspace*{-50pt}
+ {\left(\frac{\alpha_s^{\MSbar}(m)}{\pi}\right)}^{2}\bigg[ -
{\displaystyle \frac {1}{2}} \,{\beta _{0}}\,\ln^{ 3}N\,{A_{1}}
\nonumber
 + \left(\left( - {\displaystyle \frac {3}{2}} \,\gamma_E
\,{A_{1} } + {\mathit{D}_{1}} - {\displaystyle \frac {1}{2}}
\,{B_{1}}\right) \,{\beta _{0}} - {\displaystyle \frac {1}{2}}
\,{A_{2}}\right)\, \ln^{2}N \nonumber \\&& \hspace*{-50pt}
 +\left(\left( - \gamma_E \,{B_{1}} + 2\,\gamma_E \,{\mathit{D}_{1}}
 - {\displaystyle \frac {3}{2}} \,{A_{1}}\,\gamma_E ^{2} -
{\displaystyle \frac {1}{4}} \,{A_{1}}\,\pi ^{2}\right)\,{\beta
_{0}}
 - \gamma_E \,{A_{2}} - {B_{2}} + {\mathit{D}_{2}}\right)\,\ln
N \bigg] \nonumber +\cdots \bigg\}
\end{eqnarray}

To get $\bar{M}_N^{\PT,\,O_7 }$ in \eq{BXg_logs_N_2loop} one needs
the matching coefficient, which is currently known only to ${\cal
O}(\alpha_s)$.  It is given by~\cite{BDK}:
\begin{eqnarray}
\label{matching} &&C_N^{O_7} (\alpha_s(m))=1+
\frac{C_F\alpha_s}{2\pi}\,\bigg\{\bigg[ \left({\displaystyle \frac
{7}{2}}  - {\displaystyle \frac {1}{N\,(N + 1)}}  + {\displaystyle
\frac {2}{N}} \right)\,\left(\Psi (N) + \gamma_E \right)+\Psi_1(N) -
{\displaystyle \frac {\pi ^{2}}{6}}
 \nonumber \\ &&
\hspace*{70pt}- \left(\Psi (N) + \gamma_E \right)^{2}  -
{\displaystyle \frac {31}{6}}  + {\displaystyle \frac {9}{2\,N}}
 + {\displaystyle \frac {1}{(N + 1)^{2}}}  -
{\displaystyle \frac {1}{N + 2}}  - {\displaystyle \frac {1}{2\,N
 + 2}}  + {\displaystyle \frac {1}{N^{2}}} \bigg] \nonumber
 \\&&\hspace*{70pt}-\,
\bigg[-\ln^2 N+\left(\frac72-2\gamma_E\right)\ln N\bigg] \bigg\}
+{\cal O}\left(\alpha_s^2\right).
\end{eqnarray}
Let us note that having fixed the anomalous dimensions as above, the
only missing ingredient for Sudakov resummation of decay spectra
with NNLL accuracy is the $N$--independent term at ${\cal
O}(\alpha_s^2)$.

For easy comparison with future two--loop calculations of the decay
process, we summarize below the log--enhanced terms to order
$\alpha_s^2$. Expanding the exponent and taking into account the
constant terms at order $\alpha_s$ in~\eq{matching} we obtain the
following log-enhanced terms at ${\cal O}\left(\alpha_s^2\right)$:
\begin{eqnarray}
\label{N_space} &&\hspace*{-15pt} \bar{M}_N^{\PT,\,O_7 }=1   +
\bigg\{\left({\displaystyle \frac {7}{4}}  - {\displaystyle \frac
{1}{2N\,(N + 1)}}  + {\displaystyle \frac {1}{N}}
\right)\,\left(\Psi (N) + \gamma_E \right)+\frac12\Psi_1(N) -
\frac12\left(\Psi (N) + \gamma_E \right)^{2}
 \nonumber \\ &&
\hspace*{40pt}  - {\displaystyle \frac {\pi ^{2}}{12}}-
{\displaystyle \frac {31}{12}}  + {\displaystyle \frac {9}{4\,N}}
 + {\displaystyle \frac {1}{2(N + 1)^{2}}}  -
{\displaystyle \frac {1}{2(N + 2)}}  - {\displaystyle \frac
{1}{4\,(N
 + 1)}}  + {\displaystyle \frac {1}{2N^{2}}}\bigg\}
 \,{\frac{\alpha_s^{\MSbar}(m)}{\pi}}
\nonumber \\&+& \bigg\{{   \frac {C_F ^{2}}{8}}\,\ln^{4} N  + \bigg[
{   \frac {1}{12}}\,C_F \,N_f  + \left({   \frac {\gamma_E }{2}}  -
{   \frac {7}{8} } \right)\,C_F ^{2} - {   \frac {11}{24}}\,C_F \,
C_A  \bigg]\,\ln^{3} N \nonumber \\&+& \bigg[ \left( - {   \frac
{13}{144}}  + {   \frac { \gamma_E }{4}} \right)\,C_F \,N_f + \left(
- { \frac {21}{8}} \,\gamma_E  + {   \frac {3}{4}} \,\gamma_E ^{2} +
{   \frac {271}{96}}  + {   \frac {1 }{24}} \,\pi ^{2}\right)\,C_F
^{2} \nonumber \\&+& \left({   \frac {95}{ 288}}  - {   \frac
{11\,\gamma_E }{8}}  + {   \frac {\pi ^{2}}{24}} \right)\,C_A \,C_F
\bigg]\ln^{2} N  + \bigg[\left( - {   \frac {13}{72}} \,\gamma_E  -
{   \frac {85}{144}}  + {   \frac {1}{4}} \,\gamma_E ^{2} + { \frac
{5}{72}} \,\pi ^{2}\right)\, C_F \,N_f
  \nonumber \\&+& \left( - {   \frac {13\,\pi ^{2}}{48}}  +
{   \frac {3}{2}} \,\zeta_3 + {   \frac { \gamma_E \,\pi ^{2}}{12}}
- {   \frac {425}{96}}  + {   \frac {\gamma_E ^{3}}{2}}  - {   \frac
{ 21\,\gamma_E ^{2}}{8}}  + {   \frac {271\,\gamma_E }{48}}
\right)\,C_F ^{2}
   \\ \nonumber&+& \left( - {   \frac {1}{4}} \,\zeta_3 -
{   \frac {67\,\pi ^{2}}{144}}  + { \frac {\gamma_E \,\pi ^{2}}{12}}
+ {   \frac {95\, \gamma_E }{144}}  + {   \frac {905}{288}}  - {
\frac {11\,\gamma_E ^{2}}{8}} \right)\,C_A \, C_F \bigg]\ln
N\bigg\}\left({\frac{\alpha_s^{\MSbar}\left(m\right)}{\pi}}\right)^{2}
+\cdots .
\end{eqnarray}
Note that the normalization of the rate influences not only constant
terms at ${\cal O}(\alpha_s)$ but also the contributions to the
$\ln^2N$ and the $\ln^1N$ terms at ${\cal O}(\alpha_s^2)$ that are
proportional to $C_F^2$. Here we present the expansion of
$\bar{M}_N^{\PT,\,O_7 }$, where the $N=1$ moment is exactly 1 by
definition. It is straightforward to convert this expansion to other
normalization conventions (Eqs. (\ref{from_M_Nbar_to_M_N}) and
(\ref{Gamma_NLO}) may be useful).

Converting \eq{N_space} to $x$ space, the $x\to 1$ singular terms in
the $b\longrightarrow X_s\gamma$ distribution (from $O_7$) up to
${\cal O}(\alpha_s^2)$ are:
\begin{eqnarray}
\label{x_space} \left.
\frac{1}{\Gamma_{\tot}}\frac{d\Gamma}{dx}\right\vert_{\PT,\, O_7}&=&
\delta(1-x)+\bigg\{-  \,\frac{\ln (1-x)}{1-x} - {   \frac {7}{4}} \,
\frac{1}{1-x}
-\frac{1}{2}(1-x)^2\nonumber \\
&+&\left(\frac{1}{2}\ln(1-x)+\frac{3}{4}\right)(1-x)-\ln(1-x)+\frac{3}{2}
\bigg\}_{+}\,C_F\,{\frac{\alpha_s^{\MSbar}(m)}{\pi}}
\nonumber \\
&+& \bigg\{{   \frac {C_F ^{2} } {2}\,\frac{\ln^{3} (1-x)}{1-x}}  +
\bigg[ - {   \frac {1}{4}} \,C_F \,N_f + {   \frac {11}{8}} \,C_F
\,C_A  + {   \frac {21}{8}} \,C_F ^{2}\bigg]\,\frac{\ln^{2}
(1-x)}{1-x}
  \nonumber \\&+& \bigg[ - {   \frac {13
}{72}}\,C_F \,N_f  + \left( - {   \frac {\pi ^{2}}{6}}  + {   \frac
{271}{48}} \right)\,C_F ^{2} + \left( {   \frac {95}{144}}  + {
\frac {\pi ^{2}
}{12}} \right)\,C_A \,C_F \bigg]\,\frac{\ln (1-x)} {1-x} \nonumber \\
&+&
 \bigg[\left({   \frac {85}{144}}  - {   \frac {\pi ^{2}}{36}} \right)\,C_F
\,N_f    + \left( - {   \frac {1}{2}} \,\zeta_3 - {   \frac {\pi
^{2}}{6}} + {   \frac {425 }{96}} \right)\,C_F ^{2}  \\ \nonumber
&+& \left({   \frac {1}{4}} \,\zeta_3 - {   \frac {905}{288}}  + {
\frac { 17\,\pi ^{2}}{72}} \right)\,C_A \,C_F
\bigg]\frac{1}{1-x}\bigg\}_{+}
\left({\frac{\alpha_s^{\MSbar}(m)}{\pi}}\right)^{2}+\cdots .
\end{eqnarray}
We checked that the terms that are leading in $\beta_0$ in
\eq{x_space} agree with the large--$x$ expansion of
Ref.~\cite{Ligeti:1999ea}.

In \eq{x_space} $\{\,\,\}_{+}$ stand for the standard ``+''
prescription, i.e.
\begin{equation}
\label{plus_presc}
\left\{p(z)\right\}_{+}=p(z)-\delta(1-z)\,\int_0^1 p(z) dz,
\end{equation}
so, writing
\begin{equation}
\frac{1}{\Gamma_{\tot}}\frac{d\Gamma^{\PT}}{dx}(x)
=\delta(1-x)+\left\{p(x)\right\}_{+}
\end{equation}
the convolution with a smooth test function $f(x)$ (representing
non-perturbative corrections) takes the form:
\begin{eqnarray}
\frac{1}{\Gamma_{\tot}}\frac{d\Gamma}{dx}(x) =f(x)\left[1-\int_0^x
p(z) dz\right] +\int_x^1 dz
p(z)\,\left(\frac{f(x/z)}{z}-f(x)\right),
\end{eqnarray}
where both terms are well defined. In moment space, this is
equivalent to multiplying the moments of the perturbative
distribution $M_N^{\PT}$ by the moments on the test function:
$F_N=\int_0^1 dz z^{N-1} f(z)$.

\section{The $u=\frac12$ renormalon in the pole mass\label{sec:pole_mass_scheme_inv}}

In this Appendix we determine the normalization of the $u=\frac12$
renormalon in the pole mass. This result is used for the calculation
of the pole mass in the Principal--Value prescription in Sec.
\ref{sec:Lambda_PV} and for the comparison with the soft Sudakov
exponent in Sec.~\ref{sec:residue}.

Remarkably, the structure of the leading renormalon ambiguity in the
pole mass is known exactly. Ref.~\cite{Beneke:1994rs} has shown
that, owing to the vanishing of the anomalous dimension of the
$\bar{h}_vh_v$ operator, the linear ultraviolet divergence in the
self energy of a static quark has a very simple structure: it is
just proportional to the QCD scale $\Lambda$ without any logarithmic
corrections. Consequently, the imaginary part associated with the
$u=\frac12$ infrared renormalon in the pole mass has the same
property:
\begin{equation}
\label{Im_Lambda} {\rm Im}\left\{ m \right\} ={\rm
const}\times\Lambda.
\end{equation}
This means that, up to an overall normalization constant, the
large--order behavior of the relation between the pole mass and any
short--distance mass (which is induced by this renormalon) depends
only on the coefficients of the $\beta$ function.

Owing to the dominance of the $u=\frac12$ renormalon, which sets in
already at low orders, the normalization constant can be computed
from the perturbative expansion with accuracy of a few
percent~\cite{Pineda:2001zq,Lee:2003hh,Cvetic:2003wk,Niko}.

\subsection*{The renormalon in the standard Borel representation}

Let us first briefly review the standard analysis (more details can
be found in \cite{Beneke:1998ui}). To make use of~\eq{Im_Lambda} one
first integrates the renormalization--group equation,
\begin{equation}
\frac{da}{d \ln \mu^2}=-a^2\Big[1+\delta
a+\sum_{k=2}^{\infty}\delta_{k} a^k\Big],
\end{equation}
where $a(\mu)=\beta_0\alpha_s(\mu)/\pi$ and
$\delta_k=\beta_k/\beta_0^{k+1}$, writing $\Lambda$ as:
\begin{eqnarray}
\label{Lambda_mu}
\Lambda=\mu\,\exp\left\{-\frac{1}{2a}-\frac{\delta}{2}
\ln(a\delta)\right\} \,\times\,\Bigg[1+\sum_{k=1}^{\infty}\bar{c}_k
a^k \Bigg],
\end{eqnarray}
where $\bar{c}_k$ are specific combinations of the coefficients of
the $\beta$ function; for example:
\begin{eqnarray}
\bar{c}_1&=&\frac12 \,\Big[- {\delta_{2}} + \delta^{2}\Big]
\nonumber \\
\bar{c}_2&=&\frac{1}{8}\,\Big[- 2\,{\delta _{3}} + 4\,{\delta
_{2}}\,\delta  - 2\,\delta ^{3}
 + {\delta _{2}}^{2} - 2\,{\delta _{2}}\,\delta ^{2} + \delta ^{4} \Big]
\nonumber\\ \nonumber \bar{c}_3&=&\frac{1}{48}\,\Big[ - 8\,{\delta
_{4}} + 6\,{\delta _{2}}\,{\delta _{3}} - {\delta _{2}}^{3} +
8\,{\delta _{2}}^{2} + (16\,{\delta _{3}} - 12\,{
\delta _{2}}^{2})\,\delta  \\
&&\hspace*{30pt}+ (3\,{\delta _{2}}^{2} - 6\,{\delta _{3}} -
24\,{\delta _{2}})\,\delta ^{2} + 18\,\delta ^{3}\,{ \delta _{2}} +
(8 - 3\,{\delta _{2}})\,\delta ^{4} - 6\,\delta ^{5}\Big].
\end{eqnarray}
Then, writing the perturbative relation between the pole mass and
the ${\overline {\rm MS}}$ mass in the standard Borel representation
(where the Borel variable $z$ is conjugate to $a$),
\begin{eqnarray}\label{m_over_m_standard_Borel}
{m}/{m_{\MSbar}(m_{\MSbar})}&=&1+\frac{C_F}{\beta_0}\,
\int_0^{\infty}dz B(z)\,{\rm e}^{-z/a},
\end{eqnarray}
the singular part of $B(z)$ near $z=\frac12$ can be explicitly
written using Eqs. (\ref{Im_Lambda}) and (\ref{Lambda_mu}):
\begin{eqnarray}
\label{B_of_z_sing} B(z)={\rm regular}\,\,
+\,\frac{q}{(1-2z)^{1+\frac12\delta}}
\left[1+\sum_{k=1}^{\infty}c_k(1-2z)^k \right],
\end{eqnarray}
where
\begin{equation}
c_k\equiv
\frac{\Gamma(1+\frac{\delta}{2}-k)}{2^k\,\Gamma(1+\frac{\delta}{2})}\,\bar{c}_k.
\end{equation}
To see this one inserts \eq{B_of_z_sing} into
\eq{m_over_m_standard_Borel} and takes the imaginary part of the
integral, getting:
\begin{equation}
\label{Im_ratio_general} {\rm
Im}\Big\{{m}/{m_{\MSbar}(m_{\MSbar})}\Big\}=
\frac{C_F}{\beta_0}\,\frac{q}{\Gamma(1+\frac{\delta}{2})}\frac{\pi}{2}\left(2a\right)^{-\delta/2}{\rm
e}^{-1/(2a)}\,\times \left[1+\sum_{k=1}^{\infty} \bar{c}_k a^k
\right].
\end{equation}
In this calculation one uses the following formula:
\begin{equation}
I_{z_0}(a)\equiv {\rm Im} \int_0^{\infty}dz \frac{{\rm
e}^{-z/a}}{(1-z/z_0)^{1+\kappa}}= \pi
\left(\frac{z_0}{a}\right)^{\kappa}\,\frac{z_0\,{\rm
e}^{-z_0/a}}{\Gamma(1+\kappa)},
\end{equation}
which is derived assuming that $z_0$ has an infinitesimally small
positive imaginary part. \eq{Im_ratio_general} clearly has exactly
the same dependence on the coupling as \eq{Lambda_mu}, so the
requirement of \eq{Im_Lambda} is satisfied.

\begin{figure}[t]
\begin{center}
\epsfig{file=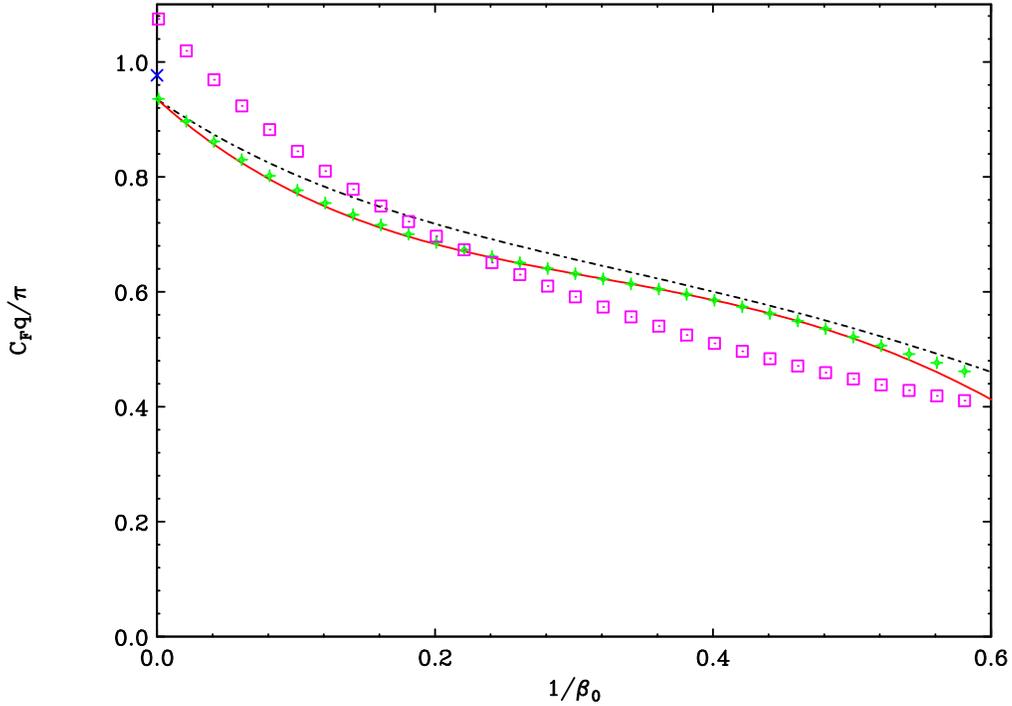,angle=90,width=13.4cm}
\caption{\label{fig:residue} The normalization constant of the
$u=\frac12$ renormalon of the pole mass ($C_F q/\pi$) as a function
of $1/\beta_0$. The {\em exact} result of the large--$\beta_0$ limit
is denoted by a cross. Calculations relying on the perturbative
expansion of $m/m_{\MSbar}(m_{\MSbar})$: standard Borel transform of
the mass ratio in ${\overline {\rm MS}}$ (dotdashes);
scheme--invariant Borel transform of the mass ratio (full line); the
latter optimized using \eq{R_opt_f} (plus signs). Squares represent
the calculation relying on the scheme--invariant Borel transform of
the soft Sudakov exponent and on the cancellation of the ambiguity
(i.e. on \eq{Bs_ren_relation}). This calculation uses an
optimization procedure similar to the one of \eq{R_opt_f}. }
\end{center}
\end{figure}

The pole mass (and its imaginary part) is renormalization--scheme
invariant. Several authors have computed the normalization constant
$q$ using the perturbative expansion of
${m}/{m_{\MSbar}(m_{\MSbar})}$ in the ${\rm {\overline{MS}}}$
scheme~\cite{Pineda:2001zq,Lee:2003hh,Cvetic:2003wk,Niko}. This
perturbative relation is available to ${\cal O}(\alpha_s^3)$ --- see
Eq. (10) in \cite{Melnikov:2000qh} (the ${\overline {\rm MS}}$ mass
anomalous dimension has been computed before
\cite{Vermaseren:1997fq}). Knowing that $z=\frac 12$ is the nearest
singularity to the origin and using \eq{B_of_z_sing} one can extract
the residue by first multiplying~\cite{Lee:1999ws,Cvetic:2001sn} the
available $n$-th order approximation of $B(z)$ by
$(1-2z)^{1+\frac12\delta}$, then expanding in powers of $z$ and
truncating at order $n$ and finally substituting $z=\frac12$. We
performed this calculation and we agree with previous results. The
residue is shown in Fig.~\ref{fig:residue} as a function of
$\beta_0$. For example, we obtained: $C_F q(N_f\longrightarrow
\infty)/\pi\simeq 0.935$ and $C_F q(N_f=4)/\pi\simeq 0.551$. We
estimate the error on this determination as $\sim 3\%$. This
estimate is based on comparison with a similar calculation using the
scheme invariant Borel transform (see below) and on the one point
where the exact value of the normalization constant has been
computed~\cite{Beneke:1994sw}, the large--$\beta_0$ limit. There
$B(z)$ is
\begin{eqnarray}
\left. B(z)\right\vert_{\rm large \beta_0}={\rm
regular}+\frac{3}{2}{\rm
e}^{\frac{5}{3}z}\,\frac{(1-z)\Gamma(z)\Gamma(1-2z)}{\Gamma(3-z)}.
\end{eqnarray}
where the regular terms are related with the renormalization of the
${\overline{\rm MS}}$ mass. Thus the exact value is $C_F q(N_f
\longrightarrow\infty)/\pi=C_F{\rm e}^{\frac56}/\pi\simeq 0.9766$.

To clarify the relation between the normalization constant in the
scheme--invariant Borel transform and in the standard one, let us
first consider the latter in different renormalization schemes.
First, let us stress that having fixed the definition of $\Lambda$
in \eq{Lambda_mu}, the scheme (and thus the value of the coupling at
the scale $\mu$) depend on all the $\beta$--function coefficients
$\beta_k$ with $k\geq 2$. We note that while the sum in
\eq{B_of_z_sing} converges, the signi\-ficance of subleading terms
there strongly depends on the scheme. In particular, we find that in
a scheme $({\rm RS})$ where the $\beta$--function coefficients are
characterized by geometrical progression, ${da_{\RS}}/{d \ln
\mu^2}=-{a_{\RS}^2}/\left({1-\delta a_{\RS}}\right)$, all the
coefficients $c_k$ vanish. On the other hand in the 't Hooft scheme
of \eq{tHooft_coupling}, one obtains
$c_k^{\tHooft}=(\delta/2)^k/k!$, so the singular part of the Borel
function sums up into:
\begin{eqnarray}
\label{B_of_z_sing_tHooft} B(z)={\rm regular}\,\, +\,\frac{q\, {\rm
e}^{\delta \left(\frac12 - z \right)}}{(1-2z)^{1+\frac12\delta}}.
\end{eqnarray}
and
\[
\bar{c}_k^{\tHooft}=\frac{\delta^k\,\Gamma(1+\frac{\delta}{2})}{\Gamma(k+1)\Gamma(1+\frac{\delta}{2}-k)},
\]
so\footnote{Note that using these coefficients in \eq{Lambda_mu} one
recovers the definition of $\Lambda$ in \eq{tHooft_coupling}.} the
imaginary part in \eq{Im_ratio_general} sums up into:
\begin{equation}
\label{Im_ratio_tHooft} {\rm
Im}\Big\{{m}/{m_{\MSbar}(m_{\MSbar})}\Big\}=
\frac{C_F}{\beta_0}\,\frac{q}{\Gamma(1+\frac{\delta}{2})}\frac{\pi}{2}\,\left(\frac{1+\delta
A}{2A}\right)^{\delta/2}\,{\rm e}^{-1/(2A)},
\end{equation}
where the 't Hooft coupling $A$ is evaluated at
$m_{\MSbar}(m_{\MSbar})$. The same result can be obtained by
substituting \eq{B_of_z_sing_tHooft} into
\eq{m_over_m_standard_Borel} (where $a$ is replaced by the 't Hooft
coupling $A$) and evaluating the integral using the relation:
\begin{eqnarray}
&&\int_0^{\infty}du T(u) \left(\Lambda^2/\mu^2\right)^u
\frac{1}{u-u_0}=-\frac{1}{u_0} \int_0^{\infty}dz {\rm e}^{-z/A}
\frac{{\rm
e}^{-z\delta}}{\left(1-\frac{z}{u_0}\right)^{1+u_0\delta}},
\label{ren_int_relation}
\end{eqnarray}
where $T(u)$ and the relation between $A$ and $\ln\mu^2/\Lambda^2$
are given in \eq{tHooft_coupling}. The imaginary part of the l.h.s.
is simply $-\pi$ times the residue at $u=\frac12$. This exercise
also shows that the scheme invariant formulation is most natural for
the problem at hand: the singular part of the corresponding Borel
function is just a simple pole.

\subsection*{The renormalon in the scheme--invariant Borel representation}

Let us turn now to consider the pole--mass renormalon in the
scheme--invariant formulation of the Borel transform. The mass ratio
is expressed as
\begin{equation}
{m}/{m_{\MSbar}(m_{\MSbar})}=1+\frac{C_F}{\beta_0}\int_0^{\infty} du
\,T(u)
B(u)\,\left(\frac{\Lambda^2}{m_{\MSbar}^2(m_{\MSbar})}\right)^u,
\label{m_over_m_SI}
\end{equation}
where as in \eq{tHooft_coupling}, $T(u)$ corresponds to the 't Hooft
scheme.

Taking the imaginary part of \eq{m_over_m_SI} we obtain:
\begin{equation}
\label{scheme_inv_Im_part} {\rm
Im}\Big\{{m}/{m_{\MSbar}(m_{\MSbar})}\Big\}=-\frac{C_F}{\beta_0} R
\times J_{\frac12}\Big(m_{\MSbar}(m_{\MSbar})\Big)
\end{equation}
where the residue is
\begin{equation}
\label{residue_formula} R\equiv\lim_{u \to \frac 12} \left(u-\frac
12\right) B(u)
\end{equation}
and
\begin{eqnarray}
\hspace*{0pt} J_{u_0}(\mu)&\equiv& {\rm Im} \int_0^{\infty} du
\,T(u) \, \left(\frac{\Lambda^2}{\mu^2}\right)^u
\frac{1}{u-u_0-i\epsilon}=\pi
T(u_0)\left(\frac{\Lambda^2}{\mu^2}\right)^{u_0}\!\!\!
=\pi \left(\frac{u_0}{\tilde{A}}\right)^{u_0\delta}\frac{{\rm e}^{-u_0/{\tilde A}}}{\Gamma (1+u_0\delta)}\nonumber \\
\end{eqnarray}
where $\epsilon>0$ is infinitesimally small; the two expressions
were obtained based on the residue theorem
and~\eq{ren_int_relation}, respectively. In the second expression
$\tilde{A}(\mu)$ is defined by
\begin{equation}
\frac{1}{\tilde{A}} = \frac{1}{A}+\delta, \label{tilde_A}
\end{equation}
or, equivalently, $\ln \left(\mu^2/\Lambda^2\right)
={1}/{\tilde{A}}\,-\,\delta\,+\,\delta\ln (\delta \tilde{A})$.
Comparing \eq{scheme_inv_Im_part} with \eq{Im_ratio_tHooft} we find
that
\begin{equation}
q=-2 R\,{\rm e}^{-\delta/2}.
\end{equation}

The approximate calculation of the residue $R$ based on the
perturbative expansion proceeds similarly to the standard Borel
transform: one multiplies by $(\frac12 -u)$ expands and uses the
truncated series to compute the value at $u=\frac12$.  The result is
shown in as a function of $1/\beta_0$ in Fig.~\ref{fig:residue},
where the comparison with the standard Borel transform is available.
One can repeat the calculation by first multiplying the truncated
series by an arbitrary function $f(u)$ (which has a Taylor expansion
at $u=0$ that converges at least for $|u|\leq \frac12$), then taking
the limit, and finally dividing by the exact value $f(\frac 12)$,
\begin{equation}
\label{R_opt_f} R_{\opt}\equiv\frac{1}{f(\frac 12)}\,\lim_{u\to
\frac 12} \left(u-\frac12\right)B(u)f(u).
\end{equation}
We have chosen the set of functions: $f(u)=\exp(\tau u)$ where
$\tau$ is arbitrary. A reliable approximation should be independent
of variations of the function $f(u)$. Therefore, the procedure can
be improved by finding a saddle point with respect to $\tau$. The
result of this optimized calculation of $R$ are shown in
Fig.~\ref{fig:residue}  as well. It turns out that the result is
quite stable so the difference between the optimized calculation and
the simple one is small, for example, for $C_F q(N_f=4)/\pi$ we get
$0.5355$ and $0.5363$ in the two, respectively.

In conclusion, the normalization of the $u=\frac12$ renormalon in
the pole mass can be determined from the perturbative relation with
the ${\overline{\rm MS}}$ mass within a few percent. The difference
between the determination using the scheme--invariant Borel
representation and the one using the standard Borel representation
is somewhat larger than the variation of the results within each of
the two approaches.

\section{NLO results for $b\longrightarrow s\gamma g$ and the matching procedure\label{sec:phi}}

We begin with the well-known expressions for the $b\longrightarrow
s\gamma g$ contributions~\cite{Chetyrkin:1996vx}:
\begin{eqnarray}
\label{phi_def} \phi_{22}(\Delta) &=& \frac{4 z}{9} \left[ \Delta
\int_0^{(1-\Delta)/z} dt \; (1-zt) \left| \frac{G(t)}{t} +
\frac{1}{2} \right|^2 \;+\; \int_{(1-\Delta)/z}^{1/z} dt \; (1-zt)^2
\left| \frac{G(t)}{t} + \frac{1}{2} \right|^2
\right], \label{phi22}\nonumber\\
\phi_{27}(\Delta) &=& -\frac{2 z^2}{3} \left[ \Delta
\int_0^{(1-\Delta)/z} dt\;    {\rm Re} \left( G(t) + \frac{t}{2}
\right) \;+\; \int_{(1-\Delta)/z}^{1/z} dt\;(1-zt) {\rm Re} \left(
G(t) + \frac{t}{2} \right) \right],
\label{phi27}\nonumber\\
\phi_{77}(\Delta) &=& \frac{5}{2} \Delta + \frac{1}{4} \Delta^2 -
\frac{1}{6} \Delta^3
+ \frac{1}{4} \Delta ( \Delta - 4 ) \ln \Delta, \label{phi77}\\
\phi_{78}(\Delta) &=& \frac{2}{3} \left[ {\rm Li}_2(1-\Delta) -
\frac{\pi^2}{6} - \Delta
\ln \Delta + \frac{9}{4} \Delta - \frac{1}{4} \Delta^2 + \frac{1}{12} \Delta^3 \right],\nonumber\\%
\phi_{88}(\Delta) &=& \frac{1}{36} \left\{ - 2 \ln \frac{m_b}{m_s}
                     \left[ \Delta^2 + 2 \Delta + 4 \ln(1-\Delta) \right]
\right. \nonumber \\ && \hspace{1cm} \left. +4{\rm Li}_2(1-\Delta)
-\frac{2\pi^2}{3} -\Delta(2+\Delta)\ln\Delta + 8\ln(1-\Delta)
-\frac{2}{3} \Delta^3 + 3 \Delta^2 + 7 \Delta \right\},
\label{phi88}\nonumber\\ \nonumber
\phi_{11}&=&\frac{1}{36}\phi_{22},~~~~~~
\phi_{12}=-\frac{1}{3}\phi_{22},~~~~~~
\phi_{17}=-\frac{1}{6}\phi_{27},~~~~~~
\phi_{18}=\frac{1}{18}\phi_{27},~~~~~~
\phi_{28}=-\frac{1}{3}\phi_{27},
\end{eqnarray}
where
\begin{eqnarray*}
G(t) = \left\{ \begin{array}{cc}
- 2 \arctan^2 \sqrt{ \frac{t}{4-t}}, & \mbox{ for $t < 4$} \vspace{0.3cm} \\
-\frac{\pi^2}{2} + 2 \ln^2\frac{\sqrt{t} + \sqrt{t-4}}{2} - 2 i \pi
\ln\frac{\sqrt{t} + \sqrt{t-4}}{2}, & \mbox{ for $t \geq 4$}.
\end{array} \right.
\end{eqnarray*}
and where, following \cite{Gambino:2001ew}, we use $z\equiv
m_c^{\MSbar}(\mu)/m_b\simeq 0.22\pm 0.04$.

As explained in Sec.~\ref{sec:matching} our matching procedure
involves splitting some of these functions,
$\phi_{ij}(\Delta)=\eta_{ij}(\Delta)+\xi_{ij}(\Delta)$, such that
$\eta_{ij}(\Delta)$ dominates the small--$\Delta$ limit. We require:
$\xi_{ij}(\Delta)={\cal O}(\Delta^2)$.

For $\phi_{77}$, which is conveniently transformed to moment space
we simply define $\eta_{77}(\Delta)=\phi_{77}(\Delta)$ and
$\xi_{77}(\Delta)=0$. It then follows from \eq{mu_ij_def} that
\begin{eqnarray}
\label{mu77}
\mu_{77}(N)&=&\frac12\left[\left(\Psi(N)+\gamma_E\right)\,\left(\frac{1}{1+N}+\frac{1}{N}\right)
-\frac{1}{2}\frac{1}{N+1}-\frac{1}{N+2}+\frac{9}{2N}
+\frac{1}{(N+1)^2}+\frac{1}{N^2}\right].\nonumber
\end{eqnarray}
Note that in \eq{M_N_full} the full $O_7$ contribution is reproduced
by $f(N)+\mu_{77}(N)$ plus the single and double logs from the
expansion of ${\rm Sud}(m,N)$ (these are the ${\cal O}(\alpha_s)$
terms in \eq{N_space}). Note also that $f(N)+\mu_{77}(N)$ differs
from $C_N^{O_7}$ of~\eq{matching} by the constant term
$\frac{31}{12}$, the reason being that the latter corresponds to the
moments $\bar{M}^{\PT,\, O_7}_N$ (of the normalized rate) while the
former corresponds to the moments $M^{\PT,\, O_7}_N$; indeed
$\mu_{77}(N=1)=\phi_{77}(\Delta=1)=\frac{31}{12}$ so
\begin{eqnarray}
\label{from_M_Nbar_to_M_N} M_{N}^{O_7}=\bar{M}_N^{O_7}\times
\left(1+\frac{C_F\alpha_s}{\pi} \,\frac{31}{12}+{\cal
O}(\alpha_s^2)\right).
\end{eqnarray}

Similarly, for $\phi_{78}$, we define:
$\eta_{78}(\Delta)=\phi_{78}(\Delta)$ and $\xi_{78}(\Delta)=0$, so
\begin{eqnarray}
\label{mu78}
\mu_{78}(N)&=&\frac{2}{3}\left[-\frac{1}{(N-1)N}(\Psi(N)+\gamma_E)
+\frac{1}{N^2}+\frac{1}{4}\frac{1}{N+2} +\frac{1}{N}\right].
\end{eqnarray}

As explained in the text, the treatment of the other contributions
involves splitting the real--emission terms into two. For all the
contributions associated with the operators $O_1$ and $O_2$ and
their interference with $O_7$ and $O_8$ (this includes $\phi_{22}$,
$\phi_{27}$, $\phi_{11}$, $\phi_{12}$ $\phi_{17}$, $\phi_{18}$ and
$\phi_{28}$) we define
\[
\dot{\phi}_{ij}\equiv
\left.\frac{d\phi_{ij}(\Delta)}{d\Delta}\right\vert_{\Delta=0}
\]
and write:
\begin{eqnarray*}
\eta_{ij}(\Delta)&=&\dot{\phi}_{ij}\,\Delta, \\
\xi_{ij}(\Delta)&=&\phi_{ij}(\Delta)-\dot{\phi}_{ij}\,\Delta.
\end{eqnarray*}
Since the derivative of $\eta_{ij}(\Delta)$ is a constant, the
moments in \eq{mu_ij_def} are pure $1/N$ terms:
\begin{equation}
\mu_{ij}(N)=\frac{\dot{\phi}_{ij}}{N}. \label{muij}
\end{equation}

Finally, the contribution of $\phi_{88}$ it taken into account by
extracting the leading term at $\Delta\longrightarrow 0$ (which
makes the corresponding contribution to the differential rate
singular) and incorporating it in moment space:
\begin{eqnarray*}
\eta_{88}(\Delta)&=&\frac{1}{9}\left[\ln\frac{m_b}{m_s}-\frac{5}{4}+\frac{1}{2}\ln \Delta\right]\,\Delta,\\
\xi_{88}(\Delta)&=&\phi_{88}(\Delta)-\eta_{88}(\Delta),
\end{eqnarray*}
so
\begin{equation}
\mu_{88}(N)=\frac{1}{3N}\left[\frac{1}{3}\ln
\frac{m_b}{m_s}-\frac14-\frac16(\Psi(N)+\gamma_E)-\frac16\frac{1}{N}\right].
\label{mu88}
\end{equation}
As follows from their definition all $\mu_{ij}(N)$ vanish at large
$N$.

Finally, let us summarize the result for the first moment which
enters the expression for the total rate (\eq{Gamma_N_NLO} with
$\Delta\longrightarrow 1$):
\begin{equation}
\label{M_N1} M_{N=1}^{\PT}=
1+\frac{C_F\alpha_s(m_b)}{\pi} \!\!\sum_{\scriptsize \begin{array}{c}{i,j=1..8}\\
{ i\leq j}\end{array} }c_{ij} \,\mu_{ij}(N=1)
\end{equation}
with
\begin{eqnarray}
\label{mu_N1}
\mu_{77}(N=1)&=&\frac{31}{12}\nonumber \\
\mu_{78}(N=1)&=&\frac{25}{18}-\frac{\pi^2}{9}\nonumber \\
\mu_{88}(N=1)&=&\frac{1}{9}\left(\ln\frac{m_b}{m_s}-\frac54\right)\nonumber \\
\mu_{22}(N=1)&=&\frac{4z}{9}\int_0^{1/z}dt\, (1-zt)\left\vert\frac{G(t)}{t}-\frac12\right\vert^2\nonumber \\
\mu_{27}(N=1)&=&-\frac{2z^2}{3}\int_0^{1/z}dt\, {\rm
Re}\left(G(t)+\frac{t}{2}\right);
\end{eqnarray}
The remaining $\mu_{ij}(N=1)$ values can be obtained from the above
using the relations in the last line in \eq{phi_def}.

\end{document}